\documentclass[a4paper,fleqn,usenatbib]{mnras}

\usepackage{mn2e-breakabs}

\usepackage[T1]{fontenc}
\usepackage{ae,aecompl}

\usepackage{graphicx}	
\usepackage{amsmath}	
\usepackage{amssymb}	
\usepackage{mathrsfs}   
\usepackage[english]{babel}
\usepackage{xspace} 
\usepackage{units}
\usepackage{wasysym} 

\usepackage{txfonts}

\usepackage[dvipsnames]{xcolor} 

\usepackage{cosmodefs}

\renewcommand*{\sciexp}[1]{\times 10^{#1}} 
\newcommand{\mathpi}{\mathrm{\pi}}

\newcommand*{\panel}[1]{\emph{#1 panel:}}

\newcommand*{\LambdaCDM}[1][\xspace]{$\Lambda$CDM#1}
\newcommand*{\Planck}[1][\xspace]{\emph{Planck}#1}
\newcommand*{\WMAP}[1][\xspace]{\emph{WMAP}#1}
\newcommand*{\Euclid}[1][\xspace]{\emph{Euclid}#1}
\newcommand*{\Nbody}[1][\xspace]{$N$-body#1}
\newcommand*{\Minerva}[1][\xspace]{\textsc{Minerva}#1}
\newcommand*{\QPM}[1][\xspace]{\textsc{QPM}#1}
\newcommand*{\Patchy}[1][\xspace]{\textsc{Patchy}#1}
\newcommand*{\MDPatchy}[1][\xspace]{\textsc{MD-Patchy}#1}
\newcommand*{\code}[1]{\textsc{#1}}

\newcommand*{\Inprep}{\emph{in prep{.}}}

\newcommand*{\deltaK}{\delta^\mathrm{K}} 
\newcommand*{\Lp}[1][\ell]{\mathcal{L}_{#1}}

\newcommand*{\RA}{\alpha}
\newcommand*{\DEC}{\delta}
\newcommand*{\COMP}{C}

\newcommand*{\wo}{w_0}
\newcommand*{\wa}{w_a}
\newcommand*{\OmegaX}[1]{\Omega_{#1}}
\newcommand*{\OK}{\OmegaX{K}}
\newcommand*{\Ol}{\OmegaX{\Lambda}}
\newcommand*{\Om}[1][M]{\OmegaX{\mathrm{#1}}}

\newcommand*{\zeff}{z_\mathrm{eff}}
\newcommand*{\DV}[1][V]{D_\mathrm{#1}}

\newcommand*{\Hz}[1][\zeff]{H(#1)}
\newcommand*{\DAz}[1][\zeff]{\DV[A](#1)}
\newcommand*{\DMz}[1][\zeff]{\DV[M](#1)}
\newcommand*{\DVz}[1][\zeff]{\DV(#1)}
\newcommand*{\FAPz}[1][\zeff]{F_\mathrm{AP}(#1)}

\newcommand*{\zd}{z_\mathrm{d}}
\newcommand*{\rs}{r_\mathrm{s}}
\newcommand*{\rd}{r_\mathrm{d}}

\newcommand*{\wabs}[1][\cdot]{w_{#1}}
\newcommand*{\wabsrm}[1]{\wabs[\mathrm{#1}]}

\newcommand*{\wsys}{\wabsrm{sys}}
\newcommand*{\wfc}{\wabsrm{fc}}
\newcommand*{\wrf}{\wabsrm{rf}}
\newcommand*{\wc}{\wabsrm{c}}
\newcommand*{\wtot}{\wabsrm{tot}}
\newcommand*{\PFKP}{P_w}
\newcommand*{\wFKP}{w_\mathrm{FKP}}
\newcommand*{\ftp}{f_\mathrm{tp}}

\newcommand*{\alphar}{\alpha_\mathrm{r}}
\newcommand*{\nbar}{\bar n}
\newcommand*{\ngal}{n_\mathrm{g}}
\newcommand*{\nexp}{n_\mathrm{exp}}
\newcommand*{\nrnd}{n_\mathrm{r}}
\newcommand*{\Ngal}{N_\mathrm{gal}}

\newcommand*{\Nrnd}{N_\mathrm{rnd}}
\newcommand*{\Anorm}{A}
\newcommand*{\Pshot}{S}

\newcommand*{\gammaEm}{\gamma_3^-}
\newcommand*{\avir}{a_\mathrm{vir}}
\newcommand*{\SN}{N}

\newcommand*{\fsig}{f \sigma_8}
\newcommand*{\Fsig}[1][\zeff]{\fsig(#1)}

\newcommand*{\Nmocks}{N_\mathrm{m}}
\newcommand*{\Nparams}{N_\mathrm{p}}
\newcommand*{\Nbins}{N_\mathrm{b}}

\newcommand*{\CA}{A_\mathrm{M}}
\newcommand*{\CB}{B_\mathrm{M}}
\newcommand*{\CD}{D}
\newcommand*{\CM}{M}

\newcommand*{\Pobs}[1][3\mathrm{w}]{\V P_{#1}}
\newcommand*{\PAobs}[1][3\mathrm{w},n]{P_{#1}}

\newcommand*{\PAtheo}[1][3\mathrm{w},n]{\tilde P_{#1}}
\newcommand*{\Like}{\mathscr{L}}
\newcommand*{\precmat}{{\boldsymbol \Psi}}
\newcommand*{\deltaP}{\Delta \V P}
\newcommand*{\Ppred}{\Vn P_{3\mathrm{w}}}
\newcommand*{\PApred}[1][3\mathrm{w},n]{\hat P_{#1}}

\newcommand*{\qpara}{q_\parallel}
\newcommand*{\qperp}{q_\perp}
\newcommand*{\apara}{\alpha_\parallel}
\newcommand*{\aperp}{\alpha_\perp}

\newcommand*{\kmin}{k_\mathrm{min}}
\newcommand*{\kmax}{k_\mathrm{max}}

\newcommand*{\ns}{n_\mathrm{s}}
\newcommand*{\Neff}{N_\mathrm{eff}}

\newif\ifhighlight 

\ifhighlight

\newcommand*{\changed}[1]{{\bf #1}}

\else

\newcommand*{\changed}[1]{{#1}}

\fi

\title[BOSS DR12 Fourier space wedges]{The clustering of galaxies in the completed SDSS-III Baryon Oscillation Spectroscopic Survey: Cosmological implications of the Fourier space wedges of the final sample}

\author[J. N. Grieb et al.]{Jan Niklas Grieb,$^{1,2}$
Ariel G. S{\'a}nchez,$^{2}$\thanks{E-mail: arielsan@mpe.mpg.de (corresponding author)}
Salvador Salazar-Albornoz,$^{1,2}$\newauthor
Rom{\'a}n Scoccimarro,$^{3}$
Mart{\'\i}n Crocce,$^{4}$
Claudio Dalla Vecchia,$^{5,6}$\newauthor
Francesco Montesano,$^{2}$
H{\'e}ctor Gil-Mar{\'\i}n,$^{7,8,9}$
Ashley J. Ross,$^{10,7}$\newauthor
Florian Beutler,$^{11,7}$
Sergio Rodr{\'i}guez-Torres,$^{12,13,14}$
Chia-Hsun Chuang,$^{14,15}$\newauthor
Francisco Prada,$^{16,14,12}$
Francisco-Shu Kitaura,$^{15}$
Antonio J. Cuesta,$^{17}$\newauthor
Daniel J. Eisenstein,$^{18}$
Will J. Percival,$^{7}$
Mariana Vargas-Maga{\~n}a,$^{19}$\newauthor
Jeremy L. Tinker,$^{3}$
Rita Tojeiro,$^{20}$
Joel R. Brownstein,$^{21}$
Claudia Maraston,$^{7}$\newauthor
Robert C. Nichol,$^{7}$
Matthew D. Olmstead,$^{22}$
Lado Samushia,$^{7,23,24}$\newauthor
Hee-Jong Seo,$^{25}$
Alina Streblyanska,$^{5}$
and Gong-bo Zhao$^{7,26}$\\
$^{1}$Universit\"ats-Sternwarte M\"unchen, Ludwig-Maximilians-Universit\"at M\"unchen, Scheinerstra\ss{}e 1, 81679 M\"unchen, Germany\\
$^{2}$Max-Planck-Institut f\"ur extraterrestrische Physik, Postfach 1312, Giessenbachstr., 85741 Garching, Germany\\
$^{3}$Center for Cosmology and Particle Physics, Department of Physics, New York University, New York, NY 10003\\
$^{4}$Institut de Ci\`encies de l'Espai, IEEC-CSIC, Campus UAB, F.~de Ci\`encies, Torre C5 par-2, Barcelona 08193, Spain\\
$^{5}$Instituto de Astrof\'\i{}sica de Canarias, C/ V\'\i{}a L\'actea s/n, 38205 La Laguna, Tenerife, Spain\\
$^{6}$Departamento de Astrof\'\i{}sica, Universidad de La Laguna, Av.~del Astrof\'\i{}sico Francisco S\'anchez s/n, 38206 La Laguna, Tenerife, Spain\\
$^{7}$Institute of Cosmology \& Gravitation, University of Portsmouth, Dennis Sciama Building, Portsmouth PO1 3FX, UK\\
$^{8}$Sorbonne Universit\'es, Institut Lagrange de Paris (ILP), 98 bis Boulevard Arago, 75014 Paris, France\\
$^{9}$Laboratoire de Physique Nucl\'eaire et de Hautes Energies, Universit\'e Pierre et Marie Curie, 4 Place Jussieu, 75005 Paris, France\\
$^{10}$Center for Cosmology and AstroParticle Physics, The Ohio State University, Columbus, OH 43210, USA\\
$^{11}$Lawrence Berkeley National Lab, 1 Cyclotron Rd, Berkeley, CA 94720, USA\\
$^{12}$Instituto de F\'{\i}sica Te\'orica, (UAM/CSIC), Universidad Aut\'onoma de Madrid,  Cantoblanco, E-28049 Madrid, Spain\\
$^{13}$Campus of International Excellence UAM+CSIC, Cantoblanco, E-28049 Madrid, Spain\\
$^{14}$Departamento de F\'{\i}sica Te\'orica, Universidad Aut\'onoma de Madrid, Cantoblanco, 28049, Madrid, Spain\\
$^{15}$Leibniz-Institut f{\"u}r Astrophysik (AIP), An der Sternwarte 16, D-14482 Potsdam, Germany\\
$^{16}$Instituto de Astrof\'{\i}sica de Andaluc\'{\i}a (CSIC), Glorieta de la Astronom\'{\i}a, E-18080 Granada, Spain\\
$^{17}$Institut de Ci\`encies del Cosmos (ICCUB), Universitat de Barcelona (IEEC-UB), Mart\'\i i Franqu\`es 1, E08028 Barcelona, Spain\\
$^{18}$Harvard-Smithsonian Center for Astrophysics, 60 Garden St., Cambridge, MA 02138, USA\\
$^{19}$Instituto de F{\'\i}sica, UNAM, P.O. Box 20-364, 01000 M\'exico D.F., Mexico\\
$^{20}$School of Physics and Astronomy, University of St Andrews, North Haugh, St Andrews KY16 9SS, UK\\
$^{21}$Department of Physics and Astronomy, University of Utah, 115 S 1400 E, Salt Lake City, UT 84112, USA\\
$^{22}$Department of Chemistry and Physics, King's College, 133 North River St, Wilkes Barre, PA 18711, USA\\
$^{23}$Kansas State University, Manhattan KS 66506, USA\\
$^{24}$National Abastumani Astrophysical Observatory, Ilia State University, 2A Kazbegi Ave., GE-1060 Tbilisi, Georgia\\
$^{25}$Department of Physics and Astronomy, Ohio University, 251B Clippinger Labs, Athens, OH 45701, USA\\
$^{26}$National Astronomy Observatories, Chinese Academy of Science, Beijing, 100012, P.R.China}

\date{Accepted XXX. Received YYY; in original form ZZZ}

\pubyear{2016}

\begin{document}
\label{firstpage}
\pagerange{\pageref{firstpage}--\pageref{lastpage}}
\maketitle

\begin{abstract}
We extract cosmological information from the anisotropic power spectrum measurements from the recently completed Baryon Oscillation Spectroscopic Survey (BOSS), extending the concept of clustering wedges to Fourier space.
Making use of new FFT-based estimators, we measure the power spectrum clustering wedges of the BOSS sample by filtering out the information of Legendre multipoles $\ell>4$.
Our modelling of these measurements is based on novel approaches to describe non-linear evolution, bias, and redshift-space distortions, which we test using synthetic catalogues based on large-volume \Nbody simulations.
We are able to include smaller scales than in previous analyses, resulting in tighter cosmological constraints.
Using three overlapping redshift bins, we measure the angular diameter distance, the Hubble parameter, and the cosmic growth rate, and explore the cosmological implications of our full shape clustering measurements in combination with CMB and SN Ia data.
Assuming a \LambdaCDM cosmology, we constrain the matter density to $\Om = 0.311_{-0.010}^{+0.009}$  and the Hubble parameter to $H_0 = 67.6_{-0.6}^{+0.7} \Unit{km \, s^{-1} \, Mpc^{-1}}$, at a confidence level (CL) of 68 per cent.
We also allow for non-standard dark energy models and modifications of the growth rate, finding good agreement with the \LambdaCDM paradigm.
For example, we constrain the equation-of-state parameter to $w = -1.019_{-0.039}^{+0.048}$.
This paper is part of a set that analyses the final galaxy clustering dataset from BOSS.
The measurements and likelihoods presented here are combined with others in \citet{Alam:2016hwk} to produce the final cosmological constraints from BOSS.
\end{abstract}

\begin{keywords}
cosmology: observations -- cosmological parameters -- dark energy -- large-scale structure of Universe
\end{keywords}



\section{Introduction}
\label{sec:intro}

Together with observations of the cosmic microwave background (CMB) \changed{and} type-Ia supernova (SN) samples, the analysis of the large-scale structure (LSS) of the Universe based on galaxy redshift surveys has been a prolific source of cosmological information over the past few decades \citep{Davis:1983,Maddox:1990,Tegmark:2004,Cole:2005sx,Eisenstein:2005su,Anderson2012,Anderson:2013oza,Anderson:2013zyy}.
\changed{These datasets have helped to stablish the \LambdaCDM model as the current standard 
cosmological paradigm, and to determine the values of its basic set of parameters with high precision.}
The \LambdaCDM model assumes that the energy density of the observable universe is dominated by (pressureless) cold dark matter (CDM) and  \changed{a mysterious `Dark Energy' (DE) component that drives the accelerated expansion of the late-time universe, which can be described by a cosmological constant $\Lambda$ or vacuum energy.}
Observations of the clustering of galaxies can shed light onto the underlying physical nature of this energy component by probing the growth of structure and the expansion history of the Universe.
\changed{Thus, important} recent and ongoing spectroscopic galaxy-redshift surveys, such as the Baryon Oscillation Spectroscopic Survey \citep[BOSS;][]{Dawson:2012va} and its extension eBOSS \citep{Dawson:2015wdb} \changed{are} very valuable probes of the late-time evolution of the Universe.

\changed{A major goal of galaxy surveys is to obtain precise measurements of the expansion history of the Universe by means of a feature imprinted into the two-point clustering statistics, the baryonic acoustic oscillations \citep[BAO; for a review see \eg][]{Bassett:2009mm}.
The BAO are relics of pressure waves that propagated through the photon-baryon plasma prior to recombination and froze in at the time of last scattering.
The interaction between dark and baryonic matter after recombination resulted in a signal of enhanced correlation of density peaks separated by a well defined physical scale, the sound horizon at the drag redshift.}
This \changed{scale can be used} as a robust standard ruler for measurements of cosmic distances \citep{Eisenstein:2004an,Seo:2005ys,Angulo:2007fw,Sanchez:2008iw}.
\changed{The first detections of the BAO feature \citep{Eisenstein:2005su,Cole:2005sx} relied on angle-averaged clustering statistics.
However, separate measurements of the BAO signal along the directions parallel and perpendicular to the line of sight (LOS) can be used to obtain separate constraints on the Hubble parameter $H(z)$ at and the angular diameter distance $D_{\rm A}(z)$ to the mean redshift of the survey by means of the Alcock-Paczynski \citep[AP;][]{AP:1979} test.
In this way, anisotropic clustering measurements can break the degeneracy obtained from angle-averaged quantities, which are only sensitive to the average distance $D_{\rm V}(z) \propto(D_{\rm A}(z)^2/H(z))^{1/3}$ \citep*{Hu:2003ti,Wagner:2007in,Shoji:2008xn}.}

\changed{The dominant source of anisotropy of the measured clustering signal are the redshift-space distortions (RSD), which are due to the impact of the LOS component of the peculiar velocities of the galaxies on the observed galaxy redshifts. 
The pattern of RSD provides} additional cosmological information beyond that of the BAO signal.
\changed{As, to linear order, peculiar velocities are related to the infall of matter} into 
gravitational potential wells \citep{Kaiser:1987qv}, the RSD are a probe of the growth of structure.
As modifications to general relativity (GR) can change \changed{the growth rate of density fluctuations}, RSD can be used to constrain the theory of gravity \citep[\eg,][]{Guzzo:2008}.
However, the galaxy velocity field is highly non-linear even on large scales so that \changed{a} detailed modelling is required \citep[\eg,][]{Scoccimarro:2004tg}.

One way to \changed{characterize the anisotropies in the clustering of galaxies} is to use the concept of clustering wedges \changed{introduced by \citet{Kazin:2011xt}, which} correspond to the average the correlation function over wide bins of the LOS parameter, $\mu$, \changed{defined as the cosine of the angle between the total separation vector between two galaxies and the LOS direction.}
Anisotropic BAO distance measurements \changed{obtained using clustering wedges were first presented} in \cite{Kazin:2013rxa} as part of the BOSS DR9 CMASS analysis \citep{Anderson:2013oza}, \changed{while \citet{Sanchez:2013uxa,Sanchez:2013tga} performed an analysis of the full shape of the wedges measured from the BOSS DR9 and DR11 galaxy catalogues, respectively.} 
An alternative \changed{tool to wedges are the Legendre multipole moments} of the two-point statistics \citep{Padmanabhan:2008ag}.
The multipoles of the correlation function measured from BOSS DR11 galaxy catalogues were used in several recent galaxy clustering analyses \citep[\eg;][]{Samushia:2013yga,Alam:2015qta,Reid:2014iaa}.
In Fourier space, the first anisotropic clustering \changed{studies} \citep[\eg,][]{Blake:2011rj,Beutler:2013yhm} were \changed{performed} on measurements of the Legendre multipoles of the power spectrum \changed{obtained by means of the Yamamoto-Blake estimator \citep{Yamamoto:2005dz,Blake:2011rj}.
In this work we extend the concept of clustering wedges to Fourier space and adapt the Yamamoto-Blake estimator to provide a measurement of these statistics.}

\changed{We perform an analysis of the full-shape of the Fourier-space clustering wedges measured from the final BOSS galaxy samples \citep{Reid:2015gra}, corresponding to SDSS data release 12 \citep[DR12;][]{Alam:2015mbd}.}
In order to make use of new estimators based on fast Fourier transforms \citep[FFT;][]{Bianchi:2015oia, Scoccimarro:2015bla}, we measure the power spectrum clustering wedges of the BOSS sample by filtering out the information of Legendre multipoles $\ell>4$.
\changed{Exploiting the signature of BAO and RSD in these measurements, we} derive distance and \changed{growth-of-structure} constraints.
We also explore the implications of the full shape of our measurements on the parameters of the standard \LambdaCDM model, as well as its most important extensions, making use also of complementary cosmological information from CMB and \changed{SN} samples.

This work is part of a series of papers that analyse the clustering properties of the final BOSS sample.
\nocite{Sanchez:2016a}
Besides the approach of this work, the analogous full-shape analysis using configuration space wedges is discussed in \citet{Sanchez:2016b}.
\changed{Complementary RSD measurements using Fourier and configuration space multipoles are presented in \citet{Beutler:2016arn} and \citet{Satpathy:2016tct}, respectively.
Tinker et al. (\Inprep) compares the performance of the different methodologies to extract cosmological information from the full shape of anisotropic clustering measurements.
Anisotropic} BAO distance measurements are presented in \citet{Ross:2016gvb} and \citet{Beutler:2016ixs} for configuration and Fourier space, respectively, making use of the linear density-field reconstruction technique \citep{Eisenstein:2006nk,Cuesta:2015mqa}.
\changed{\citet{VargasMagana:2016}} investigates the potential sources of theoretical systematics in the anisotropic BAO analysis for the final BOSS galaxy BAO analysis in configuration space.
\changed{All final BOSS} analyses are summarised \changed{in \citet{Alam:2016hwk}, where they are combined} into a set of consensus measurements following the methodology described in \citet{Sanchez:2016a}. 
A different approach is followed in \citet{Salazar-Albornoz:2016psd}, who perform a tomographic analysis by means of angular correlation functions in thin redshift shells. 

This paper is organised as follows:
Section~\ref{sec:boss} describes the \changed{final BOSS DR12 galaxy catalogue and the optimal 
estimator we use to measure the Fourier-space clustering wedges of this sample, which are the basis for our cosmological constraints.}
\changed{This section describes also the methodology we follow to estimate the covariance matrix of our measurements (section~\ref{sec:covariance_matrix}) and to account for the window function of the survey (section~\ref{sec:win_func}).}
The model for the Fourier space wedges is discussed in section~\ref{sec:model} \changed{where we} describe the recipe for the non-linear gravitational dynamics, galaxy bias and RSD and analyse the performance of the model using \changed{\Nbody} simulations and synthetic catalogues mimicking the clustering properties of the BOSS galaxy sample.
Anisotropic BAO and RSD constraints derived from the full-shape analysis of the DR12 clustering wedges analysis in Fourier space \changed{are} described in section~\ref{sec:BAO_and_RSD_measurements}.
In section~\ref{sec:cosmological_implications}, we present the cosmological results from combining the measurements of the Fourier-space wedges with complementary data sets and infer cosmological \changed{constraints for different parameter spaces}.
Finally, in section~\ref{sec:conclusions} we conclude our analysis with a summary and discussion of the results.

\section{Clustering measurements from the Baryon Oscillation Spectroscopic Survey}
\label{sec:boss}

\subsection{The final DR12 sample of BOSS}
\label{sec:boss_dr12_combined}

This work is based on the final galaxy catalogue of the BOSS program \citep{Dawson:2012va}, which is one of the four spectroscopic surveys of the third iteration of the Sloan Digital Sky Survey program \citep[SDSS-III;][]{Eisenstein:2011sa}.
The catalogue is constructed from the spectra of ca{.}~1{.}5 million galaxies \changed{from the SDSS data release 12} \citep[DR12;][]{Alam:2015mbd}.
The galaxies were selected from multi-colour SDSS imaging \citep{Fukugita:1996qt,Smith:2002pca,Doi:2010rf} that 
was obtained with a drift-scanning mosaic CCD camera \citep{Gunn:1998vh}.
The spectra were measured using the \changed{BOSS} multi-fibre spectrograph \citep{Smee:2012wd}.
The camera and spectrographs are installed on a dedicated 2.5-meter wide-field telescope \changed{at} the Apache Point Observatory \citep{Gunn:2006tw}.
The spectral classification and redshift fitting pipeline was specially written for the BOSS program \citep{Bolton:2012hz}.
The survey consists of two large patches in the sky that are located in the \changed{northern and southern galactic caps} (or NGC and SGC, for short).
The final footprint of the spectroscopic survey covers ca{.}~10{,}400 square degrees\changed{with a mean sector completeness of $0.98$ \citep{Reid:2015gra}}, corresponding to an increase in effective area of ca{.}~10 per cent over the internal DR11 release.

Previous works \changed{based on BOSS data} have used two galaxy catalogues, LOWZ and CMASS.
The LOWZ catalogue ($0.15 \le z \le 0.43$) extends the selection of the luminous red galaxy (LRG) population of SDSS-II to higher redshifts and to fainter galaxies in order to achieve a higher number density up to $z \leq 0.43$.
The CMASS sample ($0.43 \le z \le 0.7$) is \changed{nearly complete down to a stellar mass of
$M\simeq10^{11.3}\,{\rm M}_{\odot}$ for $z>0.45$ \citep{Maraston:2012jf}}.
The selection criteria for both samples were chosen to achieve a homogeneous comoving number density of $\nbar \approx 3 \sciexp{-4} \, h^3 \, \unit{Mpc}^{-3}$ \citep{Dawson:2012va} \changed{in the redshift range $0.15<z<0.7$}.
The galaxies of both samples are a highly biased \changed{tracers of the matter density field with a linear bias parameter of} $\sim 2.0$ \citep{Nuza:2012mw}, which is ideal for clustering analysis as the power spectrum can be measured with a high signal-to-noise ratio.

\changed{The DR12 LOWZ and CMASS samples have previously been analysed separately \citep[\eg,][]{Cuesta:2015mqa,Gil-Marin:2015sqa,Gil-Marin:2015nqa,Chuang:2013wga}.
In this work we use the joint information of these samples by combining them into a final BOSS `combined sample' as described in \citet{Reid:2015gra}, covering the redshift range $0.2 \le z \le 0.75$. 
The BOSS combined sample includes 1000 $\unit{deg}^2$ of additional `early' data based on slightly different selection criteria that have been included in the low-redshift part of the catalogue, leading to a final effective volume of $V_\mathrm{eff} = 2.4 \; h^{-3} \, \unit{Gpc}^3$.
These data are} publicly available at the SDSS-III \changed{web site}.\footnote{\changed{\url{https://www.sdss3.org/science/boss_publications.php}}}

\changed{The observed galaxy number density is affected by incompleteness that originates in the targeting and observing strategies of the survey.
In order to account for such systematics, different weights are assigned to the galaxies in the catalogue.
A source of incompleteness are the so-called fibre collisions, which are caused by the fact that due to the physical size of the fibres it is not possible to simultaneously take the spectra of two target galaxies that are separated by less than 62'' in the sky.
Thus, missing targets are accounted for by a weight $\wfc \ge 1$ that is applied to observed neighbouring galaxies.
In a similar way, the weight $\wrf \ge 1$ is used to up-weight a near-by galaxy in case of a failure of the spectroscopic redshift determination}.
These two weights are combined into the `counting weight', $\wc = \wfc + \wrf - 1$.
An additional weight $\wsys$ is assigned to each galaxy to correct for the systematic effects introduced by the local stellar density and the seeing during the photometric observations \citep{Ross:2012qm,Anderson:2013zyy,Reid:2015gra}.
The final weight, \changed{$\wtot$}, of a galaxy is given by
\begin{align}
 \label{eq:galaxy_weight}
 \wtot &= \wsys \, \wc.
\end{align}

\begin{table}
 \centering
 \caption{The redshift ranges, effective volumes and effective redshifts of the redshift bins used in this work and its companion papers.
  The volumes $V_\mathrm{eff}$ (in units of $h^{-3} \, \unit{Gpc}^3$) of the two galactic caps (NGC and SGC) are computed for the fiducial cosmology defined in Table~\ref{tab:dr12_cosmologies}.
  }
 \label{tab:dr12_z_ranges}
 \begin{tabular}{lllllll}
  \hline
  \multicolumn{2}{l}{Bin no. and label} &
                      Redshift range         & $\zeff$ & $V_\mathrm{eff}^\mathrm{NGC}$ & $V_\mathrm{eff}^\mathrm{SGC}$ \\
  \hline
  1 &  low          & $0.2 \leq z \leq 0.5$  & $0.38$  & $0.821$                       & $0.317$                       \\
  2 &  intermediate & $0.4 \leq z \leq 0.6$  & $0.51$  & $0.961$                       & $0.351$                       \\
  3 &  high         & $0.5 \leq z \leq 0.75$ & $0.61$  & $0.915$                       & $0.332$                       \\
  \hline
 \end{tabular}
\end{table}

The redshift binning for the analysis of the combined sample is tuned for optimal extraction of cosmological information from the two-point clustering statistics.
We analyse the final sample in two wide, non-overlapping redshift bins -- referred to as `low' ( $0.2 \le z < 0.5$) and `high' ($0.5 \le z < 0.75$) -- while consistency checks are performed with an overlapping, `intermediate' redshift bin ($0.4 \le z < 0.6$).
The definitions of the redshift ranges, their effective redshift and effective volumes in the two galactic caps (NGC and SGC) are \changed{given} in Table~\ref{tab:dr12_z_ranges}.

The angular and radial survey selection function is described by the set of $\Nrnd$ random points, which sample the survey volume more densely than the galaxies ($\Nrnd \simeq 50 \, \Ngal$).
Within the geometrical boundaries of the survey, galaxies \changed{cannot} been observed in certain small regions, such as the centre posts of the observational plates or the surroundings of a bright star.
Despite the small angular size of each individual `masked' region, they are not randomly distributed across the sky and \changed{their total effect adds up} to a non-negligible area.
Thus, they are excluded from any analysis by the use of veto masks removing points of the random catalogue that fall within these masked regions \changed{\citep[see][for more details]{Reid:2015gra}}.

\begin{table}
 \centering
 \caption{The set of cosmological parameters used in this work and its companion papers.
  \changed{Except for the `template' cosmology, all cosmologies are flat, $\Ol = 1 - \Om$, so that $\Om[c] h^2$ can be derived from $\Om[c] h^2 = \Om h^2 - \Om[b] h^2$.}
  For the template cosmology, there is a massive neutrino component in addition, $\OmegaX{\nu} h^2 = 0.00064$ (corresponding to $\sum m_\nu = 0.06 \Unit{eV}$) --- just as for the \Planck 2015 reference \LambdaCDM cosmology \citep{Adam:2015rua}.}
 \label{tab:dr12_cosmologies}
 \addtolength{\tabcolsep}{-1pt}
 \begin{tabular}{llllllll}
  \hline
  Name       & $\Om$      & $h$      & $\Om[b] h^2$ & $\sigma_8$ & $n_\mathrm{s}$ \\
  \hline
  Fiducial   & $0.31$     & $0.676$  & $0.022$      & $0.8$      & $0.97$         \\
  \Minerva   & $0.285$    & $0.695$  & $0.02104$    & $0.828$    & $0.9632$       \\
  \QPM       & $0.29$     & $0.7$    & $0.02247$    & $0.8$      & $0.97$         \\
  \MDPatchy  & $0.307115$ & $0.6777$ & $0.02214$    & $0.8288$   & $0.96$         \\
  Template   & $0.315298$ & $0.6726$ & $0.022204$   & $0.828$    & $0.9648$       \\
  \hline
 \end{tabular}
 \addtolength{\tabcolsep}{1pt}
\end{table}

The spectroscopic redshifts are converted \changed{into} distances \changed{adopting the same fiducial cosmology as in all BOSS DR12 clustering analyses \citep{Alam:2016hwk}, which is specified in Table~\ref{tab:dr12_cosmologies} and is characterized by a matter density parameter close to the central value measured from the latest analysis of the CMB data from the \Planck satellite} \citep{Planck:2015xua}.

\subsection{Optimal clustering wedges measurements in Fourier space}
\label{sec:clustering_wedges_fourier_space}

Let $P(\mu, k)$ be the anisotropic power spectrum in terms of the wavenumber $k$ and the LOS parameter $\mu$.
\changed{In Fourier space,} the latter parameter is defined as the cosine of the separation angle $\theta$ between the Fourier mode $\V k$ and the LOS direction $\Vn r$,
\begin{equation}
 \mu \equiv \cos \theta = \abs{\V k \cdot \Vn r} \, \abs{k}^{-1}.
\end{equation}
\changed{In principle, $\mu$ can take values in the range $-1$ to $1$. However, due to the symmetry along the line of sight direction the power spectrum is an even function of $\mu$ and only the range from 0 to 1 needs to be considered.
The concept of clustering wedges \citep{Kazin:2011xt} can be extended to Fourier space by defining the power spectrum wedge, as the average of the two-dimensional power spectrum, $P(\mu, k)$, over a number of wide, non-intersecting bins in $\mu$, that is}
\begin{equation}
 P_{\mu_1}^{\mu_2}(k) \equiv \frac{1}{\mu_2 - \mu_1} \int_{\mu_1}^{\mu_2} P(\mu, k) \dint \mu,
 \label{eq:Fourier_space_wedge}
\end{equation}
where $\mu_1$ ($\mu_2$) is the lower (upper) limit for the LOS parameter.
\changed{The wedges are usually defined by dividing up the full range of $\mu \in [0, 1]$ into $n$ intervals of equal width, $\mu_2 - \mu_1 = n^{-1}$.}

The Fourier-space wedges can be estimated from a galaxy catalogue by means of an analogue of the Yamamoto-Blake estimator \citep{Yamamoto:2005dz,Blake:2011rj,Beutler:2013yhm} \changed{used} to measure the power spectrum multipoles.
\changed{In this estimator, the LOS direction for each pair of galaxies is approximated by the distance vector to one of them.}
This \changed{method, dubbed `moving-LOS'} significantly reduces the computational costs compared to the original estimator of \citet{Yamamoto:2005dz}, while preserving most of the LOS information.
\changed{The more simplifying assumption of a fixed (global) plane-parallel approximation for the LOS, the `fixed-LOS' method \citep*{Yoo:2013zga,Samushia:2015wta}, would significantly bias the anisotropic clustering measurement for wide-angle surveys such as BOSS.}

The Feldman-Kaiser-Peacock (FKP) estimator for the power spectrum monopole \citep*{Feldman:1993ky} assigns an additional weight $\wFKP$ to each galaxy in order to minimize the variance of the estimator.
Here, \changed{we extend} the optimal-variance estimator is to wedges.
We define the weighted wedge overdensity field,
\begin{equation}
 F_{\mu_1}^{\mu_2}(\V k) = \frac{1}{(\mu_2 - \mu_1) \, \sqrt{\Anorm}} \left[ D_{\mu_1}^{\mu_2}(\V k) - \alphar \, R_{\mu_1}^{\mu_2}(\V k) \right],
 \label{eq:overdensity_field}
\end{equation}
where $\Anorm$ is a normalization constant and $\alphar$ is the data-to-randoms ratio (both are discussed later in this section).
Further, the individual density fields of the galaxies\changed{, $D_{\mu_1}^{\mu_2}(\V k)$, and the randoms, $R_{\mu_1}^{\mu_2}(\V k)$,} are given by
\begin{align}
 D_{\mu_1}^{\mu_2}(\V k) &= \sum_{i=1}^{\Ngal} \wtot(\V x_i) \, \wFKP(\V x_i) \, \EE{\ii \V k \cdot \V x_i} \, \Theta_{\mu_1}^{\mu_2} \left( \frac{\V k \cdot \V x_i}{\abs{\V k} \, \abs{\V x_i}} \right) \text{ and}
 \label{eq:data_grid} \\
 R_{\mu_1}^{\mu_2}(\V k) &= \sum_{j=1}^{\Nrnd} \wFKP(\V x_j) \, \EE{\ii \V k \cdot \V x_j} \, \Theta_{\mu_1}^{\mu_2} \left( \frac{\V k \cdot \V x_j}{\abs{\V k} \, \abs{\V x_j}} \right),
 \label{eq:randoms_grid}
\end{align}
respectively.
Here $\Theta_{\mu_1}^{\mu_2}(\mu)$ is the top-hat function equal to one inside the range $\mu_1 \le \mu \le \mu_2$ and to zero outside of it.
The weight $\wtot$ for the galaxies is given in equation~\eqref{eq:galaxy_weight}.
As derived in appendix~\ref{app:FKP}, the weight $\wFKP$ \changed{that minimises the variance of the measured power spectrum wedges} depends on the expected number density of galaxies $\nexp(\V x)$ in addition to the systematic weights,
\begin{equation}
 \wFKP^{-1}(\V x) = \ftp \, \wsys(\V x) + (1 - \ftp) \, \wtot(\V x) + \nexp(\V x) \, \PFKP,
 \label{eq:FKP_weight}
\end{equation}
\changed{generalizing the original FKP weight given in equation~\eqref{eq:orig_FKP_weight} to take into account} our treatment of fibre collisions (see appendix~\ref{app:shot_noise}).
In equation~\eqref{eq:FKP_weight}, $\ftp$ is the fraction of true fibre collision pairs and is fiducially set to $\ftp = 0.5$ \changed{in agreement} with the value used in \citet{Gil-Marin:2014sta}.
In order to optimize the variance for the power spectrum at the position of the BAO peaks of a CMASS-like sample, the fiducial power spectrum amplitude is set to $\PFKP = 10^4 \; h^{-3} \, \unit{Mpc}^3$ \changed{\citep[consistently with the rest of the series of companion papers lead by][]{Alam:2016hwk}}.
\changed{This choice is motivated by the fact that this value is close to the amplitude of the power spectrum of the BOSS combined sample at $k=0.14\,h{\rm Mpc}^{-1}$, which is the effective scale suggested by \citet{Font-Ribera:2014} to use for BOSS BAO measurements}.

The effective data-to-randoms ratio $\alphar$ is defined by
\begin{equation}
 \label{eq:alphar}
 \alphar \equiv \textstyle \left( \sum_i^{\Ngal} \wtot(\V x_i) \, \wFKP(\V x_i) \right) \left( \sum_j^{\Nrnd} \wFKP(\V x_j) \right)^{-1}
\end{equation}
This expression is further discussed in appendix~\ref{app:pspect_est}, where we also derive the normalization constant to be
\begin{equation}
 \Anorm = \alphar \sum_j^{\Nrnd} \nexp(\V x_j) \, \wFKP^2(\V x_j).
 \label{eq:normalization}
\end{equation}
\changed{Here,} $\nexp(\V x_j)$ is the expected number density, which already entered the FKP-weight definition in equation~\eqref{eq:FKP_weight}.

The wedge power spectrum is estimated from the wedge overdensity field using
\begin{equation}
 P_{\mu_1}^{\mu_2}(\V k) = F_{\mu_1}^{\mu_2}(\V k) \, \left[ F_{-1}^1(\V k) \right]^\ast - \Pshot_{\mu_1}^{\mu_2}(\V k),
 \label{eq:wedge_estimation}
\end{equation}
where $[\cdot]^\ast$ denotes complex conjugation \changed{and $\Pshot_{\mu_1}^{\mu_2}$ is the shot-noise term. 
Following a derivation analogous to the one of the multipole analysis in \citet{Gil-Marin:2015sqa}, it is easy to see that the shot-noise term can be computed as\footnote{\changed{\citet{Beutler:2013yhm} use a slightly different approach that also incorporates a sum over the observed galaxies, which provides similar results than the one we use.}}}
\begin{equation}
 \Pshot_{\mu_1}^{\mu_2}(\V k) = \frac{\alphar \, (\alphar + 1)}{(\mu_2 - \mu_1) \, \Anorm} \sum_j^{\Nrnd} \wFKP^2(\V x_j) \, \Theta_{\mu_1}^{\mu_2} \left( \frac{\V k \cdot \V x_j}{\abs{\V k} \, \abs{\V x_j}} \right).
 \label{eq:shot_noise_wedges}
\end{equation}
\changed{However}, this treatment does not account for deviations from a Poisson distributed galaxy and random sample in a real survey such as BOSS.
In order to account for exclusion effects caused by the fibre collisions, we split the shot noise in separate sums over the galaxies and the random points as discussed in appendix~\ref{app:shot_noise},
\begin{align}
 \nonumber
 \Pshot &= \sum_i^{\Ngal} \frac{\wFKP^2(\V x_i)}{\Anorm} \left[ \ftp \, \wtot(\V x_i) \, \wsys(\V x_i) + (1 - \ftp) \, \wtot^2(\V x_i) \right] \\
 \label{eq:shot_noise_final}
 &\quad + \frac{\alpha^2}{\Anorm} \sum_j^{\Nrnd} \wFKP^2(\V x_j).
\end{align}
\changed{We remind the reader that the fiducial true-pair} fraction is set to $\ftp = 0.5$.
\changed{In equation~\eqref{eq:shot_noise_final}, we dropped the indices on $\Pshot$ to highlight the fact that we assume a constant shot-noise contribution to all wedges.
Given that our wedges are defined using equal-width $\mu$ bins, the shot-noise contribution is also equally distributed among the wedges.}

\subsection{FFT-based estimators}
\label{sec:pseudo_clustering_wedges}

Even though the computing time of the Yamamoto-Blake estimator has been significantly reduced by adopting the
moving-LOS approximation, time efficiency is still a concern as the power spectrum wedges
must be estimated for thousands of synthetic catalogues (\Cf, section~\ref{sec:covariance_matrix}).
As shown recently by \citet{Bianchi:2015oia} and \citet{Scoccimarro:2015bla}, the estimation of power spectrum multipoles can 
be sped up significantly by use of multiple FFTs. The Legendre polynomials $\Lp(\mu)$ can be expressed as a sum of power-law terms $\mu^\ell = (\Vn x \cdot \Vn k)^\ell$, so that the $\Vn x$ and $\Vn k$ components can be factored out.
The multipole-analogue of the weighted density field of equation~\eqref{eq:overdensity_field} is
\begin{equation}
 F_\ell(\V k) = \frac{(2 \ell + 1)}{2} \int F(\V x) \, \EE{\ii \V k \cdot \V x} \, \Lp\left( \frac{\V k \cdot \V x}{\abs{\V k} \, \abs{\V x}} \right) \, \dnx{3}{x},
\end{equation}
where $F(\V x)$ is the usual FKP-weighted density field defined in equation~\eqref{eq:FKP_density_field}.
The power spectrum multipoles can be estimated using
\begin{equation}
 \label{eq:yamamoto_blake_pk_ell}
 P_\ell(\V k) = F_\ell(\V k) \, \left[ F(\V k) \right]^\ast - \Pshot \, \deltaK_{\ell0},
\end{equation}
where $\deltaK_{\ell0}$ is the Kronecker delta ensuring that the shot-noise contribution is only subtracted from the monopole.

\begin{figure*}
 \centering
 \includegraphics[width=1\textwidth]{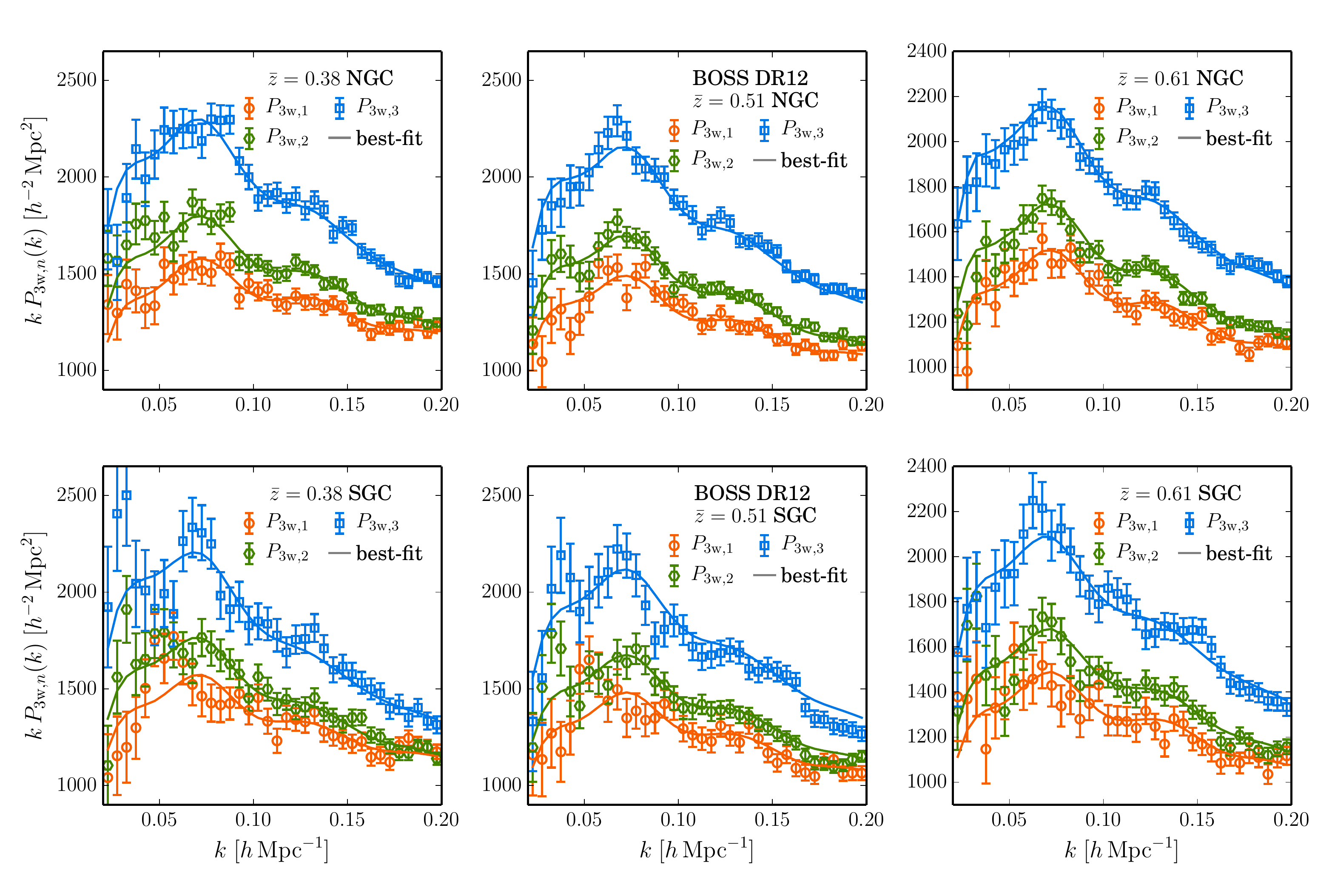}
 \caption{The power spectrum wedges computed by filtering out the information of Legendre multipoles $\ell>4$ for NGC (upper panels) 
 and SGC (lower panels) of the BOSS DR12 combined sample in the low (left-hand panels), intermediate (centre panels), and high 
 (right-hand panels) redshift bins defined in Table~\ref{tab:dr12_z_ranges}.
 The error bars are derived as the square root of the diagonal entries of \MDPatchy covariance matrix 
 (see section~\ref{sec:covariance_matrix}). The theoretical predictions are based on the model described in section~\ref{sec:model} and for the maximum-likelihood BAO+RSD parameters using a best-fit \Planck 2015 input power spectrum.
  The low redshift bin fits use separate bias, RSD, and shot noise parameters for NGS and SGC, whereas the intermediate and high bins use only one
  set of nuisance parameters.}
 \label{fig:pw_dr12_comb_model_bestfit}
\end{figure*}

The weighted quadrupole and hexadecapole density fields can be written as
\begin{align}
 F_2(\V k) &= \frac{3}{2} \sum_{i,j} \Vn k_i \, \Vn k_j \, Q_{ij}(\V k) - \frac 1 2 F(\V k) \quad \text{and} \\
 \nonumber
 F_4(\V k) &= \frac{35}{8} \sum_{i,j,k,l} \Vn k_i \, \Vn k_j \, \Vn k_k \, \Vn k_l \, Q_{ijkl}(\V k) - \frac{15}{4} F_2(\V k) + \frac 3 8 F(\V k),
\end{align}
where $Q_{ij}(\V k)$ and $Q_{ijkl}(\V k)$ are the Fourier transforms of
\begin{equation}
 \label{eq:Q_tensors}
 Q_{ij}(\V x) = \Vn x_i \, \Vn x_j \, F(\V x) \quad \text{and} \quad Q_{ijkl}(\V x) = \Vn x_i \, \Vn x_j \, \Vn x_k \, \Vn x_l \, F(\V x),
\end{equation}
respectively.
Due to the symmetries of the $Q_{\cdot}$ tensors, the calculation of $\hat F_2(\V k)$ needs six FFTs in addition to the one of the original FKP estimator.
Calculating $\hat F_4(\V k)$ requires 15 additional transforms.
Because of the low computational costs of FFTs, \changed{the computing time is negligible compared to the runtime of the original Yamamoto-Blake estimator even for large grid sizes.}

The FFT estimators cannot be directly applied to clustering wedges because of the non-polynomial dependency of the wedge 
top-hat kernel on the LOS parameter $\mu$. 
However, the FFT-Yamamoto scheme can be applied to compute an accurate approximation of the wedges.
The relation between wedges and multipoles is given by
\begin{equation}
 \label{eq:Pell2Pw}
 P_{\mu_1}^{\mu_2}(k) = \sum_{\ell} T_{n\ell} \, P_\ell(k),
\end{equation}
where, $ T_{n\ell}$ are the elements of the transformation matrix
\begin{equation}
 \label{eq:Tnell}
 T_{n\ell} \equiv \frac{1}{\mu_2 - \mu_1} \int_{\mu_1}^{\mu_2} \Lp(\mu) \dint \mu.
\end{equation}
While the FFT-based estimator can be defined for any multipole order in principle, we \changed{only} compute the power spectrum multipoles up to the hexadecapole.
\changed{The power spectrum wedges are approximated from the combined multipole measurements by} truncating the series in equation~(\ref{eq:Pell2Pw}) at the $\ell=4$ term.
The resulting ``pseudo-wedges" correspond to the result of filtering out the information of 
multipoles $\ell>4$ of the full two-dimensional power spectrum.
Even in the case in which the intrinsic power spectrum multipoles for $\ell>4$ could be neglected, the AP 
distortions caused by the assumption of different fiducial cosmologies would generate higher order multipoles that 
would not be included in this approximation, leading to small differences with the direct measurement 
of the wedges.

For our tests using \Nbody simulations we use the full definition of the clustering wedges. 
However, for time efficiency, in the analysis of the BOSS data and the different sets of mock catalogues 
we use the pseudo-wedges derived from the power spectrum multipoles $P_{\ell=0,2,4}(k)$.
Appendix~\ref{app:fft_yamamoto_estimator} presents a comparison of the full power spectrum wedges obtained using the estimator 
of equation~\eqref{eq:wedge_estimation} and their approximation from the multipoles derived from the FFT approach for a 
CMASS-like catalogue. This comparison shows that, up to wavenumbers $k\leq 0.2\; h \, \unit{Mpc}^{-1}$, the
pseudo-wedges computed using equation~(\ref{eq:Pell2Pw}) provide an accurate approximation of the full result.
Note that, as the pseudo-wedges correspond to the linear transformation of equation~(\ref{eq:Pell2Pw}), 
they contain the same information as the original multipoles and result in an identical likelihood function. 
However, we prefer to present our measurements in terms of this linear combination instead of multipoles 
directly, as they more closely represent the average of the full anisotropic power spectrum in the different 
$\mu$ bins. For simplicity, we will refer to these measurements as {\it wedges}, but the 
fact that they contain exactly the same information as the combination of the multipoles $P_{\ell=0,2,4}(k)$
should be taken into account when interpreting our results.
We leave the quantification of the precise loss of information to a future analysis.

Before applying the FFTs, $F(\V x)$ is calculated on a mesh using $1200^3$ grid cells applying the triangular-shaped 
cloud (TSC) scheme to assign galaxies and randoms to the cells. The side length of the grid is $4000 \; h^{-1} \, \unit{Mpc}$.
After the FFT, the mass-assignment scheme is corrected for by using the approximative anti-aliasing correction that was 
used in \citet{Montesano:2010qc}: each Fourier mode is divided by the corrective term $C_1(\V k)$ given 
in \citet[equation~20]{Jing:2004fq}. This yields a more precise power spectrum estimate than dividing by the Fourier 
transform of the mass assignment function.

The final measurements are estimated by averaging equation~\eqref{eq:Pell2Pw} over spherical $k$-space shells.
We adopt wavenumber bins with $\Delta k = 0.005 \; h \, \unit{Mpc}^{-1}$ from $\kmin = 0 \; h \, \unit{Mpc}^{-1}$ 
to $\kmax = 0.25 \; h \, \unit{Mpc}^{-1}$ and label the central wavenumbers of each bin as $k_i$.
With this binning scheme, already the smallest central wavenumber is much larger than the fundamental mode of the grid, 
$k_\mathrm{fund} = 1.57 \cdot 10^{-3} \; h \, \unit{Mpc}^{-1}$.
Also, $\kmax$ is always much smaller than the Nyquist frequency of the grid, $k_\mathrm{Ny} = 0.942 \; h \, \unit{Mpc}^{-1}$. 
Using the predictions in \citet{Sefusatti:2015aex}, we expect the error from aliasing to be less than $0.01$ per cent.

We consider configurations of two and three bins in $\mu$ defined by dividing the $\mu$ range from 0 to 1 into 
equal-width intervals. In each case we denote the measurements corresponding to the $n$-th $\mu$ bin as 
$P_{2\mathrm{w},n}$ and $P_{3\mathrm{w},n}$. For general references, we combine all measurement bins into the vectors 
$\Pobs[2\mathrm{w}] = \left( \PAobs[2\mathrm{w},n](k_i) \right)$ and $\Pobs = \left( \PAobs(k_i) \right)$.

Fig{.}~\ref{fig:pw_dr12_comb_model_bestfit} shows the three power spectrum wedges derived from the FFT-based multipoles 
of the NGC (upper panels) and SGC (lower panels) of the combined sample obtained in this way for the low 
(left-hand panels), intermediate (centre panels) and high (right-hand panels) redshift bins.
The predictions shown as solid lines are based on the model for the Fourier space wedges that is described in section~\ref{sec:model} 
and the maximum-likelihood parameters from the full-shape BAO+RSD fits of each redshift bin separately.
For the low redshift bin, we use two different sets of clustering nuisance parameters to account for the fact that the NGC and SGC 
samples might contain two slightly different galaxy population at low redshifts (see discussion in appendix~\ref{sec:NGC_vs_SGC}).

\subsection{Covariance matrix estimates from mock catalogues}
\label{sec:covariance_matrix}

\begin{figure}
 \includegraphics[width=.95\columnwidth]{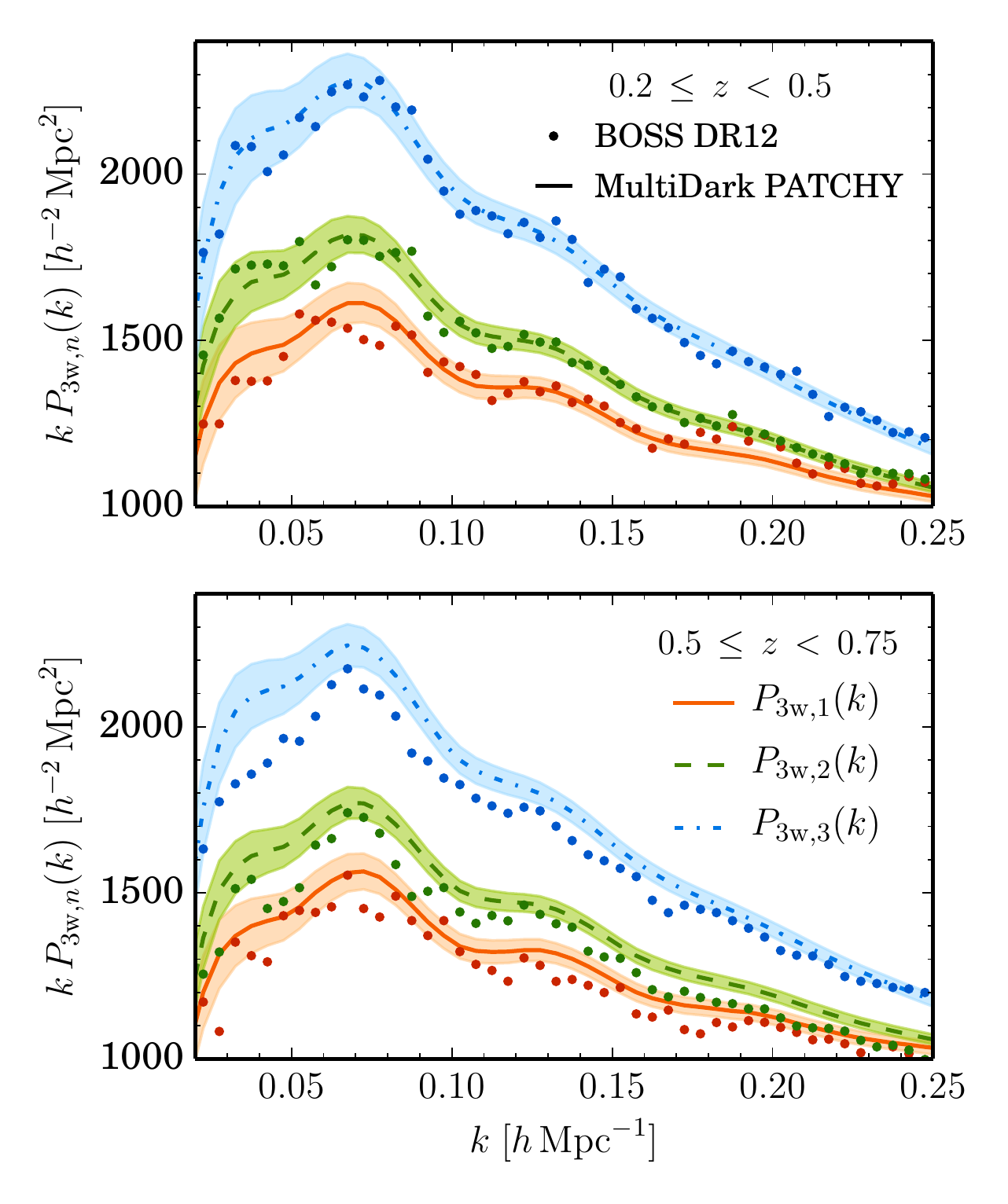}
 \caption{\MDPatchy power spectrum wedges derived from the multipoles $P_{\ell=0,2,4}(k)$ compared against the results 
 of the BOSS DR12 combined sample for the low (upper panel) and high (lower panel) redshift bin.
  These measurements correspond to 2045 full survey (combining NGC and SGC) mocks and have been performed 
  assuming the fiducial cosmology.}
 \label{fig:patchy_ps_wedges}
\end{figure}

\begin{figure}
 \includegraphics[width=.9\columnwidth]{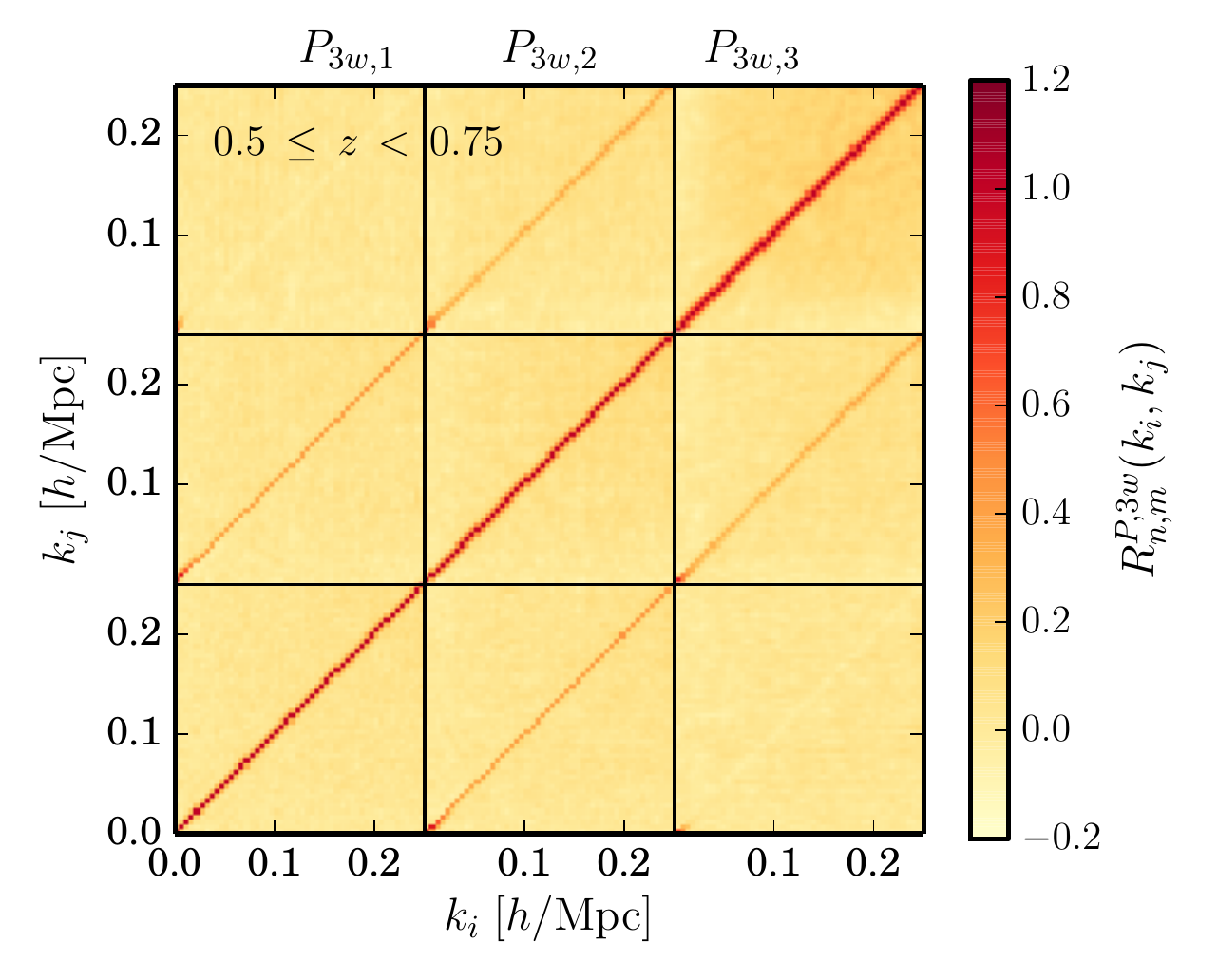}
 \caption{Correlation matrix of the \MDPatchy power spectrum wedges derived from the power spectrum 
 multipoles $P_{\ell=0,2,4}(k)$ for the high redshift bin.
  As in Fig{.}~\ref{fig:patchy_ps_wedges}, for this measurement NGC and SGC have been combined for simplicity.
  The correlation matrix for the low redshift bin looks similar.}
 \label{fig:patchy_ps_wedges_cov}
\end{figure}

As current theoretical predictions of the anisotropic clustering covariance cannot account for the observational systematics of the BOSS survey with the required accuracy, the covariance matrix for the analysis of the BOSS DR12 combined sample is estimated from large sets of synthetic catalogues.
These mock catalogues are based on large-scale haloes that are generated using fast, approximate solvers for the gravitational evolution equations.
Phenomenological small-scale models are used to populate these haloes with synthetic galaxies basing the calibration of the model on a few \Nbody simulations.
We use two sets of mock catalogues mimicking the DR12 combined sample, both with a large number of realizations to overcome the sample noise in the precision matrix estimate.
All synthetic survey catalogues incorporate the survey geometry (selection window, veto mask) and the most important observational systematics such as fibre collisions.

Here we focus on the set of \textsc{MultiDark}-\Patchy \citep[\MDPatchy,][]{Kitaura:2015uqa} mocks that are based on the \Patchy \citep*{Kitaura:2013cwa} recipe to generate mock halo catalogues.
In appendix~\ref{app:covariance_cross_check}, we also use an alternative set of mocks, based on the quick-particle-mesh \citep*[\QPM;][]{White:2013psd} technique, to cross-check our reference covariance matrix.

The first step of the \MDPatchy recipe is to generate a DM density and velocity field using the Augmented Lagrangian Perturbation Theory \citep[ALPT;][]{Kitaura:2012tj} formalism.
This algorithm splits the Lagrangian displacement field into a large-scale component, which is derived by 2-LPT, and a small-scale component that is modelled by the spherical collapse approximation.
The initial conditions are generated with cosmological parameters that are matched to the \textsc{Big-MultiDark} \Nbody simulations \citep{Klypin:2014kpa}.
These parameters are given as `\MDPatchy' in Table~\ref{tab:dr12_cosmologies}.
The halo density field is then modelled using perturbation theory and nonlinear stochastic biasing with parameters calibrated against the fully non-linear simulations \citep{Rodriguez-Torres:2015vqa}.

The second step populates the haloes with galaxies by abundance matching between the DR12 combined sample and simulations using \code{Hadron} \citep{Zhao:2015jga}.
The clustering of the \MDPatchy catalogues reproduces the DR12 two- and three-point statistics \citep{Rodriguez-Torres:2015vqa}.
The survey selection is applied to a light-cone interpolation of the galaxy snapshots at 10 different intermediate redshifts.

A set of $\Nmocks = 2045$ realizations exists from which we obtain the reference covariance matrix for the fits of the clustering model to the data.
The elements of this matrix are estimated from the covariance of the $\PAobs(k_i)$ measurements,
\begin{align}
 C_{nm,ij} &= \average{ \PAobs(k_i) \, \PAobs[3\mathrm{w},m](k_j) } - \average{ \PAobs(k_i) } \, \average{ \PAobs[3\mathrm{w},m](k_j) },
 \label{eq:covariance}
\end{align}
where $\average{\cdot}$ represents the average over the $\Nmocks$ mock realizations.

The mean \MDPatchy power spectrum wedges show good agreement with the clustering of the DR12 combined sample as shown by the comparison in Fig.~\ref{fig:patchy_ps_wedges}.
For a better visualization of the structure in the covariance matrix, we plot the correlation matrix, defined by
\begin{equation}
 \label{eq:correlation_matrix}
  R_{nm,ij} = C_{nm,ij} \, \left( C_{nn,ii} \, C_{mm,jj} \right)^{-\frac 1 2},
\end{equation}
for the high redshift bin in Fig.~\ref{fig:patchy_ps_wedges_cov} (the correlation matrix for the two other redhift bins are similar).
The effect of the window function (discussed later in section~\ref{sec:win_func}) introduces a correlation between neighbouring bins and wedges that can be seen as non-vanishing sub-diagonal entries.
Especially in the correlation for the most-parallel wedge in the high-redshift bin, cross-covariance between all bins is increased by the fibre collisions between pairs too close in angular separation \citep[the CMASS sample is more affected by this problem than LOWZ;][]{Reid:2015gra}.

Our power spectrum measurements and their corresponding covariance matrices \changed{are} publicly available.\footnote{\changed{\url{https://sdss3.org/science/boss_publications.php}}}

\subsubsection{The precision matrix}
\label{sec:precision_matrix}

\begin{table}
 \centering
 \caption{The correction factors for the precision matrix as given by equation~\eqref{eq:rescaled_covariance} for our configurations of measurement bins and numbers of realizations used to estimate the covariance matrix.
  $\kmin$ and $\kmax$ are given in units of $h \, \unit{Mpc}^{-1}$.}
 \label{tab:Hartlap_correction}
 \begin{tabular}{lllllll}
  \hline
  $\Nmocks$ & $\kmin$ & $\kmax$ & $\#(k_i)$ & $\#(\text{wedges})$ & $\Nbins$ & $\CD$    \\
  \hline
  $1000$    & $0.02$  & $0.2$   & $36$      & $3$                 & $108$    & $0.1091$ \\
  $2045$    & $0.02$  & $0.2$   & $36$      & $3$                 & $108$    & $0.0533$ \\
  \hline
 \end{tabular}
\end{table}

We denote a point in the parameter space of a theoretical model as $\zeta \in \mathcal{X}$ and the model predictions of the observed Fourier-space wedges as $\Ppred(\zeta) = \left( \PApred(k_i) \right)$.
The comparison of model predictions with the data $\Pobs$ relies on the calculation of the likelihood function.
Assuming that the number of modes observed is large enough, the power spectrum wedges follow a multi-variate Gaussian distribution with a fixed covariance.
This approximation is justified on quasi-linear scales \citep{Manera:2012sc,Ross:2012sx} and, thus, the likelihood is given by
\begin{equation}
 \Like[\Ppred(\zeta)] = \frac{\abs{\precmat}}{\sqrt{2 \mathpi}} \ee{- \frac 1 2 \chi^2(\Ppred(\zeta),\Pobs,\precmat)},
 \label{eq:likelihood}
\end{equation}
where the precision matrix $\precmat$ is the inverse of the exact covariance matrix.
The log-likelihood function $\chi^2$ makes use of the difference vector, $\deltaP(\zeta)\equiv \Ppred(\zeta) - \Pobs$, as
\begin{equation}
 \chi^2(\Ppred(\zeta),\Pobs,\precmat) = \transp{\deltaP(\zeta)} \cdot \precmat \cdot \deltaP(\zeta),
 \label{eq:chi_sqr}
\end{equation}
where $\transp{\V P}$ denotes the transpose of $\V P$.

\begin{table}
 \centering
 \caption{The correction factors for the parameter constraints as given by equation~\eqref{eq:rescaled_constraints} for our configurations of measurement bins, numbers of realizations used to estimate the covariance matrix, and number of fitting parameters.
  $\kmin$ and $\kmax$ are given in units of $h \, \unit{Mpc}^{-1}$.}
 \label{tab:Percival_correction}
 \begin{tabular}{lllllll}
  \hline
  $\Nmocks$ & $\kmin$ & $\kmax$ & $\Nbins$ & $\Nparams$ ($z$-bin) & $\CM$    \\
  \hline
  $1000$    & $0.02$  & $0.2$   & $108$    & $8$  (int,high)      & $1.0494$ \\
  $1000$    & $0.02$  & $0.2$   & $108$    & $13$ (low)           & $1.0439$ \\
  $2045$    & $0.02$  & $0.2$   & $108$    & $8$  (int,high)      & $1.0231$ \\
  $2045$    & $0.02$  & $0.2$   & $108$    & $13$ (low)           & $1.0206$ \\
  \hline
 \end{tabular}
\end{table}

The exact covariance matrix is not known.
Hence, the precision matrix is estimated as the inverse of the covariance matrix \changed{inferred from our mock catalogues}, ${\mathbfss C} = \left( C_{AB,ij} \right)$, whose elements are given by equation~\eqref{eq:covariance}.
This estimate is affected by noise due to the finite number of mocks.
Consequently, the precision matrix and \changed{the resulting parameter constraints} are biased \citep{Taylor:2012kz,Dodelson:2013uaa,Percival:2013sga}.
In the following, we account for this bias by a rescaling \citep*{Hartlap:2006kj},
\begin{equation}
 \label{eq:rescaled_covariance}
 \precmat = (1 - \CD) \, {\mathbfss C}^{-1}, \quad \text{where} \quad \CD = \frac{\Nbins + 1}{\Nmocks - 1},
\end{equation}
where $\Nbins$ is the total number of bins in the measurements $\Pobs(k_i)$.
In addition, the effect of the noise propagates to the parameter constraints, so that the obtained variance of each parameter needs to be rescaled by \citep{Percival:2013sga}
\begin{equation}
 \label{eq:rescaled_constraints}
 \CM = \sqrt{\frac{1 + \CB \, (\Nbins - \Nparams)}{1 + \CA + \CB \, (\Nparams + 1)}},
\end{equation}
where $\Nparams$ is the number of fitting parameters and \changed{the two factors $\CA$ and $\CB$ are given as}
\begin{align}
 \CA &\equiv \frac{2}{(\Nmocks - \Nbins - 1) \, (\Nmocks - \Nbins - 4)}, \\
 \CB &\equiv \frac{\Nmocks - \Nbins - 2}{(\Nmocks - \Nbins - 1) \, (\Nmocks - \Nbins - 4)}.
\end{align}

As $\Nmocks$ is large, the correction factors for the covariance of the $\PAobs(k)$ measurements and the fitted parameters, listed in Tables~\ref{tab:Hartlap_correction} and \ref{tab:Percival_correction}, respectively, are small despite the large number of measurement bins used.

\subsection{The window function}
\label{sec:win_func}

\begin{figure}
 \centering
 \includegraphics[width=.9\columnwidth,page=1]{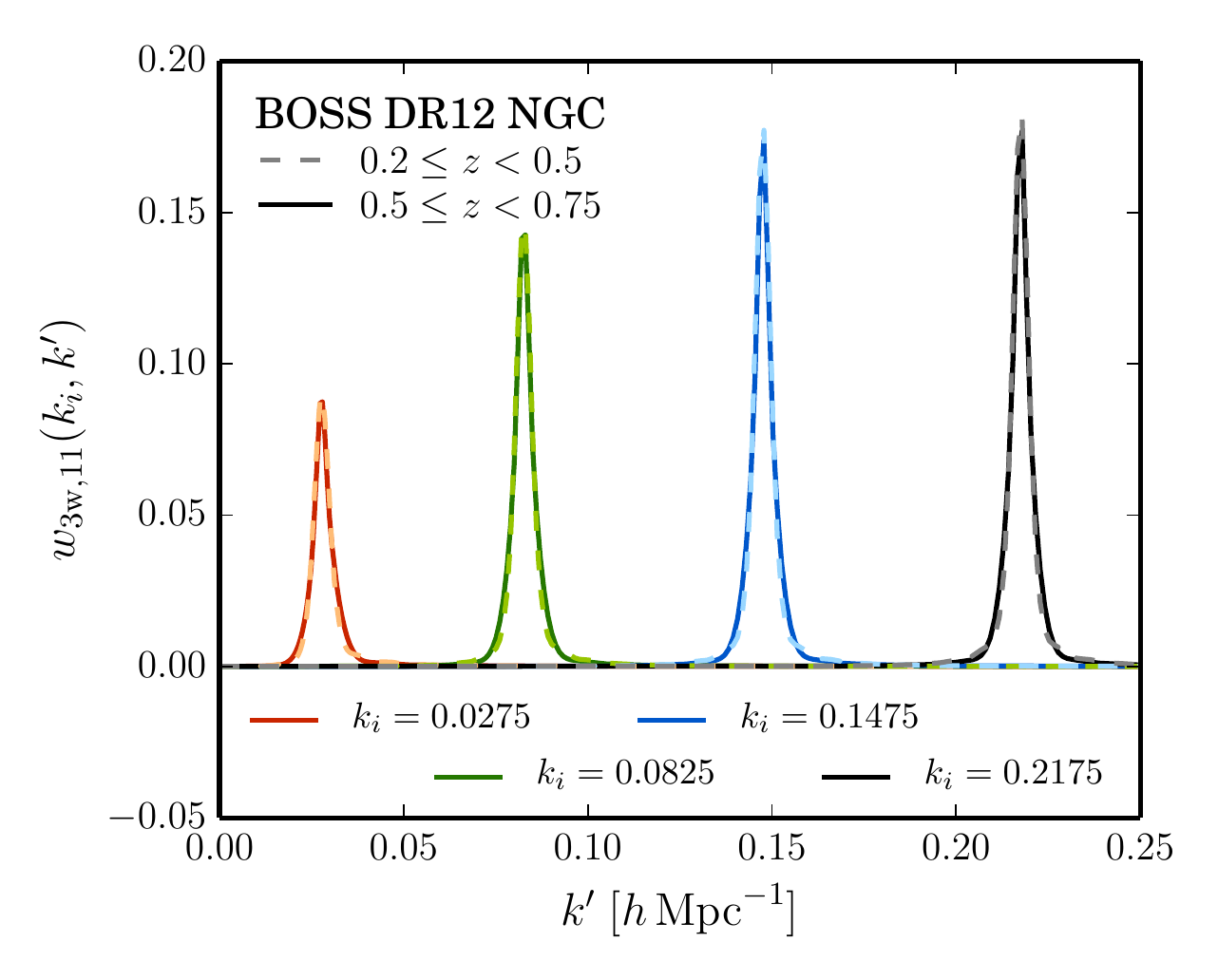}
 \includegraphics[width=.9\columnwidth,page=4]{./images/dr12_comb_winmat_w3.pdf}
 \caption{The window matrix $w_{3\mathrm{w},nm}(k_i,k')$ of the DR12 combined sample for the most-perpendicular wedge in the upper panel and for all wedges in the lower panel.
  The upper panel shows the dependency of $w_{3\mathrm{w},11}$ on the redshift range and the mean $k_i$ (given in $h \, \unit{Mpc}^{-1}$) of the output bin.
  The window matrices of each redshift bin are similar (dashed lines -- low redshift bin, solid lines -- high redshift bin).
  The lower panel shows the contributions of the different input wedges to the output wedges for the bin $k_i = 0.0275$ (from left to right, the $x$-axis is split into repeating intervals for better visibility).
  The SGC window matrix resembles that of the NGC, but the suppression of power is slightly stronger as the volume is smaller (see also Fig{.}~\ref{fig:pk_conv_dr12_comb}).}
 \label{fig:win_mat_dr12_comb}
\end{figure}

A non-trivial survey geometry distorts the shape of the power spectrum estimator presented in section~\ref{sec:clustering_wedges_fourier_space}.
For scales of sizes close to or larger than the distances between the boundaries of the survey, the power spectrum is suppressed as the modes within the survey fail to resolve the perturbations at their full length.
\changed{Conversely, they are enhanced at small scales due to mode coupling.}
As discussed in \citet{Beutler:2013yhm} and \citet{Gil-Marin:2015sqa}, this effect is stronger for higher-order 
multipoles in a survey like BOSS that covers a large angular area on the sky.\footnote{See also the discussion of 
the binning effect due to a finite grid in $k$ in \citet{Beutler:2016arn}}.
We describe this effect by the convolution of a theoretical prediction $\tilde P(\V k)$ with \changed{the survey} window function,
\begin{equation}
 \label{eq:win_conv}
 \hat P(\V k) = \int \abs{W(\V k - \VPrime k)}^2 \, \tilde P(\VPrime k) \, \dnx{3}{k'}.
\end{equation}
As already done in \citet{Gil-Marin:2015sqa}, we neglect the integral constraint \citep[section~5.2]{Beutler:2013yhm} due to its marginal effect for large-volume surveys.

The window function $W(\V k)$ is given by
\begin{equation}
 W(\V k) = \frac{1}{\sqrt{\Anorm}} \int \nexp(\V x) \, \ft{\V k}{\V r} \, \dnx{3}{r},
\end{equation}
where \changed{$\Anorm$ is the normalization factor given by equation~\eqref{eq:normalization}.}
The expected number density can be expressed by the random field, $\nexp(\V x) = \alphar \, \nrnd(\V x)$ (see details in appendix~\ref{app:FKP}).

As described in section~\ref{sec:pseudo_clustering_wedges}, we approximate the clustering wedges as a linear combination of the power spectrum multipoles $P_{\ell=0,2,4}(k)$ computed using the Yamamoto-FFT estimator.
We can then apply the formalism of the multipole window functions described in \citet[section 5.1]{Beutler:2013yhm} to our clustering measurements.
The pseudo-wedge window function can be written in terms of the multipole window functions, which we measure using
\begin{multline}
 \label{eq:window_func_ell}
 \abs{W(k,k')}^2_{\ell L} = 2 \ii^\ell (-\ii)^L \, (2 \ell + 1) \sum_{i j, i \neq j}^{\Nrnd} \wFKP(\V x_i) \wFKP(\V x_j) \\
 \times j_\ell(k\abs{\Delta \V x}) \, j_L(k'\abs{\Delta \V x}) \, \Lp(\Vn x_h \cdot \Vn x) \, \Lp[L](\Vn x_h \cdot \Vn x),
\end{multline}
where $\Delta \V x = \V x_i - \V x_j$, $\V x_h = \frac 1 2 (\V x_i + \V x_j)$, and $j_\ell(x)$ represents the spherical Bessel function of order $\ell$.
Due to its immense computation time, this double sum is only performed for a subset of ca{.}~65{,}000 of the randoms.
We performed a convergence test and did not find improvement if a larger subset of randoms is used.
\changed{In a second step}, these window functions are transformed into pseudo-wedge window functions by use of the transformation matrix ${\mathbfss T}$, 
whose elements are given in equation~\eqref{eq:Tnell},
\begin{equation}
 \label{eq:window_func_w}
 \abs{W(k,k')}^2_{3\mathrm{w},nm} = \sum_{\ell,L \in \{ 0, 2, 4 \}} T_{n\ell} \, T^{-1}_{Lm} \, \abs{W(k,k')}^2_{\ell L}.
\end{equation}
Here, $T^{-1}_{Lm}$ are the elements of the inverse ${\mathbfss T}^{-1}$.

In practice, the convolution of equation~\eqref{eq:win_conv} is described by a window matrix multiplication.
The normalized elements $w_{3\mathrm{w},nm}(k_i,k')$ of this window matrix are precomputed using
\begin{equation}
 \label{eq:window_matrix_def}
 w_{3\mathrm{w},nm}(k_i,k') = W_{k_i}^{-1} \, w_{k'} \, \abs{W(k_i,k')}^2_{3\mathrm{w},nm} \, (k')^2.
\end{equation}
Here, the input wavenumbers $k'$ and their weights $w_{k'}$ are determined using the Gauss-Legendre quadrature.
The normalization $W_{k_i}$ is chosen such that $\sum_{n,m} \sum_{k'} w_{3\mathrm{w},nm}(k_i,k') = 1$ for each $k_i$.
The final prediction for the vector $\Ppred = (\PApred[3\mathrm{w},n](k_i))$ is then given by
\begin{equation}
 \label{eq:window_matrix_conv}
 \PApred[3\mathrm{w},n](k_i) = \sum_{k'} w_{3\mathrm{w},nm}(k_i,k') \, \PAtheo[3\mathrm{w},m](k'),
\end{equation}
where $\PAtheo[3\mathrm{w},n](k')$ are the wedges of the underlying power spectrum at the input wavenumbers $k'$.

\begin{figure}
 \centering
 \includegraphics[width=\columnwidth]{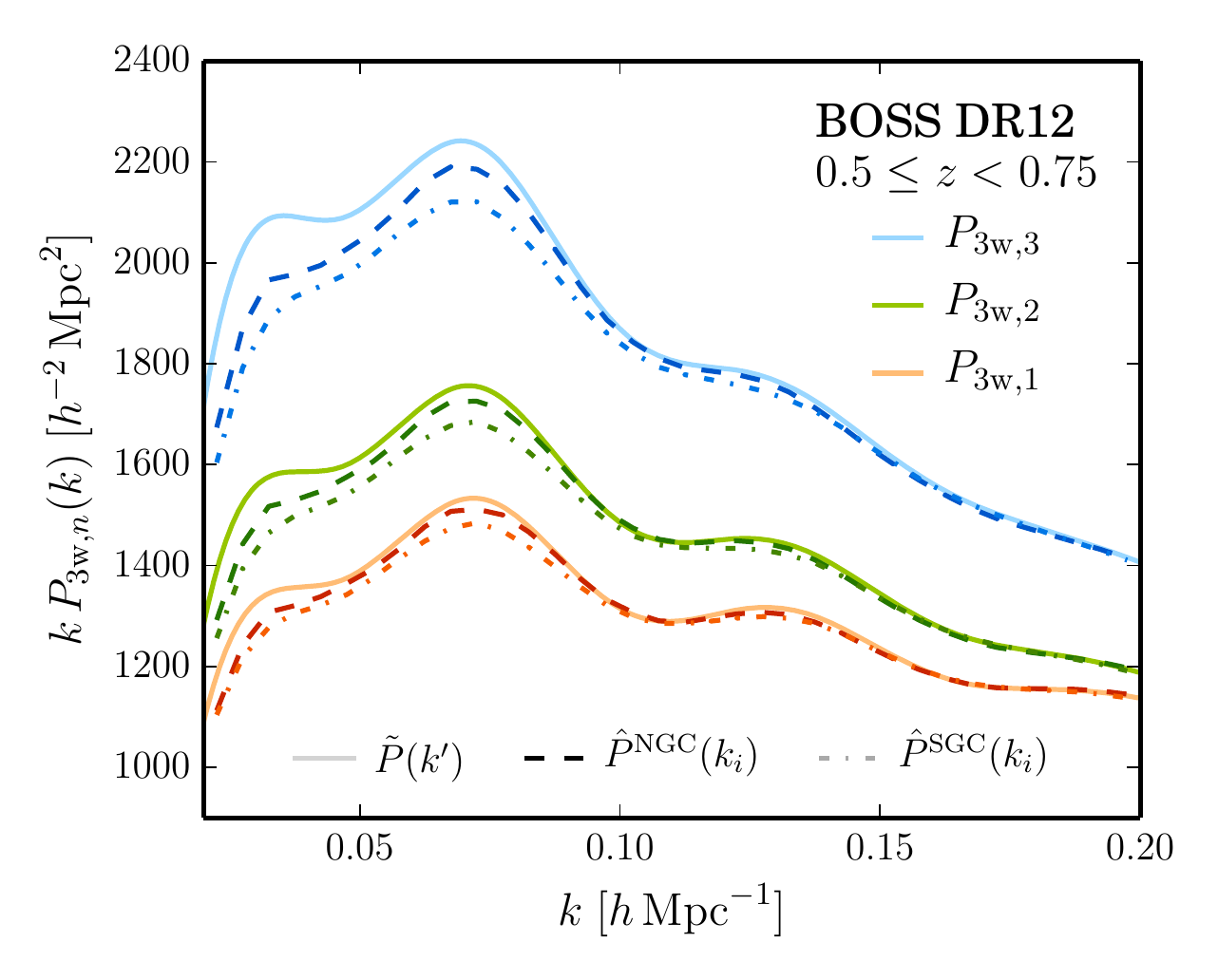}
 \caption{The effect of the window matrix $w_{3\mathrm{w},nm}$ for the DR12 combined sample on the Fourier space wedges in the high redshift bin.
  The solid lines are the theoretical predictions $\PAtheo[3\mathrm{w},n](k')$ (using the model described in section~\ref{sec:zspace_clustering_model} for best-fitting \LambdaCDM parameters), and the dashed (dash-dotted) lines corresponds to the prediction convolved with the window function, $\PApred[3\mathrm{w},n](k_i)$, for the Northern (Southern) galactic cap.}
 \label{fig:pk_conv_dr12_comb}
\end{figure}

To illustrate the features of the window matrix, we plot its elements $w_{3\mathrm{w},nm}(k_i,k')$ \changed{for the NGC subsample} in Fig{.}~\ref{fig:win_mat_dr12_comb}.
In the upper panel, we show that the window matrices for the low and high redshift bin do not significantly differ.
Further, this plot shows the narrow range in which the window function is non-zero around each $k_i$.
The window matrices for the NGC and the SGC have slightly different normalizations due to the smaller volume of the South, but otherwise follow the same trends with $k_i$ and $k'$.
The lower panel shows the cross- and auto-contributions of the three power spectrum wedges for $k_i = 0.0275$.
This plot illustrates that the cross-talking induced by the anisotropic window matrix is non-negligible for the most-parallel wedge.
As an illustration of the effect of the window function, Fig{.}~\ref{fig:pk_conv_dr12_comb} shows the theoretical power spectrum wedges corresponding to the best-fitting \LambdaCDM model to our BOSS measurements in the high-redshift bin (see Section~\ref{sec:LCDM_model}) together with their convolution with the NGC and SGC window functions.
\changed{While the suppression of power caused by the window function is stronger for the SGC subsample, the window functions computed for the other redshift bins are very similar to each other.}

\changed{Comparing the results of our analysis on simulated galaxy catalogues with the results on periodic boxes, we do not see a significant loss of constraining power caused by the treatment of the window function.
An alternative, but mathematically identical technique to account for the anisotropic window function effect using a plane-parallel approximation was presented in \citet{Wilson:2015lup}. That method has the advantage that the results of the window function convolutions can be computed much faster by means of 1D FFTs.
\citet{Beutler:2016arn} show that this technique can be extended to wide surveys such as BOSS.
However, as the window matrix is computed only once and this calculation does not represent a significant fraction of the total computing time of our analysis, switching to this new technique would not represent a significant improvement in our methodology.}

\section{The Modelling of redshift-space clustering wedges}
\label{sec:model}

An accurate model of the redshift-space galaxy clustering statistics is a key element for precise cosmological constraints from galaxy clustering analysis.
Our \changed{power spectrum} fits make use of a state-of-the-art description of the effects of the non-linear evolution of density fluctuations, bias and RSD that allowed us to extract information from the full shape of our clustering measurements including smaller scales than in previous studies. 
The analyses of our companion papers \citet{Sanchez:2016b} and \citet{Salazar-Albornoz:2016psd} are based on the same model.
The modelling of the non-linear matter power spectrum is described in section~\ref{sec:gRPT}.
The galaxy bias model and the theoretical framework for RSD are summarized in sections~\ref{sec:galaxy_bias} and \ref{sec:RSD_model}, respectively.
The parameter space of our model for the Fourier-space wedges is summarized in section~\ref{sec:model_summary}.
In section~\ref{sec:model_performance}, we present performance tests of this model based on a set of large-volume \Nbody simulations, as well as synthetic catalogues for the DR12 combined sample.
Within the BOSS collaboration, the performances of the various \changed{full-shape clustering} analysis techniques used for the DR12 combined sample are compared with each other and checked for systematics by means of the analysis of a set of `challenge' catalogues.
\changed{Details on the generation of these catalogues and the accuracy with which each method recovers the simulated distance and growth parameters can be found in Tinker et al. (\Inprep).}
Our RSD challenge results are described in section~\ref{sec:challenge_mocks}.

In order to test the model on artificial catalogues that match the clustering properties of the BOSS combined sample, we also performed fits of the wedges $P_{3{\rm w}}(k)$ obtained from the set of \MDPatchy mocks.
These fits also serve as a basis for the estimation of the cross covariance between the results of the different analysis approaches that are applied to the BOSS combined galaxy sample, as described in \citet{Sanchez:2016a}.
This estimate is needed to generate the consensus distance and growth measurements of \citet{Alam:2016hwk}.
\citet{Sanchez:2016b} present complementary tests of the model using the correlation function wedges.

\subsection{The modelling of the redshift-space clustering}
\label{sec:zspace_clustering_model}

\subsubsection{Non-linear gravitational dynamics}
\label{sec:gRPT}

\changed{The constraining power of galaxy clustering measurements increases as smaller scales are included in the analysis.
However,} this requires a careful modelling of the real- and redshift-space galaxy two-point statistics beyond the linear regime.

\changed{Our model of} the non-linear matter power spectrum wedges \changed{is based on} gRPT (Blas, Crocce \& Scoccimarro, \Inprep), a new version of RPT \citep{Crocce:2005xy} and later developments such as RegPT \citep{Bernardeau:2008fa}.
This approach uses the symmetries of the equations of motion to resum the mode-coupling power spectrum consistently with the resummation of the propagator in order to avoid symmetry-breaking one-loop approximations of the mode-coupling term.
The one-loop gRPT approximation allows us to to predict the matter power spectrum inferred from $N$-body simulations with an accuracy sufficient for our analysis up to $k \sim 0.25 \; h \, \unit{Mpc}^{-1}$.
This corresponds to a significant improvement over previous fast implementations along these lines \citep[e.g. `{\sc MPTbreeze}';][]{Crocce:2012fa}.
A more detailed description of the theoretical framework for the non-linear gravitational dynamics of the model is given in Blas et al. (\Inprep).
\citet{Sanchez:2016b} describe the implementation of this model in our analysis pipeline in more detail.

\subsubsection{The modelling of galaxy bias}
\label{sec:galaxy_bias}

As galaxies are biased tracers of the total matter, we consider the non-linear and non-local contributions to the galaxy bias in order to achieve improved accuracy.
Assuming the velocity field to be bias free, \changed{our} galaxy bias prescription consistently includes terms \changed{up to} second-order Lagrangian bias \citep*{Chan:2012jj}.
\changed{T}he galaxy density contrast $\delta_\mathrm{g}$ is given by
\begin{align}
 \delta_\mathrm{g} = b_1 \, \delta_\mathrm{m} + \frac{b_2}{2} \, \delta^2_\mathrm{m} + \gamma_2 \, \mathcal{G}_2[\phi_v]  + \gammaEm \, \Delta_3 \mathcal{G}[\phi,\phi_v] + \ldots,
\end{align}
\changed{Here,} $\delta_\mathrm{m}$ is the matter density contrast, $b_1$ and $b_2$ are the linear and second-order local bias, respectively, and $\gamma_2$ and $\gammaEm$ are non-local bias terms of second order.
The `Galileon' operators $\mathcal{G}_2$ and $\Delta_3 \mathcal{G}$ of the gravitational potential $\phi$ and the velocity potential $\phi_v$ are given by
\begin{align}
 \mathcal{G}_2[\phi_v] &= \left( \nabla_{ij} \phi_v \right)^2 - \left( \nabla^2 \phi_v \right)^2, \\
 \Delta_3 \mathcal{G}[\phi,\phi_v] &= \mathcal{G}_2[\phi] - \mathcal{G}_2[\phi_v].
\end{align}
In principle, the non-local bias terms can be expressed in terms of the first-order bias assuming a local bias in Lagrangian coordinates \citep{Chan:2012jj},
\begin{align}
 \label{eq:local_Lagrangian}
 \gamma_2 &= -\frac{2}{7} \, (b_1 - 1), & \gammaEm &= \frac 3 2 \times \frac{11}{63} \, \left( b_1 - 1 \right).
\end{align}
\changed{Our tests on N-body simulations show that treating $\gammaEm$ as a free parameter yields more accurate results than fixing it to the local-Lagrangian prediction.
This is consistent with recent studies showing that Eulerian bias is not necessarily compatible to local-Lagrangian bias in the non-linear regime \citep{Matsubara:2011ck}.
Thus, we vary $\gammaEm$ independently of $b_1$ in our fits.
However, we notice that the precise value of $\gamma_2$ has little impact on our theoretical predictions and we use the local-Lagrangian relation of equation~\eqref{eq:local_Lagrangian} to relate this parameter to a given $b_1$.  
These choices are further discussed in \citet[section~3.1.2]{Sanchez:2016b}.}

\subsubsection{Modelling redshift-space distortions}
\label{sec:RSD_model}

To linear order in Lagrangian perturbation theory \citep[1-LPT;][]{Zeldovich:1970:G}, the effect of RSD is given by a velocity field whose divergence is proportional to the density contrast.
The coefficient of this dependence is the growth-rate parameter $f(z)$, defined by
\begin{equation}
 \label{eq:f_growth}
 f(z) \equiv \Diff{\ln D}{\ln a}(z).
\end{equation}
Here, $D(z)$ is the linear growth function and $a(z)$ the scale factor.
Thus, the redshift-space clustering signal can be used as a probe of the growth of structure \citep{Guzzo:2008}.

Quasi-linear perturbative approaches for the RSD have been developed in \citet{Scoccimarro:2004tg}, 
\citet{Percival:2008sh} and \citet*{Taruya:2010mx}.
For a more advanced modelling of the non-linear effects, we use the one-loop approximation of the Gaussian generation function approach in \citet*{Scoccimarro:1999ed} for the redshift-space power spectrum, $\tilde P_\mathrm{zs}(k,\mu)$, which yields \citep[compare to equations~(19) and (20) in][]{Sanchez:2016b}
\begin{multline}
 \label{eq:Pzs}
 \tilde P(k,\mu) = \bigg\{ \int \frac{\dnx{3}{r}}{(2 \pi)^3} \rft{\V k}{\V r} \big[ \average{ D_s D_s\Prime }_\mathrm{c} + \lambda \average{ \Delta u_z D_s D_s\Prime }_\mathrm{c} \\
 + \lambda^2 \average{ \Delta u_z D_s }_\mathrm{c} \, \average{ \Delta u_z D_s\Prime }_\mathrm{c} \big]
 \bigg\} \, W(k, \mu),
\end{multline}
where $\lambda = \ii f k \mu$, $D_s = \delta_\mathrm{g} + f \Delta_z u_z$, and a prime denotes evaluation at $\VPrime x \equiv \V x + \V r$ instead of $\V x$.
Defining the velocity divergence $\theta \equiv \nabla \cdot \V v$ and assuming no velocity bias, the first term is the non-linear version of the Kaiser formula in Fourier space,
\begin{equation}
 \label{eq:nonlin_Kaiser}
 P^{(1)}_\mathrm{zs}(k,\mu) = P_\mathrm{gg}(k) + 2 f \mu^2 P_{\mathrm{g}\theta}(k) + f^2 \mu^4 P_{\theta\theta}(k),
\end{equation}
depending on $P_\mathrm{gg} = \average{ \delta_\mathrm{g} \, \delta_\mathrm{g} }$, $P_{\mathrm{g}\theta} = \average{ \delta_\mathrm{g} \, \theta }$, and $P_{\theta\theta} = \average{ \theta \, \theta }$.
The other two terms are given by a three-level PT bispectrum contribution between the densities and velocities,
\begin{multline}
 \label{eq:first_one_loop_term}
 P^{(2)}_\mathrm{zs}(k,\mu) = \int \frac{q_z}{q^2} \big[
  B_{\theta D_s D_s}(\V q, \V k - \V q, -\V k) \\
  + B_{\theta D_s D_s}(\V q, -\V k, \V k - \V q)
 \big],
\end{multline}
and a quadratic linear-theory power spectrum expression,
\begin{multline}
 \label{eq:second_one_loop_term}
 P^{(3)}_\mathrm{zs}(k,\mu) = \int \frac{q_z \, (k_z - q_z)}{q^2 (\V k - \V q)^2} \, (b_1 + f \mu_q^2) \, (b_1 + f \mu_{k-q}^2) \\
  \times P_{\delta\theta}(k - q) \, P_{\delta \theta}(q) \, \dnx{3}{q} \big],
\end{multline}

Further, $W(k, \mu)$ is the generating function of velocity differences \changed{which, in the large-scale limit,} we describe as
\begin{equation}
 \label{eq:Wvir}
 W(k,\mu) \equiv \frac{1}{\sqrt{1 + f^2 \, \mu^2 \, k^2 \, \avir^2}} \ee{\frac{-f^2 \, \mu^2 \, k^2}{1 + f^2 \, \mu^2 \, k^2 \, \avir^2}},
\end{equation}
where $\avir$ is a free parameter that describes the kurtosis of the small-scale velocity distribution.
The factor $W(k,\mu)$ is usually associated with the `Fingers-of-God' (FOG) effect caused by the non-linear velocity component due to virialization. 

The power spectrum multipoles can be obtained by integrating equation (\ref{eq:Pzs}) against the Legendre polynomials $\Lp(\mu)$.
From now on we refer to our model as `gRPT+RSD'.
More details on the implementation of this model can be found in \citet{Sanchez:2016b}.

A similar description for the non-linear RSD effect, dubbed the `eTNS model' \citep{Taruya:2010mx,Nishimichi:2011jm}, is based on the same approach and was used in previous analyses of \changed{galaxy clustering measurements from BOSS} \citep{Oka:2013cba,Beutler:2013yhm,Gil-Marin:2014sta,Gil-Marin:2015sqa}.
That model differs from our method in certain aspects:
first, the second-order bias contributions (\changed{depending on} $b_2$ and $\gamma_2$) to the first corrective one-loop term in equation~\eqref{eq:first_one_loop_term} are dropped, \changed{while in} our approach, these terms are kept in order to consistently include \changed{all} second-order bias terms.
Second, our FOG term in equation~\eqref{eq:Wvir} is non-Gaussian.
Third, \changed{we treat $\gamma_3$ as a free parameter instead of fixing its value according to the local-Lagrangian relation}.
Fourth, \changed{our the predictions of the nonlinear matter power spectrum are computed using gRPT} instead of RegPT.

\subsubsection{Modelling the Alcock-Paczynski effect}
\label{sec:AP_modelling}

\begin{table}
 \centering
 \caption{The sound horizon scale $\rd \equiv \rs(\zd)$ at the drag redshift (in units of $\unit{Mpc}$) and the derived angular diameter distances $\DAz$ and Hubble parameters $\Hz$ (in units of $\unit{Mpc}$ and $\unit{km \, s^{-1} \, Mpc^{-1}}$, respectively) for the cosmologies specified in Table~\ref{tab:dr12_cosmologies} at the effective redshifts $\zeff$ of the ranges defined in Table~\ref{tab:dr12_z_ranges}.}
 \label{tab:distance_quantities}
 \addtolength{\tabcolsep}{-1pt}
 \begin{tabular}{lllllllll}
  \hline
  Cosmology      & $\rd$ & \multicolumn{3}{c}{$\DAz$} & \multicolumn{3}{c}{$\Hz$} \\
  \multicolumn{1}{r}{$\zeff$} &       & $0.38$ & $0.51$ & $0.61$   & $0.38$ & $0.51$ & $0.61$  \\
  \hline
  Fiducial       & $147.8$      & $1109$ & $1313$ & $1433$ & $82.9$ & $89.6$ & $95.2$ \\
  \Minerva       & $148.5$      & \multicolumn{3}{c}{$\zeff = 0.57$: $1364$} & \multicolumn{3}{c}{$93.7$} \\
  \QPM           & $147.1$      & $1077$ & $1277$ & $1395$ & $84.9$ & $91.5$ & $97.0$ \\
  \MDPatchy      & $147.7$      & $1107$ & $1311$ & $1431$ & $83.0$ & $89.7$ & $95.5$ \\
  Template       & $147.3$      & $1112$ & $1316$ & $1436$ & $82.8$ & $89.5$ & $95.2$ \\
  \hline
 \end{tabular}
 \addtolength{\tabcolsep}{1pt}
\end{table}

\begin{table}
 \centering
 \caption[The model parameters of the BAO+RSD fits and their priors.]{The parameter space $\mathcal{X}$ of our full-shape fits with the gRPT+RSD model.
  BAO+RSD fits use the distortion, growth, bias, RSD, and shot-noise parameters.
  Fits for the cosmological interference use the bias, RSD, and shot-noise parameters, besides the parameters of cosmological model and the nuisance parameters of the complementary cosmological probes.
  All parameters have a flat prior that is uniform within the given limits and zero outside.}
 \label{tab:parameters_and_priors}
 \begin{tabular}{rlll}
  \hline
  Param{.}   & Function                    & Unit                     & Prior limits      \\
  \hline
  \multicolumn{4}{c}{Bias} \\
  \hline
  $b_1$      & Linear bias                 & $-$                      & $0.5$--$9$        \\
  $b_2$      & Second-order bias           & $-$                      & $(-4)$--$4$       \\
  $\gammaEm$ & Non-local bias              & $-$                      & $(-3)$--$3$       \\
  \hline
  \multicolumn{4}{c}{RSD} \\
  \hline
  $\avir$    & FoG kurtosis                & $-$                      & $0.2$--$10$       \\
  \hline
  \multicolumn{4}{c}{Shot noise} \\
  \hline
  $\SN$      & Extra shot noise$^\ast$     & $h^{-3} \, \unit{Mpc}^3$ & $(-1800)$--$1800$ \\
  \hline
  \multicolumn{4}{c}{AP Distortion} \\
  \hline
  $\qperp$   & $k_\perp$ rescaling         & $-$                      & $0.5$--$1.5$      \\
  $\qpara$   & $k_\parallel$ rescaling     & $-$                      & $0.5$--$1.5$      \\
  \hline
  \multicolumn{4}{c}{Growth} \\
  \hline
  $\fsig$    & Growth-rate factor  & $-$                      & $0$--$3$          \\
  \hline
 \end{tabular}
  \vskip 5pt
  \begin{flushleft}
  {$^\ast$ In the case of the low-redshift bin, $\SN$ is varied within $(-1000)$--$1000$ as the estimate for the Poisson shot-noise is also smaller.}
  \end{flushleft}
\end{table}

The clustering measurements inferred from real galaxy catalogues depend on the assumption of a fiducial cosmology used to transform the observed redshifts into distances. 
A mismatch between the assumed and true cosmologies leads to a geometrical distortion (the AP effect) corresponding to a rescaling of the wavenumbers transverse, $k_\perp$, and parallel, $k_\parallel$, to the LOS direction as
\begin{equation}
 \label{eq:distortion_parameters}
 k_\perp'=\qperp k_\perp  \quad \text{and} \quad k_\parallel' = \qpara k_\parallel,
\end{equation}
where the primes denote quantities observed assuming the fiducial cosmology and the two distortion parameters $\qpara$ and $\qperp$ are given by
\begin{equation}
 q_\perp = \frac{D_{\rm A}(\zeff)}{D_{\rm A}'(\zeff)} \quad \text{and} \quad
 q_\parallel = \frac{H'(\zeff)}{H(\zeff)},
\end{equation}
that is, the ratios of the angular diameter distance\changed{, $\DAz$,} and the Hubble parameter\changed{, $\Hz$,} in the true and fiducial cosmologies at the effective redshift of the sample, $\zeff$.

The theoretical prediction for the distorted power spectrum wedges, 
\changed{$\PAtheo[\mu_1]^{\mu_2}(k')$}
can be computed as
\begin{equation}
 \PAtheo[\mu_1]^{\mu_2}(k') = \frac{\qperp^{-1} \, \qpara^{-2}}{\mu_2 - \mu_1} \int_{\mu_1}^{\mu_2} \tilde P(k(k', \mu'), \mu(k', \mu')) \dint \mu',
 \label{eq:distorted_pk2d}
\end{equation}
where $\tilde P(k,\mu)$ is the model prediction of equation~\eqref{eq:Pzs} \changed{and the relations
\begin{align}
 k(\mu', k') &\equiv k' \sqrt{\qpara^{-2} \, (\mu')^2 + \qperp^{-2} \left( 1 - (\mu')^2 \right)} \\
 \mu(\mu', k') &\equiv \mu' \, \qpara^{-1} \left[ \qpara^{-2} \, (\mu')^2 + \qperp^{-2} \left( 1 - (\mu')^2 \right) \right]^{-1/2},
\end{align}
correspond to those of equation~\eqref{eq:distortion_parameters} expressed in terms of 
$k$ and $\mu$ \citep{Ballinger:1996}.}
The scaling of the power spectrum with $\qperp^{-1} \, \qpara^{-2}$ is due to the volume distortion from the AP effect.

In BAO distance measurements, the results rely on a prediction for the underlying power spectrum, computed using a fixed `template' cosmology.
\changed{$\DAz$ and $\Hz$} are measured relative to the sound horizon scale at the drag redshift, $\rd \equivß \rs(\zd)$, of the template.
The distortion parameters $\qperp$ and $\qpara$ of equation~\eqref{eq:distortion_parameters} only take into account the geometric AP effect.
Thus, results that are comparable across different analyses (using different templates) can be obtained by defining a second set of AP parameters, which also include an additional rescaling of the angular-diameter distance $\DAz$ and the Hubble parameter $\Hz$ by the fiducial sound horizon scale, $\rs'(\zd)$,
\begin{equation}
 \label{eq:AP_parameters}
 \aperp \equiv \frac{\DMz}{\DV[M]'(\zeff)} \frac{\rd'}{\rd} \quad \text{and} \quad \apara \equiv \frac{H'(\zeff)}{\Hz} \frac{\rd'}{\rd}.
\end{equation}
Table~\ref{tab:distance_quantities} lists the values of $\DAz$, $\Hz$, and $\rd$ for the different cosmologies used in this work.

\subsection{Summary of the model parameters}
\label{sec:model_summary}

\changed{We perform two kinds of cosmological clustering analyses.
For the first type, we use a fixed set of cosmological parameters (to which we refer as our `template' cosmology) to predict a template for the two-dimensional power spectrum.
Then, we distort the template according to equation~(\ref{eq:distorted_pk2d}) in order to constrain the AP parameters of equation~(\ref{eq:AP_parameters}).
We refer to this method as `BAO+RSD' fits in the following.
Second, we perform `cosmological full-shape fits', for which we explore the parameter space of a given cosmological model directly.
This means that the predictions for the power spectrum wedges are directly computed for the parameters being considered and then compared with the observed Fourier-space wedges.
Thus, the parameter spaces of the two fitting methods are not exactly the same.
We explore these parameter spaces by means of the Markov chain Monte Carlo (MCMC) technique.}

For our BAO+RSD fits, the shape of the input power spectrum is kept fixed.
Variations of the cosmology are modelled by treating the distortion parameters 
$\qperp$ and $\qpara$ and the growth rate $\fsig$ as free parameters. We account for possible 
deviations from a pure Poisson shot noise with a free, 
constant, and additive shot-noise contribution $\SN$ to all power spectrum wedges.
Thus, the full parameter space $\mathcal{X}$ for \changed{these fits} consists of 8 parameters,
\begin{equation}
 \V \zeta = \transp{( \qperp, \, \qpara, \, \fsig, \, b_1, \, b_2, \, \gammaEm, \, \avir, \, \SN )} \in \mathcal{X}.
\end{equation}
\changed{When performing fits on the real BOSS data or our mock catalogues we allow for different values of the the parameters $\{ b_1, b_2, \gammaEm, \avir, \SN \}$ for the NGC and SGC sub-samples in the low-redshift bin, increasing the total number of parameters to 13.}

For the cosmological fits of section~\ref{sec:cosmological_implications}, full model predictions must be computed \changed{for each point in the parameter space being considered}.
In this case, $\qperp$ and $\qpara$ and $\fsig$ are not treated as free parameters and are instead derived from the cosmological parameters being tested.

\changed{The MCMC are constructed} using the July 2015 \changed{version} of 
\code{CosmoMC}\footnote{\url{http://cosmologist.info/cosmomc/}} \citep{Lewis:2002ah} modified to compute the gRPT+RSD model predictions as described in section~\ref{sec:zspace_clustering_model}.
Further details can be found in our companion paper \citep{Sanchez:2016b}.
Using the MCMC technique, the choice of the prior distribution can have an influence on the accuracy of the obtained parameter constraints.
We choose flat priors on all parameters given by the limits listed in Table~\ref{tab:parameters_and_priors}.
The chains are considered converged if the Gelman-Rubin convergence criterion \citep{Gelman:1992:I} satisfies $R - 1 < 0.02$.

\subsection{Performance of the Model}
\label{sec:model_performance}

\subsubsection{Model verification with full non-linear simulations}
\label{sec:minerva_verification}

\begin{figure}
 \centering
 \includegraphics[width=\columnwidth]{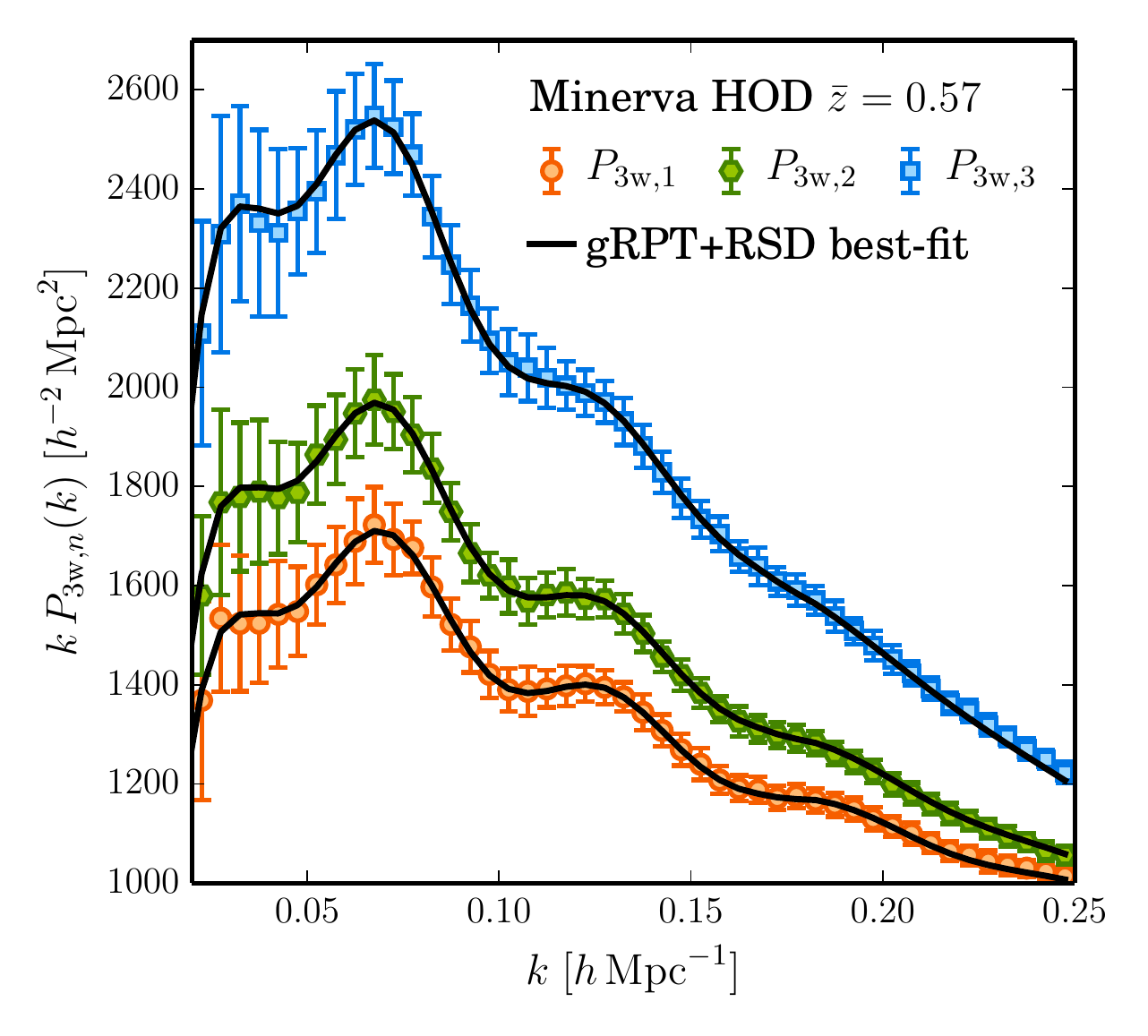}
 \caption{Best-fitting gRPT+RSD model to the mean power spectrum wedges of the \Minerva HOD sample using $\kmin = 0.01  \; h \, \unit{Mpc}^{-1}$ and $\kmax = 0.2 \; h \, \unit{Mpc}^{-1}$.
  The cosmology was fixed, \ie, $\qpara = \qperp = 1$, $\Fsig = 0.473$ (\Cf, \Minerva in Table~\ref{tab:dr12_cosmologies}).}
 \label{fig:rsd_model_best_fit_minerva}
\end{figure}

As a first test of the model described in section~\ref{sec:zspace_clustering_model}, we used the \Minerva \Nbody simulations described in \citet{Grieb:2015bia}.  
These are a set of 100 large-volume \Nbody simulations run using \code{Gadget}\footnote{The latest public release is \code{Gadget-2} \citep{Springel:2005mi} which is available at \url{http://www.gadgetcode.org/}.}.
Each realization is a cubic box of side length $1500 \; h^{-1} \, \unit{Mpc}$ with $1000^3$ dark-matter (DM) particles.
The initial conditions (at redshift $z_\mathrm{ini} = 63$) were derived using second-order Lagrangian perturbation theory (2-LPT)\footnote{A 2-LPT code for generating initial conditions is available at \url{http://cosmo.nyu.edu/roman/2LPT/}.} from a linear \code{Camb} \citep*{Lewis:1999bs} power spectrum whose cosmological parameters were chosen to match 
the best-fitting results of the \WMAP[9] + BOSS DR9 $\xi(r)$ analysis \citep[Table I]{Sanchez:2013uxa}, \changed{which are listed as `Minerva' in Table~\ref{tab:dr12_cosmologies}}.
At each redshift output $z \in \{2.0, 1.0, 0.57, 0.3, 0\}$, the DM particle positions and velocities were stored along with the halo catalogues obtained with a friends-of-friend algorithm, which were later post-processed with {\sc Subfind} \citep{Springel:2000qu} to eliminate spurious unbound objects. 
The halo mass resolution is $m_\mathrm{min} = 2.67 \cdot 10^{12} \; M_\Sun / h$.

We used the snapshot at $z = 0.57$ to obtain galaxy catalogues comparable to the CMASS sample by populating the haloes and subhaloes with galaxies according to a halo occupation distribution 
(HOD) model with suitable parameters, as described in \citet{Grieb:2015bia}.
The final synthetic galaxy catalogues have a mean galaxy density of 
$\nbar \approx 4 \sciexp{-4} \; h^3 \, \unit{Mpc}^{-3}$ and a linear bias of $b \approx 2$.

\begin{figure}
 \centering
 \includegraphics[width=\columnwidth]{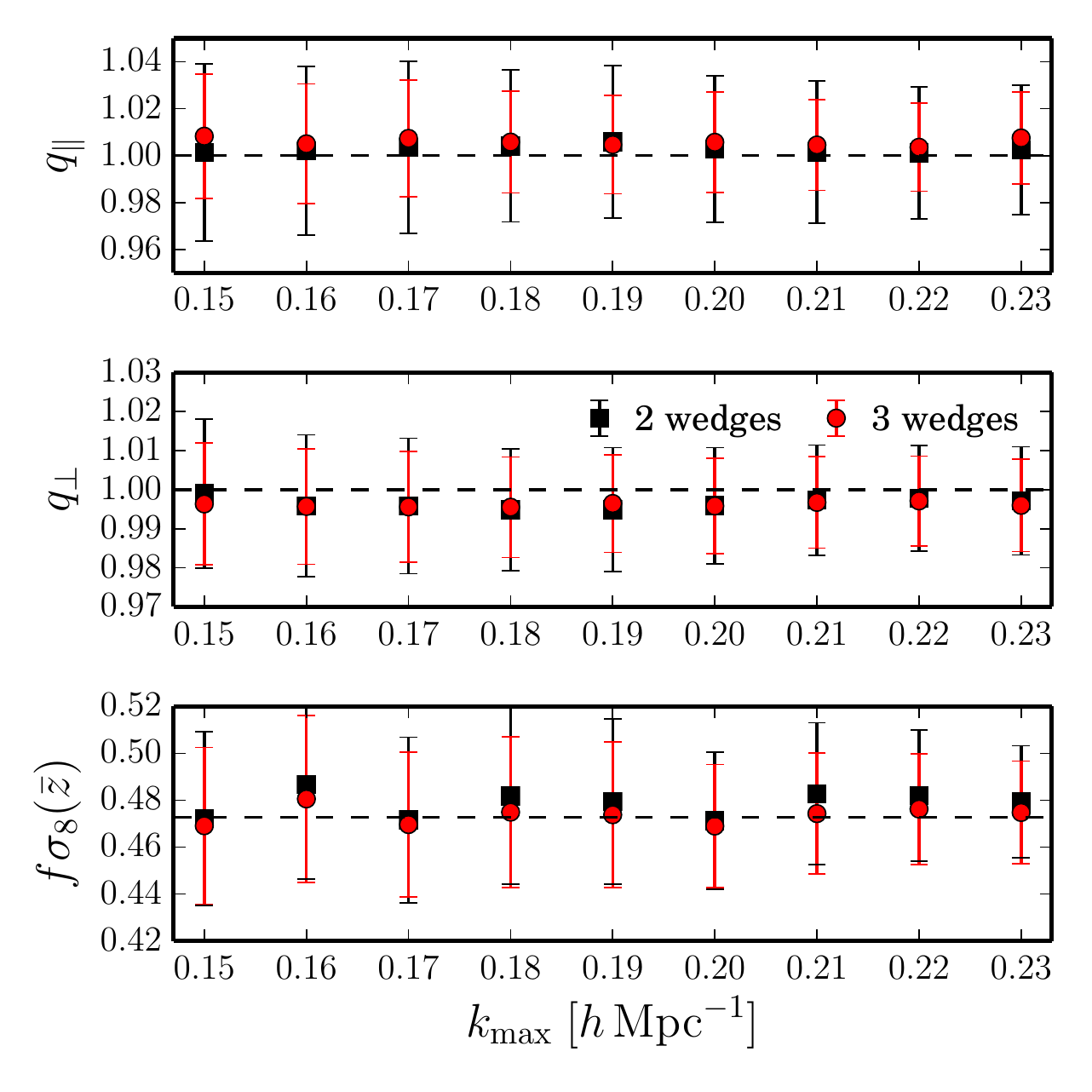}
 \caption{Marginalized results for $\qpara$, $\qperp$, and $\Fsig$ from gRPT+RSD model fits to the mean Fourier space wedges of the \Minerva HOD samples using different fitting ranges $0.01 \; h \, \mathrm{Mpc}^{-1} \leq k_i \le k_\mathrm{max}$.
  The fits using three wedges (red) has significantly smaller error bars than for two wedges.}
 \label{fig:rsd_model_fits_minerva}
\end{figure}

The points in Fig.~\ref{fig:rsd_model_best_fit_minerva} show the mean \Minerva HOD power spectrum wedges, which \changed{we use} as a test case to validate the model described in section~\ref{sec:zspace_clustering_model} using a \changed{sample with similar clustering properties as the real CMASS galaxies}.
We used these measurements in the range $\kmin = 0.01  \; h \, \unit{Mpc}^{-1}$ and $\kmax = 0.2 \; h \, \unit{Mpc}^{-1}$ to fit for the nuisance parameters of the model, while fixing all cosmological parameters to their underlying values.
For this test, we use the recipe for the theoretical covariance matrix for clustering wedges in Fourier space presented in \citet{Grieb:2015bia}.
The error bars in Fig.~\ref{fig:rsd_model_best_fit_minerva} correspond to the square root of the diagonal entries of the resulting covariance matrix.
The solid lines in Fig.~\ref{fig:rsd_model_best_fit_minerva} correspond to the model computed using the resulting values for the nuisance parameters, showing excellent agreement with the results from the \Minerva simulations that extends even into the non-linear regime outside the \changed{range of scales included in the fits}.

In order to validate the wavenumber \changed{range} for which the model provides the tightest 
unbiased estimates of the distortion and growth parameters, we use the gRPT+RSD model to perform \changed{BAO+RSD} fits to the mean \changed{power spectrum wedges} of the \Minerva HOD sample using two and three $\mu$-bins as a function of the upper limit of the fitting range, $\kmax$.
The obtained results, shown in Fig.~\ref{fig:rsd_model_fits_minerva}, are in excellent agreement with the correct values of these parameters for the case of two and three wedges\changed{, but we find a higher accuracy for the latter case}.
The marginalized confidence intervals of the distortion parameters are not exactly centred on the \changed{true} values, which we find is due to the \changed{correlation between these parameters and the additional shot noise contribution $\SN$}.
Although \changed{that} parameter is not necessary to fit the results of the \Minerva simulations, \changed{we included it to mimic the analysis that we apply to the real BOSS data, where it is required to account for the uncertainties in the shot-noise level of our clustering measurements}. 
However, the results from \Minerva show that the potential systematic errors introduced by this parameter are much smaller than the statistical error for a single \Minerva volume.\footnote{The volume of a single \Minerva realization, $V = (1500 \; h^{-1} \, \unit{Mpc})^3$, is roughly equivalent to the effective volume of the full BOSS combined sample, \Cf, Table~\ref{tab:dr12_z_ranges}.}
Thus, we do not take it into account for the RSD analyses in the following.
Due to the higher constraining power of the analyses with three wedges over using two wedges only, from now on we present results obtained using $\Pobs$ only, restricting the fitting range to $\kmax = 0.2 \; h \, \unit{Mpc}^{-1}$, as we do not see improvements in the recovered mean and error of $\fsig$ for larger $\kmax$.
Measuring a number of wedges that is larger than three from a real survey is impracticable with current methods, and thus we did not include such cases into \changed{our} analysis.

\subsubsection{Model verification with synthetic catalogues for the BOSS DR12 combined sample}
\label{sec:patchy_verification}

\begin{figure}
 \centering
 \includegraphics[width=\columnwidth]{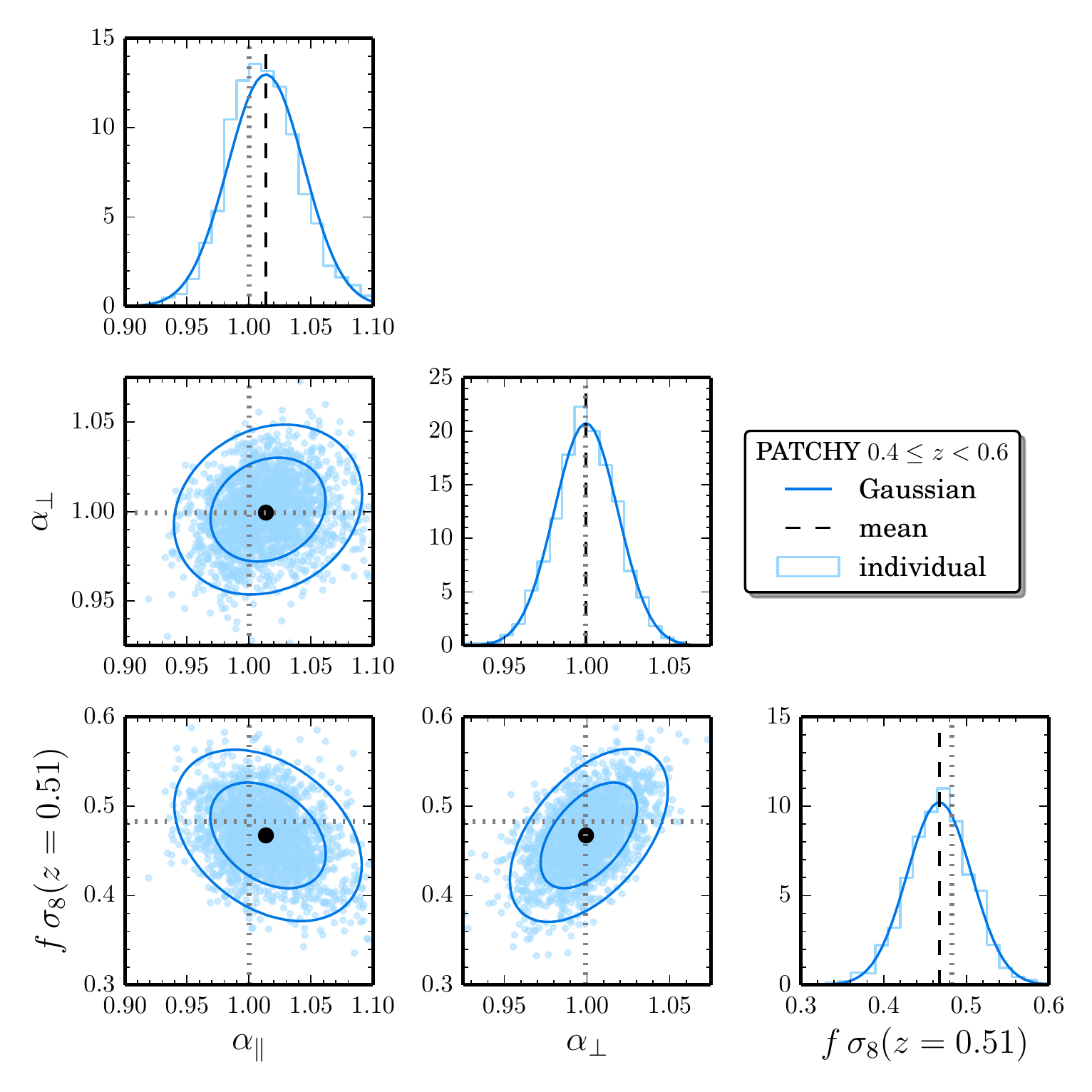}
 \caption{The best-fitting parameters for $\apara$, $\aperp$, and $\Fsig$ from gRPT+RSD model fits 
 to the three Fourier space wedges, derived from the power spectrum multipoles $P_{\ell=0,2,4}(k)$, of the 
 2045 \MDPatchy catalogues in the intermediate redshift bin are indicated as dots in the 2D plots in the panels below 
 the diagonal.
 Their histograms are plotted in the diagonal panels in fainter colours.
 In all panels, the Gaussian approximations of the parameter distributions are shown in darker colours.
 The mean of the best-fit parameters are indicated as black dashed lines in the histograms and as a 
 black filled circle in the 2D plots.
 The results for the low and high redshift bin are similar but show smaller deviations from the 
 correct values (see Table~\ref{tab:rsd_model_fits_patchy}).}
 \label{fig:rsd_model_fits_patchy}
\end{figure}

\begin{table}
 \centering
 \caption{The results for $\apara$, $\aperp$, and $\Fsig$ from gRPT+RSD model fits to the three Fourier space wedges that were measured from the \MDPatchy catalogues filtering out the information of Legendre multipoles $\ell>4$.
  In all three redshift bins, we the fitting range is $0.02 \; h \, \mathrm{Mpc}^{-1} \leq k_i \leq 0.2 \; h \, \mathrm{Mpc}^{-1}$.
  Here we report the mean and standard deviation of the best-fitting parameters of the 2045 individual fits and compare them to the expected values.
  The low redshift bin fits used separate bias, RSD, and shot noise parameters for NGS and SGC, whereas the other two bins used only one set of nuisance parameters.}
 \label{tab:rsd_model_fits_patchy}
 \begin{tabular}{llll}
  \hline
  Bin          & Parameter & Best-fitting value & Expected \\
  \hline
               & $\aperp$  & $0.996 \pm 0.023$ & $0.999$ \\
  Low          & $\apara$  & $0.998 \pm 0.037$ & $1.000$ \\
               & $\fsig$   & $0.462 \pm 0.048$ & $0.483$ \\
  \hline
               & $\aperp$  & $0.999 \pm 0.020$ & $0.999$ \\
  Intermediate & $\apara$  & $1.014 \pm 0.031$ & $1.000$ \\
               & $\fsig$   & $0.467 \pm 0.039$ & $0.483$ \\
  \hline
               & $\aperp$  & $0.004 \pm 0.020$ & $1.000$ \\
  High         & $\apara$  & $1.004 \pm 0.028$ & $1.001$ \\
               & $\fsig$   & $0.479 \pm 0.038$ & $0.478$ \\
  \hline
 \end{tabular}
\end{table}

The \MDPatchy mocks described in section~\ref{sec:covariance_matrix} can be used to test our modelling of non-linearities and RSD on a sample matching the full redshift range and survey geometry of the BOSS combined sample.
As described in section~\ref{sec:clustering_wedges_fourier_space}, \changed{we measured} the power spectrum wedges of each \MDPatchy mock catalogue by filtering out the information of Legendre multipoles $\ell>4$ \changed{for the three redshift bins} defined in Table~\ref{tab:dr12_z_ranges} \changed{taking into account the effect of the window function of the survey as} described in section~\ref{sec:win_func}.
For consistency with the treatment of the \changed{real BOSS} data, two different sets of bias, RSD, and shot-noise parameters are assumed for the low-redshift bin to account for the two potentially different galaxy populations (see appendix~\ref{sec:NGC_vs_SGC}).
As described in section~\ref{sec:precision_matrix}, the obtained parameter uncertainties must be rescaled by the correction factor $\CM$ of equation~\eqref{eq:rescaled_constraints} in order to account for the impact of sampling noise on the precision matrix.
The rescaling factor is given in Table~\ref{tab:Percival_correction} for $\Nmocks = 2045$.

The \changed{mean} constraints on $\apara$, $\aperp$, and $\Fsig$ \changed{from} the fits to the 2045 individual \MDPatchy measurements are given in Table~\ref{tab:rsd_model_fits_patchy}.
For illustration, the 2D contours and 1D histograms of the best-fitting parameters for the intermediate redshift bin are shown in Fig{.}~\ref{fig:rsd_model_fits_patchy}.
The mean and dispersion of the best-fitting values are in good agreement with the expected values.
The largest systematic deviations are found for $\fsig$ in the low redshift bin and for $\apara$ in the intermediate bin, where they correspond to $\approx 50$ per cent of the statistical errors for one realization, but are significantly smaller in all other cases.

In order to \changed{verify that our treatment of the window function does not introduce} any systematic bias into our analysis, we studied the scale-dependency of the results of the gRPT+RSD fits to the Fourier space wedges.
By varying $\kmin$, we exclude the regime of lower wavenumbers from the fitting range where the window function is more important.
An incorrect treatment of the window function effect can introduce a trend with $\kmin$ in the parameter constraints.
We do not find any dependency of the BAO+RSD results for the mean \MDPatchy $\Pobs$ measurements on $\kmin$, indicating that our window matrix formalism does not induce any systematic bias into our analysis.
We also tested for a scale-dependency of the cosmological parameters due to inaccuracies of our clustering model for the (approximative) non-linear evolution of the clustering obtained from the \MDPatchy catalogues.
We performed fits with varying $\kmax$ and find consistent results, free of systematic trends with $\kmax$, even when smaller scales than our fiducial fitting range are included in the analysis. 

As described in \citet{Sanchez:2016a}, the constraints obtained here from the individual \MDPatchy mocks can be used to compute the cross-covariance matrix between the results inferred from $\Pobs$ and those of the \changed{other} analysis methods applied to the BOSS DR12 combined galaxy sample in our companion papers.
This is a key ingredient in the estimation of the final consensus results presented in \citet{Alam:2016hwk}.

\subsubsection{Fourier Space Results on the Challenge Mocks}
\label{sec:challenge_mocks}

\begin{figure}
 \centering
 \includegraphics[width=\columnwidth]{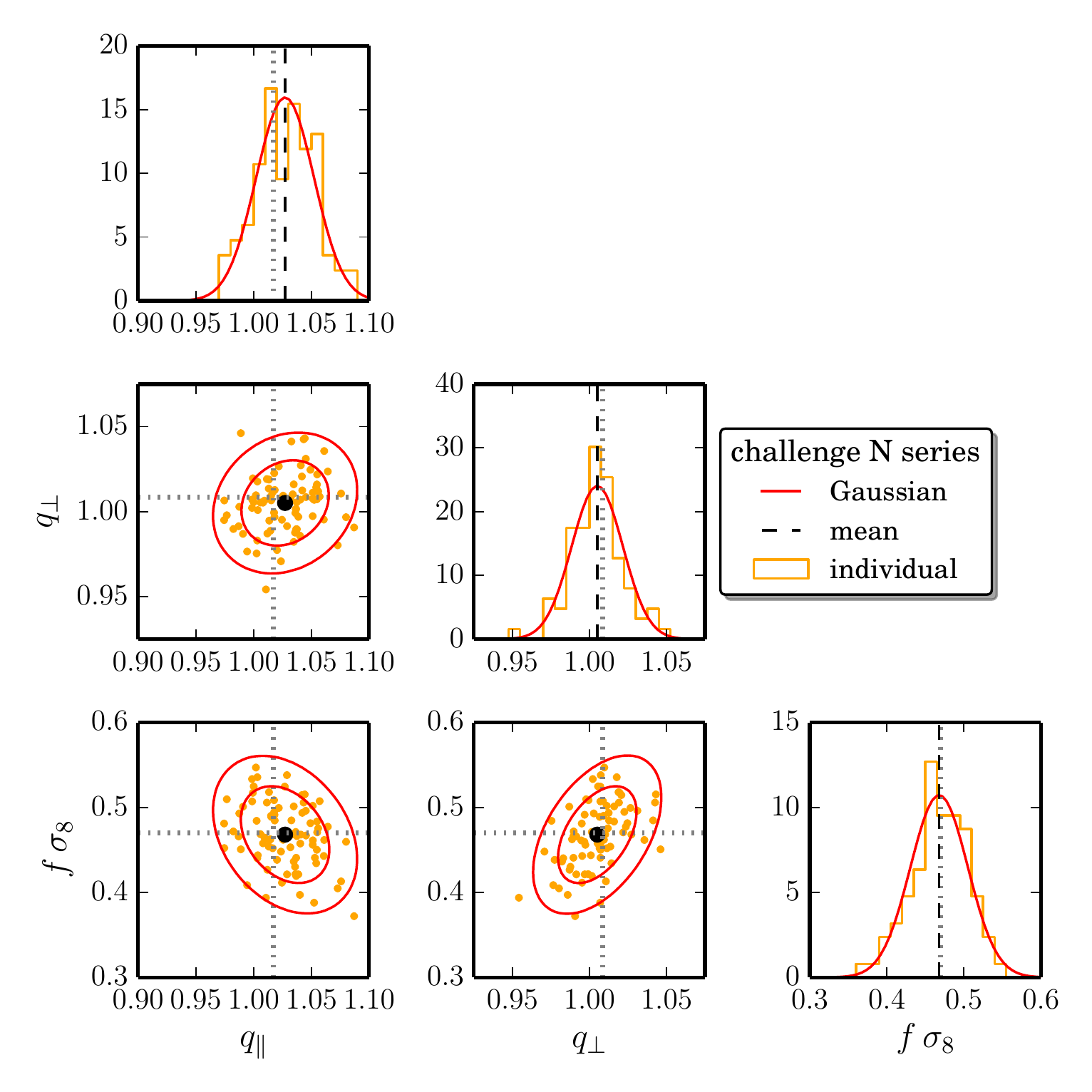}
 \caption{Best-fitting $\qpara$, $\qperp$, and $\Fsig$ parameters from gRPT+RSD model fits to the 84 challenge N 
 series measurement of three Fourier space wedges fitting wavenumbers in the range 
 $0.02 \; h \, \mathrm{Mpc}^{-1} \leq k_i \le 0.2 \; h \, \mathrm{Mpc}^{-1}$.
 The diagonal panels show the histogram of these results, with the mean best-fitting parameters indicated as black 
 dashed lines. The panels below the diagonal show 2D plots with the 84 individual best-fitting parameters 
 as orange dots and the mean as a filled circle.
  A Gaussian fit to the marginalized parameter distribution is plotted as red contours and histograms in all panels.
  The results using two wedges is similarly accurate but less precise.}
 \label{fig:rsd_model_fits_challenge_Nseries}
\end{figure}

\begin{figure*}
 \centering
 \includegraphics[width=.8\textwidth]{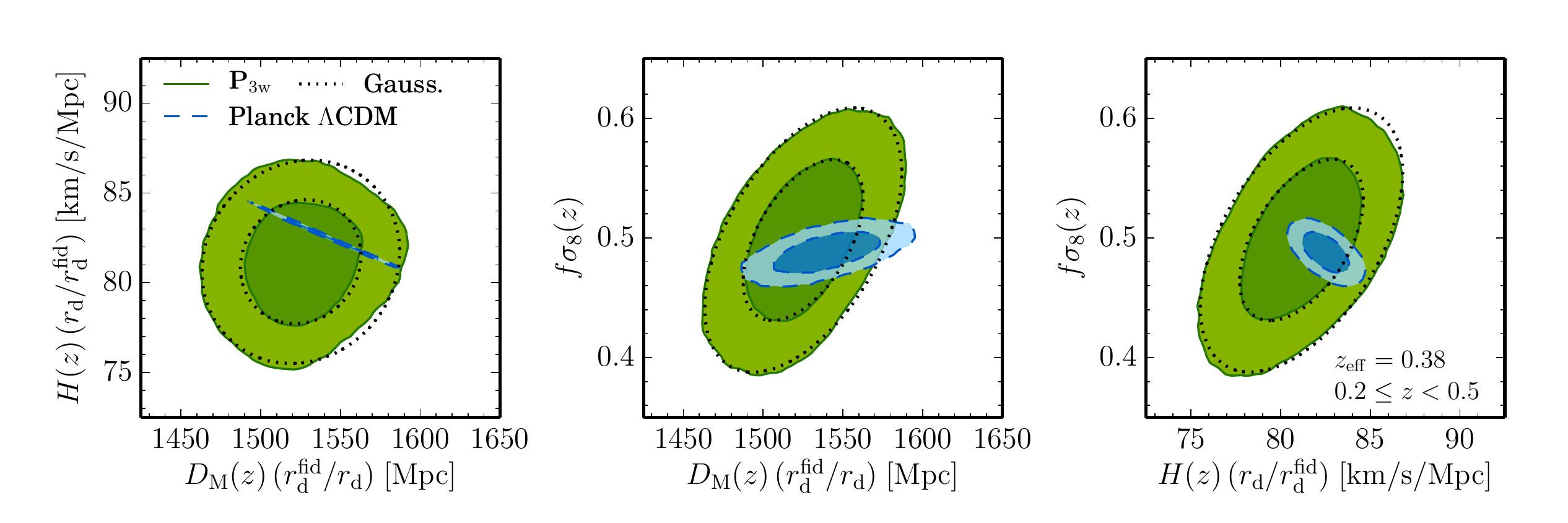}
 \includegraphics[width=.8\textwidth]{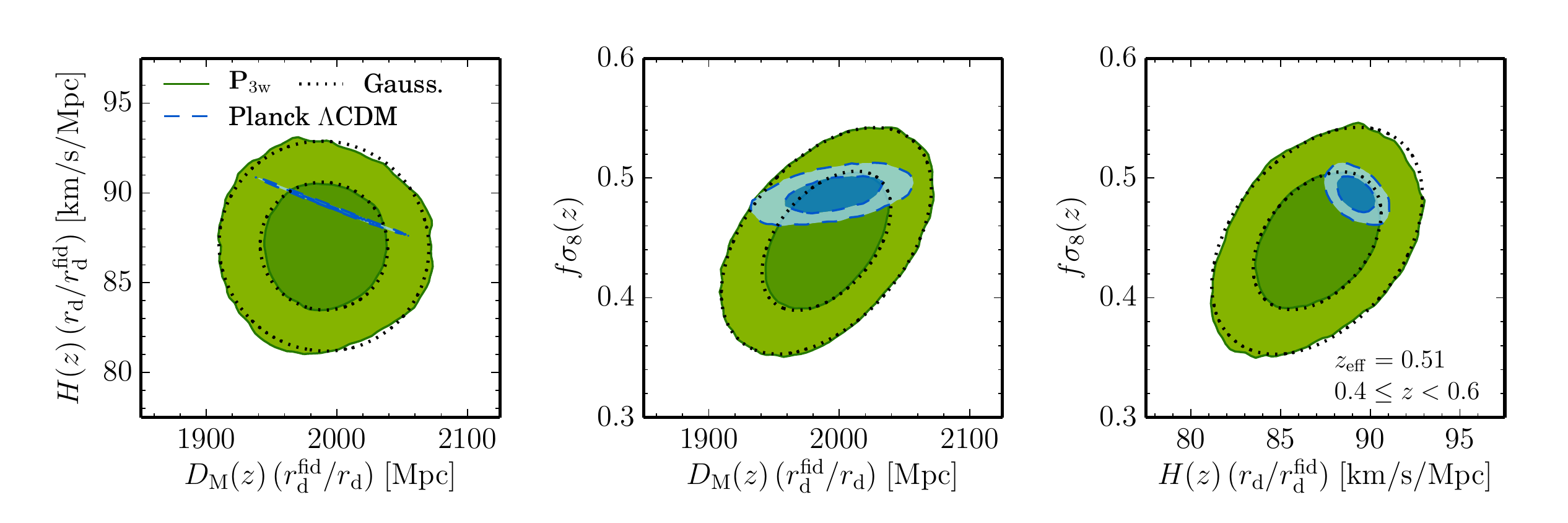}
 \includegraphics[width=.8\textwidth]{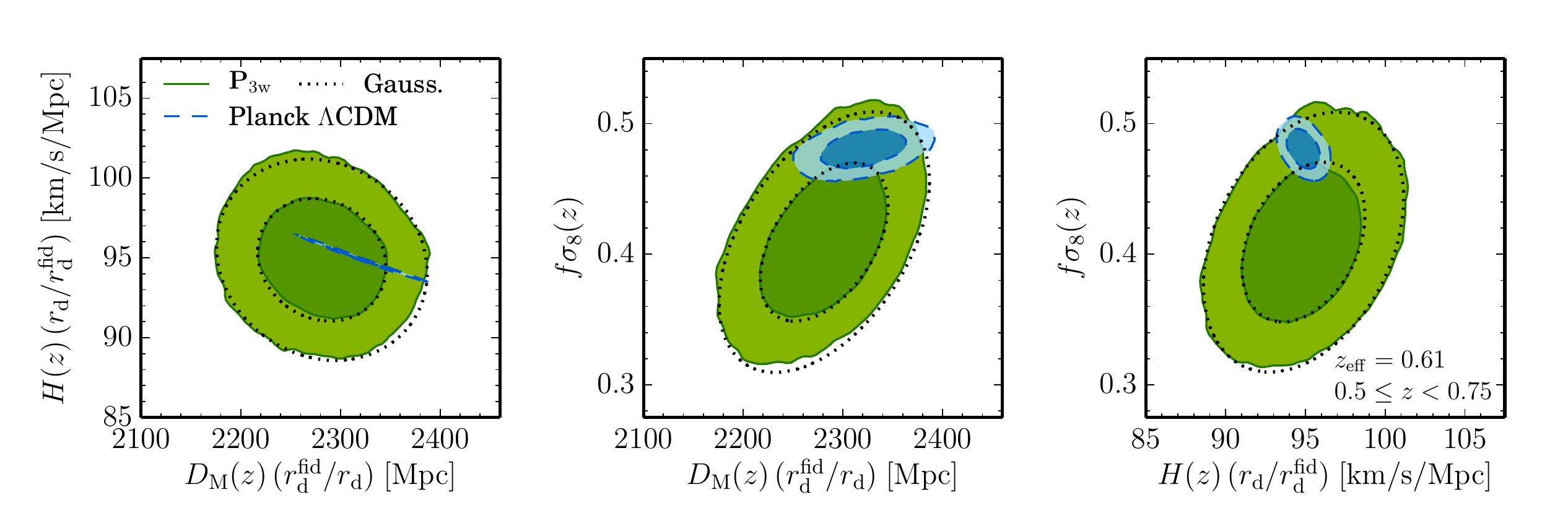}
 \caption{The regions of 68 and 95 per cent CL in the marginalized 2D posteriors of the ratio of the comoving transverse distance and the sound horizon, $\DMz \, \left[ \rs^\mathrm{fid}(\zd) / \rs(\zd) \right]$, the product of the Hubble parameter and the sound horizon, $\Hz \, \left[ \rs(\zd) / \rs^\mathrm{fid}(\zd) \right]$ (these combinations are normalized by the sound horizon of the fiducial cosmology), and the growth parameter $\Fsig$ from BAO+RSD fits to the DR12 combined sample in the low (upper panel), intermediate (middle panel), and high redshift bin (lower panel).
  For these MCMC-estimated contours plotted in green, three power spectrum wedges $\Pobs$ have been fitted in the wavenumber range $0.02 \; h \, \unit{Mpc}^{-1} \le k \le 0.2 \; h \, \unit{Mpc}^{-1}$ using the covariance from 2045 \MDPatchy mocks.
  The low redshift bin fits used separate bias, RSD, and shot noise parameters for the NGC and SGC subsamples, whereas the results in the intermediate and high $z$-bins were obtained using only one set of nuisance parameters.
  For comparison, the theoretical predictions for the standard cosmological model (\LambdaCDM) from the \Planck 2015 TT+lowP \citep{Planck:2015xua} observations are overplotted as blue confidence regions.}
 \label{fig:dr12_comb_contours}
\end{figure*}

Within the BOSS collaboration, special attention was paid to perform stringent cross checks of the different modelling and measurement techniques used in the DR12 analysis of the combined sample, especially for those approaches that are combined \changed{into} the final consensus constraints \citep{Alam:2016hwk}.
Hence the performance of the various methodologies to extract cosmological information from the full-shape approaches are compared in an RSD-fit `challenge' in which the results obtained  all contributing methods are discussed and compared with each other in on large-volume synthetic catalogues to check for possible systematics and the consistency of the results from the different analysis techniques.
\changed{The results of this comparison are described in detail in Tinker et al. (\Inprep).}

The first part of \changed{this comparison} was based on the analysis of seven different HOD galaxy samples constructed out of large-volume \Nbody simulations.
Apart from standard HOD parameters, other non-standard cases, including velocity or assembly bias, are considered.
The simulations correspond to \LambdaCDM cosmologies with slightly different density parameters.
The Fourier space results of the gRPT+RSD model reach the same level of precision as the corresponding configuration space results;
in general, the different methods show excellent accuracy and consistency in the obtained constraints on the challenge catalogues.

The second part of \changed{the model comparison was based on a} set of 84 synthetic catalogues mimicking the DR12 CMASS NGC subsample (dubbed `cut-sky' mocks).
These mocks are designed to test for systematic biases in the obtained parameter constraints as they are all generated from \Nbody simulations assuming the same cosmological parameters and HOD model.
As the full survey geometry is modelled, we measure the multipole-filtered wedges for consistency with the rest of the analysis and the window matrix prescription of section~\ref{sec:win_func} is used to take the selection function into account in our fits.
Fig{.}~\ref{fig:rsd_model_fits_challenge_Nseries} shows the best-fitting distortion and growth 
parameters from the N series fits using three Fourier space wedges.
We obtain results that are in good agreement with those inferred from \Minerva, but the mean $\aperp$ and $\apara$ results found in the light-cone catalogues deviate a little more from the true values.
These deviations are significantly smaller than the statistical uncertainty obtained from a single realization. The results \changed{obtained} using two wedges \changed{show a similar accuracy but are less precise}.

\section{BAO and RSD measurements from the DR12 Fourier space wedges}
\label{sec:BAO_and_RSD_measurements}

\begin{table}
 \centering
 \caption{The 68 per cent CL results of the BAO+RSD fits to the DR12 combined sample 
  power spectrum wedges $\Pobs$, in terms of
  $\DMz[z] \left( \rd^rm{fid} / \rd \right)$,
  $\Hz[z] \left( \rd / \rd^\mathrm{fid} \right)$,
  (expressed in units of $\unit{Mpc}$ and $\unit{km \, s^{-1} \, Mpc^{-1}}$, respectively),
  and the growth parameter $\fsig(z)$ for our three redshift bins.
  We also give the ratio of the angle-averaged BAO distance and fiducial sound horizon scale, $\DVz[z] / \rd$, and 
  the AP parameter $\FAPz[z]$.}
 \label{tab:dr12_comb_results}
 \begin{tabular}{llll}
  \hline
  Parameter & Low & Intermediate & High \\
  \hline
  $\DMz[z] \left( \rd^\mathrm{fid}/\rd \right)$  &
  $1525 \pm 24$ &
  $1990 \pm 32$ &
  $2281_{-43}^{+42}$ \\
  $\Hz[z] \left( \rd/\rd^\mathrm{fid} \right)$ &
  $81.2_{-2.3}^{+2.2}$ &
  $87.0_{-2.4}^{+2.3}$ &
  $94.9 \pm 2.5$ \\
  $\Fsig[z]$  &
  $0.498_{-0.045}^{+0.044}$ &
  $0.448 \pm 0.038$ &
  $0.409 \pm 0.040$ \\
  \hline
  $\DVz[z] / \rd$  &
  $10.05 \pm 0.13$ &
  $12.92 \pm 0.18$ &
  $14.60 \pm 0.24$ \\
  $\FAPz$ &
  $0.424 \pm 0.017$ &
  $0.578 \pm 0.018$ &
  $0.722 \pm 0.022$ \\
  \hline
 \end{tabular}
\end{table}

In this section, we present the \changed{constraints obtained from the BAO+RSD fits of our BOSS clustering measurements}.
For this analysis, the three power spectrum wedges of the DR12 combined sample of each redshift bin are fitted separately\footnote{All results in this and the following section have been obtained by fitting the power spectrum wedges filtering out the information of Legendre multipoles $\ell>4$.
Unless stated otherwise, we use the fiducial wavenumber range of $0.02 \; h \, \unit{Mpc}^{-1} \le k \le 0.2 \; h \, \unit{Mpc}^{-1}$ and the reference covariance matrix obtained from the \MDPatchy mock catalogues (see section~\ref{sec:covariance_matrix}).} using the gRPT+RSD model described in section~\ref{sec:zspace_clustering_model}.
We assume a \Planck 2015 input power spectrum, whose cosmological parameters are listed as `template' in Table~\ref{tab:dr12_cosmologies}.

Using the definitions of the AP parameters in equation~\eqref{eq:AP_parameters} and the fiducial distances given in Table~\ref{tab:distance_quantities}, our results can expressed in terms of the combinations, $\DV[M](z) \, \left( \rd^\mathrm{fid} / \rd \right)$, $H(z) \, \left( \rd / \rd^\mathrm{fid} \right)$, and $\fsig(z)$.
The green contours in Fig{.}~\ref{fig:dr12_comb_contours} correspond to the 68 and 95 per cent confidence levels (CL) of the two-dimensional posterior distributions of these parameters inferred from the BOSS DR12 power spectrum wedges for the low, intermediate, and high redshift bin (top, middle and lower panels, respectively).
The dotted lines in the same figure correspond to the Gaussian approximation of these constraints, which give a good description of the full distributions.  
The blue dashed contours correspond to the \LambdaCDM predictions from the \Planck 2015 \citep{Planck:2015xua} TT+lowP measurements \changed{to which we refer simply as \Planck}, which are in excellent agreement with our results. 
The mean values of these parameters and their associated 68 per cent confidence intervals are listed in 
Table~\ref{tab:dr12_comb_results}.
\changed{BAO distance measurements are often expressed in terms of certain derived parameters:
the ratio of the volume-averaged distance and the sound horizon scale, $\DVz[z] / \rd$ and the AP 
parameter $\FAPz[z]$, where
\begin{align}
  \DVz[z] &= \left( \DV[M]^2(z) \, c z \, H^{-1}(z) \right)^{1/3} , \\
  \FAPz[z] &= \DV[M](z) \, H(z) \, c^{-1}.
\end{align}
Thus, we gibe these quantities as well.}
Appendixes~\ref{app:P2w_P3w_validity}--\ref{app:covariance_cross_check} show various consistency tests of the results of our BAO+RSD fits, such as a change in the number of wedges used, the covariance matrix or the wavenumber ranges included in the fits.
The results from these tests show that our constraints are robust with respect to the details of our analysis methodology.

\begin{figure}
 \centering
 \includegraphics[width=0.99\columnwidth]{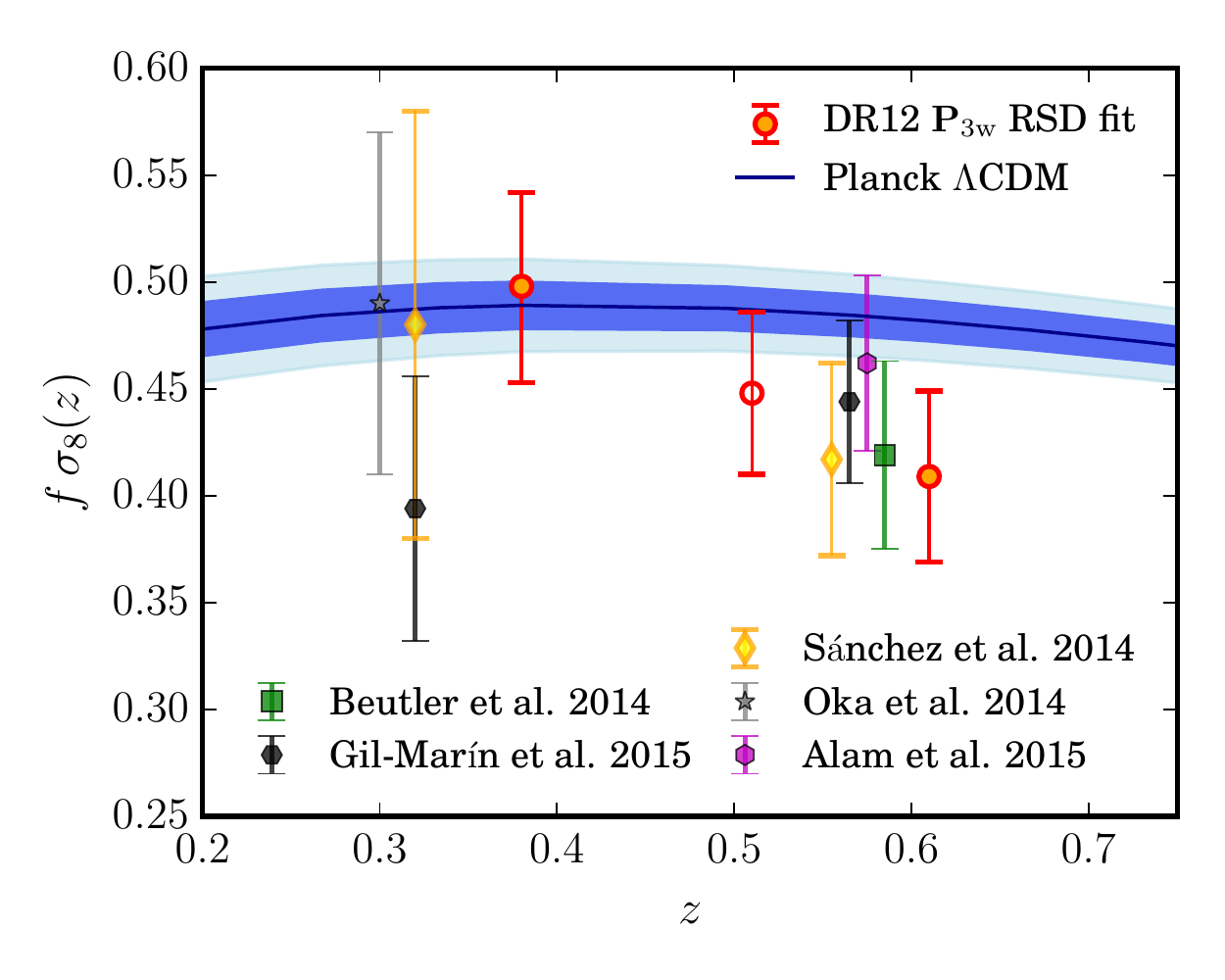}
 \caption{\changed{The red symbols show the measurements of $\Fsig$ in the three different redshift bins. The two outer redshift bins are shown as filled circles to indicate that they are independent from each other. The intermediate result has an open symbol in order to indicate the correlation with the other two results due to the overlap in the redshift ranges.}
  The \Planck \LambdaCDM predictions are shown as blue bands where the darker (lighter) shaded region indicates the 1-sigma (2-sigma) region.
  See text for the references to the previous measurements on BOSS samples;
  the measurements using the CMASS sample ($\zeff = 0.57$) are separated by a small offset for better visibility.}
 \label{fig:dr12_comb_fsig8_planck}
\end{figure}

The solid lines in Fig{.}~\ref{fig:pw_dr12_comb_model_bestfit} correspond to the model predictions for the best-fitting parameters from BAO+RSD fits to each redshift bin, which \changed{closely describe the clustering wedges measured from} the BOSS DR12 combined sample.
The model predictions were convolved with the window function as described in section~\ref{sec:win_func}.
In the low redshift bin, we use two different sets of nuisance parameters for the bias and RSD model to account for the fact that the NGC and SGC samples might contain two slightly different galaxy populations at low redshifts (as discussed in appendix~\ref{sec:NGC_vs_SGC}). 
For the intermediate and high redshift bins, the NGC--SGC difference in the model prediction is the result of the different window matrices only.

In Fig.~\ref{fig:dr12_comb_fsig8_planck} we compare our $\fsig$ measurements in the three redshift bins defined in Table~\ref{tab:dr12_z_ranges} with \Planck \LambdaCDM predictions and previous results on BOSS samples:
the Sloan DR7 LRG sample \citep[CF multipoles]{Oka:2013cba}, the configuration space clustering wedges of the LOWZ and CMASS samples \citep[CF wedges]{Sanchez:2013tga}, the most-recent analysis of the DR11 CMASS sample in configuration space \citep[CF multipoles]{Alam:2015qta} and Fourier space \citep[PS multipoles]{Beutler:2013yhm}, and the DR12 LOWZ and CMASS samples \citep[PS multipoles]{Gil-Marin:2015sqa}.
All these results are consistent with each other.
The LOWZ measurement of the last reference is lower than our constraint from the low-redshift bin at roughly $1\sigma$.
However, differences of this order can be expected as our low-redshift measurement corresponds to a larger volume than that of the LOWZ sample.
\changed{In the high-redshift bin, we measure $\fsig$ to be} lower than the \Planck \LambdaCDM prediction by roughly $1\sigma$.
This is consistent with the results of recent measurements based on the CMASS sample \citep[\eg,][]{Beutler:2013yhm,Sanchez:2013tga}.

The constraints derived here and the results of our companion BAO-only and full-shape analyses of the BOSS DR12 combined sample are summarized and compared \changed{to each other} in \citet{Alam:2016hwk}, showing the consistency of the result from various fitting methods.
All BOSS DR12 results are combined into the final set of BOSS consensus constraints in the same paper,
using the methodology described in \citet{Sanchez:2016a}.

\section{Cosmological implications of the DR12 Fourier space wedges}
\label{sec:cosmological_implications}

In this section we explore the cosmological implications of the BOSS DR12 power spectrum wedges by directly \changed{comparing the galaxy clustering measurements themselves with the predictions obtained for a given model. 
We then compare the constraints that result from combining our clustering measurements with various other cosmological data sets.
These data sets are described in section~\ref{sec:cosmo_parameter_spaces} which also contains a summary of the parameter spaces we consider.}
Sections~\ref{sec:LCDM_model}--\ref{sec:mnu_nnu_model_fit} describe our constraints on the parameters 
of the standard \LambdaCDM model as well as some of its most common extensions.

\subsection{Parameter spaces and additional data sets}
\label{sec:cosmo_parameter_spaces}

\begin{table}
 \centering
 \caption{The parameters and priors of the cosmological standard model and its extensions considered in this work.
  All parameters have flat priors defined by the given limits.
  The parameters for the extensions are set to a fiducial value for the standard \LambdaCDM model.}
 \label{tab:cosmomc_parameter_space}
 \begin{tabular}{lll}
  \hline
  Parameter [Unit]               & Prior limits     & Fiducial value \\
  \hline
  \multicolumn{3}{c}{\LambdaCDM (flat, standard $\nu$)} \\
  \hline
  $\Om[b] \, h^2$                & $0.005$--$0.1$   & --- \\
  $\Om[c] \, h^2$                & $0.001$--$0.99$  & --- \\
  $100 \, \theta_\mathrm{MC}$    & $0.5$--$10$      & --- \\
  $\tau$                         & $0.01$--$0.8$    & --- \\
  $\ns$                          & $0.8$--$1.2$     & --- \\
  $\ln(10^{10} \, A_\mathrm{s})$ & $2$--$4$         & --- \\
  \hline
  \multicolumn{3}{c}{extensions of section~\ref{sec:wCDM_model} to \ref{sec:mnu_nnu_model_fit}} \\
  \hline
  $w,\wo$                        & $(-3)$--$(-0.3)$ & $-1$ \\
  $\wa$                          & $(-2)$--$2$      & $0$ \\
  $\gamma$                       & $0$--$3$         & $0.55$ \\
  $\OK$                          & $(-0.3)$--$0.3$  & $0$ \\
  $\sum m_\nu \; [\unit{eV}]$    & $0$--$2$         & $0.06$ \\
  $\Neff$                        & $0.05$-$10$      & $3.046$ \\
  \hline
 \end{tabular}
\end{table}

A redshift survey such as BOSS probes the geometry of the Universe and the growth of structure in a limited redshift range.
\changed{In order to improve the obtained cosmological constraints we combine the information encoded in the full shape of our clustering measurements with} complementary cosmological probes, most importantly CMB observations to determine the sound horizon scale at the drag epoch.
In this work, we use the temperature and low-$\ell$ polarization measurements and derived implications \citep[denoted simply as \Planck,][]{Planck:2015xua} of the \Planck 2015 release \citep{Adam:2015rua}.
We also include the information from SN, which probe the cosmic expansion history at low redshifts via the luminosity distance scale.
We make use of the joint light-curve analysis \citep[JLA;][]{Betoule:2014frx} of the supernova samples of SDSS-II and the Supernova Legacy Survey.
In order to avoid a complex systematic error budget and measurements that are highly correlated with the ones described above, we abstain from including other cosmological probes.

We start our analysis with the standard six-parameter \LambdaCDM model.
It assumes that the energy budget of the Universe \changed{contains contributions from} (pressureless) CDM, baryonic non-relativistic matter, relativistic radiation, and DE modelled as a cosmological constant.
The \changed{upper part of Table~\ref{tab:cosmomc_parameter_space} lists the parameters of the \LambdaCDM} parameter space.
In the MCMC code \code{CosmoMC}, the baryon and CDM density are modelled by the physical density parameters $\Om[b] \, h^2$ and $\Om[c] \, h^2$, respectively.
The angular size of the sound horizon at recombination is given by $\theta_\mathrm{MC}$.
Finally, $\tau$ is the optical depth to \changed{the last-scattering surface}.
The primordial power spectrum has an amplitude given by $A_\mathrm{s})$ and a tilt given by $\ns$.
In this standard model, \changed{the Universe is assumed to be flat (\ie, $\OK = 0$) and the equation of state (EOS) parameter of DE is fixed to a constant value $w = -1$}.
The effective number of relativistic degrees of freedom (DOF) is given by $\Neff = 3.046$.
We follow \citet{Adam:2015rua} and assume also \changed{fixed contribution from} massive neutrinos of $\OmegaX{\nu} h^2 = 0.00064$.
This corresponds to a fixed sum over the neutrino masses of $\sum \mu_\nu = 0.06 \Unit{eV}$ \changed{\citep[corresponding to the minimal total neutrino mass that is consistent agreement 
with neutrino oscillation experiments;][]{Otten:2008}}.
All cosmological observations are consistent with this standard paradigm \citep[\eg,][]{Planck:2015xua,Anderson:2013zyy}.

\begin{table}
 \centering
 \caption{The 68 per cent CL intervals of the most-relevant parameters for fits using the cosmological standard model and its extensions.
  The fits include at least the \Planck 2015 TT+lowP data, which are successively combined with the power spectrum wedges $\Pobs$ of the BOSS DR12 low and high redshift bins and the JLA SN data.
  The constraints for curvature extensions are listed in Table~\ref{tab:model_constraints_curvature}, those for neutrino extensions in Table~\ref{tab:model_constraints_neutrinos}.}
 \label{tab:model_constraints}
 \begin{tabular}{lll}
  \hline
  Parameter & \Planck + BOSS $\Pobs$ & + JLA SN \\
  \hline
  \multicolumn{3}{c}{\LambdaCDM (flat, standard $\nu$)} \\
  \hline
  $\Om$     &
  $0.312_{-0.009}^{+0.008}$ &
  $0.311_{-0.010}^{+0.009}$ \\
  $h$       &
  $0.675_{-0.006}^{+0.007}$ &
  $0.676_{-0.006}^{+0.007}$ \\
  \hline
  \multicolumn{3}{c}{$w$CDM (linear EOS for DE)} \\
  \hline
  $\Om$     &
  $0.306_{-0.015}^{+0.014}$ &
  $0.307_{-0.012}^{+0.011}$ \\
  $w$       &
  $-1.029_{-0.054}^{+0.070}$ &
  $-1.019_{-0.039}^{+0.048}$ \\
  \hline
  \multicolumn{3}{c}{$\wo\wa$CDM (CPL parametrization for DE)} \\
  \hline
  $\wo$     &
  $-1.03 \pm 0.24$ &
  $-0.98 \pm 0.11$ \\
  $\wa$     &
  $-0.06_{-0.62}^{+0.77}$ &
  $-0.16_{-0.36}^{+0.46}$ \\
  \hline
  \multicolumn{3}{c}{\LambdaCDM + $\gamma$ (modified growth)} \\
  \hline
  $\Om$     &
  $0.312_{-0.009}^{+0.008}$ &
  $0.311_{-0.010}^{+0.009}$ \\
  $\gamma$  &
  $0.52 \pm 0.10$ &
  $0.52 \pm 0.10$ \\
  \hline
  \multicolumn{3}{c}{$w$CDM + $\gamma$ (linear EOS for DE, modified growth)} \\
  \hline
  $w$       &
  $-1.04_{-0.07}^{+0.10}$ &
  $-1.02_{-0.05}^{+0.06}$ \\
  $\gamma$  &
  $0.56_{-0.14}^{+0.12}$ &
  $0.54 \pm 0.11$ \\
  \hline
 \end{tabular}
\end{table}

In order to test non-standard cosmologies, we explore the most important extensions to \changed{the $\Lambda$CDM model by varying also the additional parameters} listed in the second part of Table~\ref{tab:cosmomc_parameter_space}, with flat priors in the given ranges. \changed{For all parameter spaces}, the value of the Hubble parameter $h$ was restricted to the range $0.2\le h \le 1$. 

In all \changed{cases}, the cosmological parameter spaces were extended by the nuisance parameters of the model described in section~\ref{sec:model_summary}.
The range of wavenumbers included in the fits was the same as for the BAO+RSD fits to each individual redshift bin presented in section~\ref{sec:BAO_and_RSD_measurements}.
\changed{Our clustering measurements on the intermediate $z$-bin are strongly correlated with those of the two independent bins and do not lead to a significant improvement in the obtained constraints. 
To avoid the complication of including the covariance between these measurements, in this section we only use the information from the wedges of the low and high-z bins.
While for the high redshift bin we assumed that the NGC and SGC subsamples can be described by the same nuisance parameters, in the low-redshift bin we allowed for different values of these parameters for the galaxies in these two sub-samples}. 

\subsection{The \altPdfText{\LambdaCDM}{LambdaCDM} parameter space}
\label{sec:LCDM_model}

\begin{figure}
 \centering
 \includegraphics[width=.95\columnwidth]{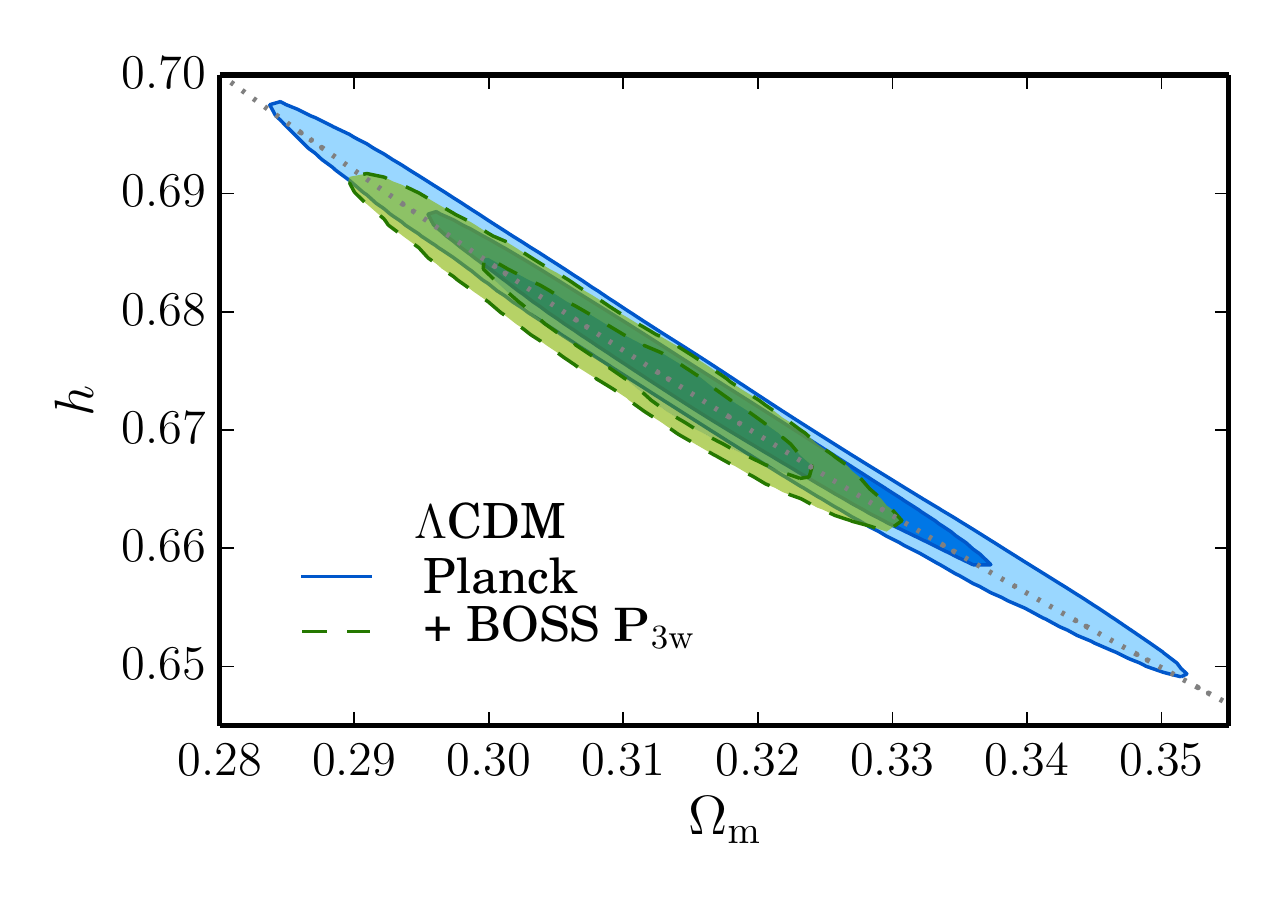}
 \caption{The marginalized 68 and 95 per cent CL in the $\Om$--$h$ plane for the \LambdaCDM parameter space from thee \Planck 2015 TT+lowP \citep{Planck:2015xua} observations (blue) and adding the DR12 combined sample $\Pobs$ (green).
  The \Planck confidence contours as well as those of the combined fits follow the $\Om \, h^3$ degeneracy \citep{Percival:2002gq} shown as dotted gray line.}
 \label{fig:dr12_comb_model_lcdm}
\end{figure}

\changed{We first focus on the} standard \LambdaCDM parameter space.
The resulting constraints on $\Om$ and $h$ from the combined \Planck[+]{}BOSS $\Pobs$ fits (green) are shown in Fig{.}~\ref{fig:dr12_comb_model_lcdm}, compared with the constraints from \Planck alone (blue).
The corresponding marginalized 68 per cent CL intervals of these parameters are listed in the upper part of Table~\ref{tab:model_constraints}.
The full-shape of our BOSS clustering measurements prefers slightly \changed{lower} values for the matter density parameter\footnote{Unless stated otherwise, all constraints given in this section correspond to a CL of 68 per cent.} 
($\Om = 0.312_{-0.009}^{+0.008}$) than the \Planck data alone, while the constraints on the Hubble parameter ($h = 0.675_{-0.006}^{+0.007}$) are centred around a similar mean value.
Adding the JLA SN data to the fits does not \changed{improve the obtained constraints}.
The confidence contours follow a degeneracy along $\Om \, h^3 = \const$, indicated by a dotted line in the plot.
This \changed{degeneracy} is given by equally good fits to the locations and relative heights of the acoustic peaks \citep{Percival:2002gq}.
\changed{In summary}, we find excellent consistency between the three different probes assuming a \LambdaCDM cosmology \changed{as could be expected from the agreement between \Planck and BOSS data that was a result of the BAO+RSD fits described in section~\ref{sec:BAO_and_RSD_measurements}}.

\subsection{The \altPdfText{$w$CDM}{wCDM} parameter space}
\label{sec:wCDM_model}

\begin{figure*}
 \centering
 \includegraphics[width=.47\textwidth]{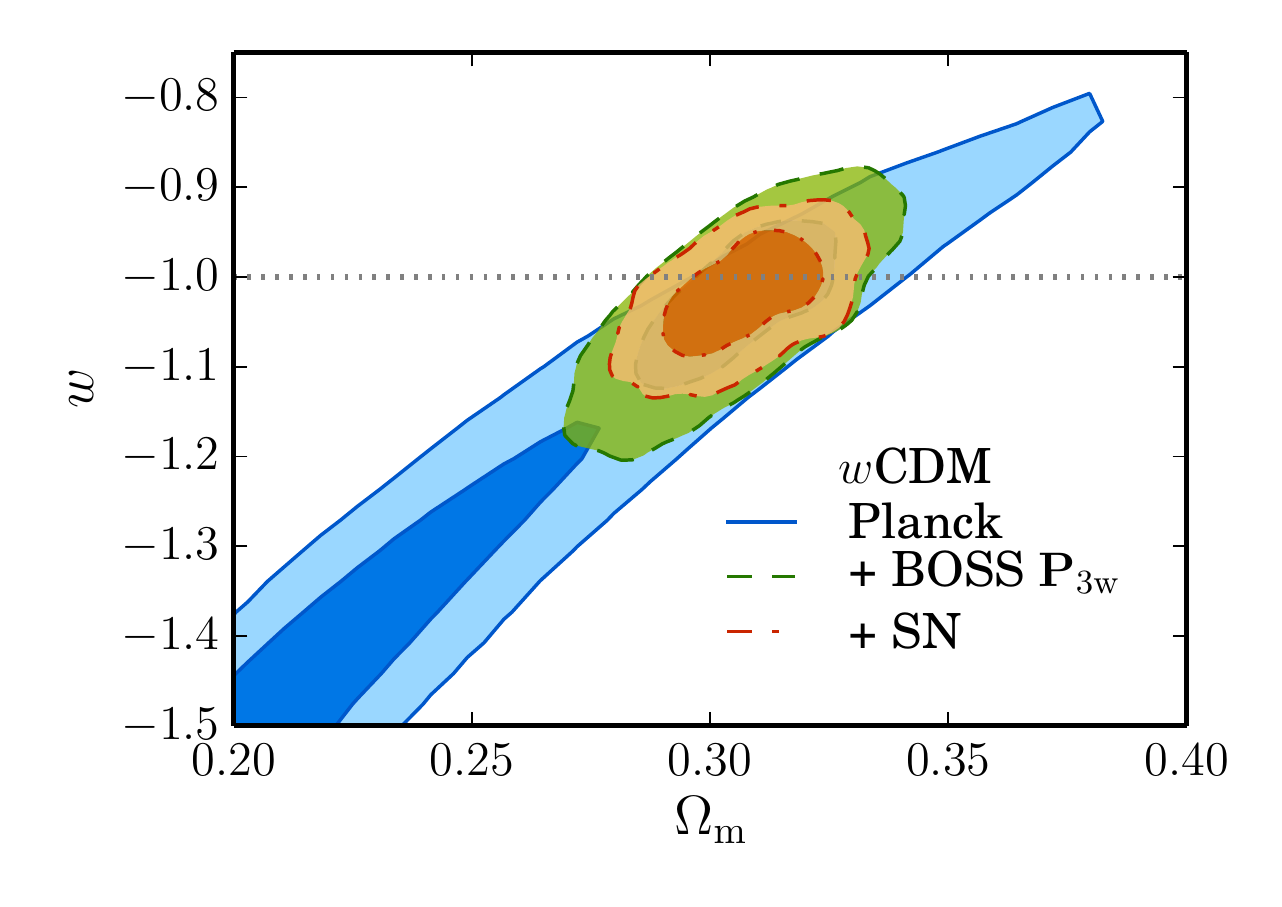}
 \includegraphics[width=.47\textwidth]{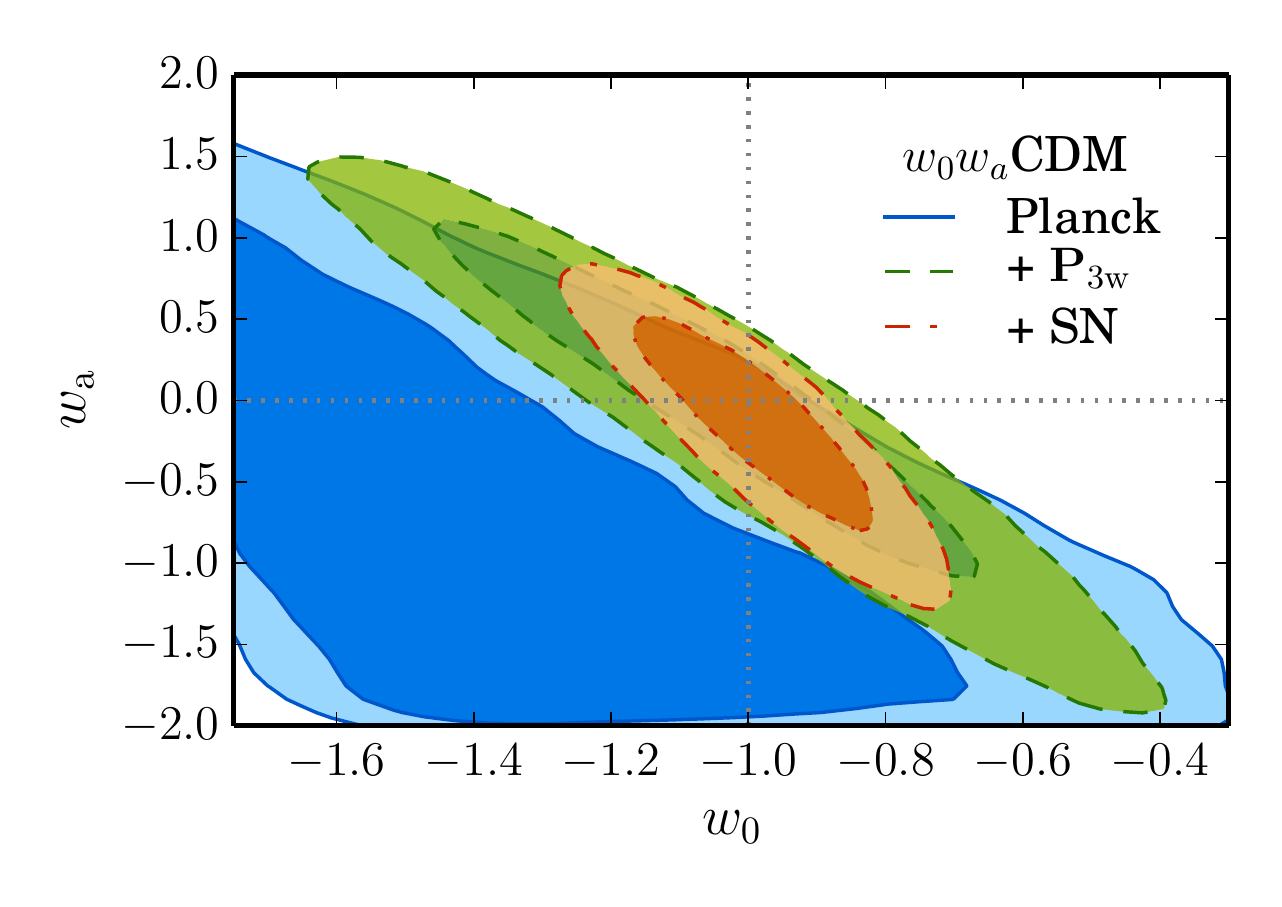}
 \caption{\panel{Left-hand} The 68 and 95 per cent CL in the $\Om$--$w$ plane of the $w$CDM parameter space from the \Planck 2015 (blue) fits, successively adding the BOSS DR12 $\Pobs$ (green) and JLA SN (orange) data.
  \panel{Right-hand} The same CL in the $\wo$--$\wa$ plane of the $\wo\wa$CDM parameter space from the \Planck (blue) fits, successively adding the BOSS $\Pobs$ (green) and SN (orange) data.}
 \label{fig:dr12_comb_model_wcdm_wacdm}
\end{figure*}

The first relaxation of the assumptions of the standard $\Lambda$CDM model is to abandon the idea that DE can be described by a cosmological constant.
The simplest case, the $w$CDM model, assumes a \changed{constant DE EOS parameter},
\begin{equation}
 p_\mathrm{DE} = w \, \rho_\mathrm{DE}.
\end{equation}
For $w = -1$, the \LambdaCDM model with a cosmological constant is recovered.

As the EOS parameter $w$ controls the late-time expansion of the Universe, galaxy clustering and SN are ideal cosmological probes to constrain DE, which is not well constrained by CMB data alone.
In this last case, $w$ follows a degeneracy with $\Om$ and values \changed{significantly lower than} $w = -1$ are preferred, resulting in poor constraints of $w = -1.55_{-0.30}^{+0.32}$.
As shown in the left-hand panel of Fig{.}~\ref{fig:dr12_comb_model_wcdm_wacdm}, including the power spectrum wedges in the fits results in confidence regions that are centred around the standard \LambdaCDM value of $w = -1$ (indicated by a dotted line in the figure), with $w = -1.029_{-0.062}^{+0.066}$.
Including also SN data, the late-time expansion is even better probed so that $w$ is measured to $5$ per cent accuracy, $w = -1.019_{-0.045}^{+0.043}$, in good agreement with \LambdaCDM at $1\sigma$.

\subsection{The \altPdfText{$\wo\wa$CDM}{w0waCDM} parameter space}
\label{sec:waCDM_model}

\begin{figure*}
 \centering
 \includegraphics[width=.45\textwidth]{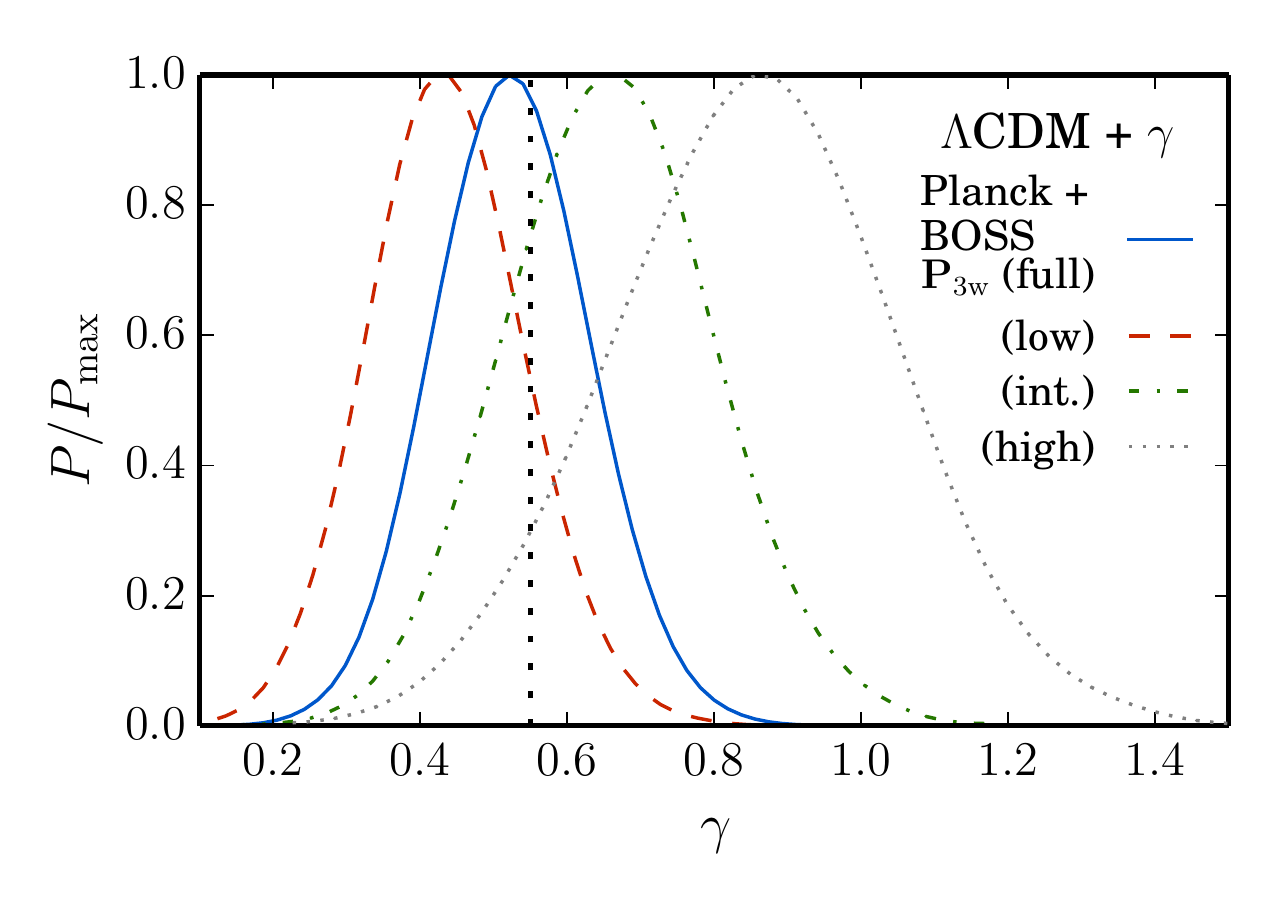}
 \includegraphics[width=.45\textwidth]{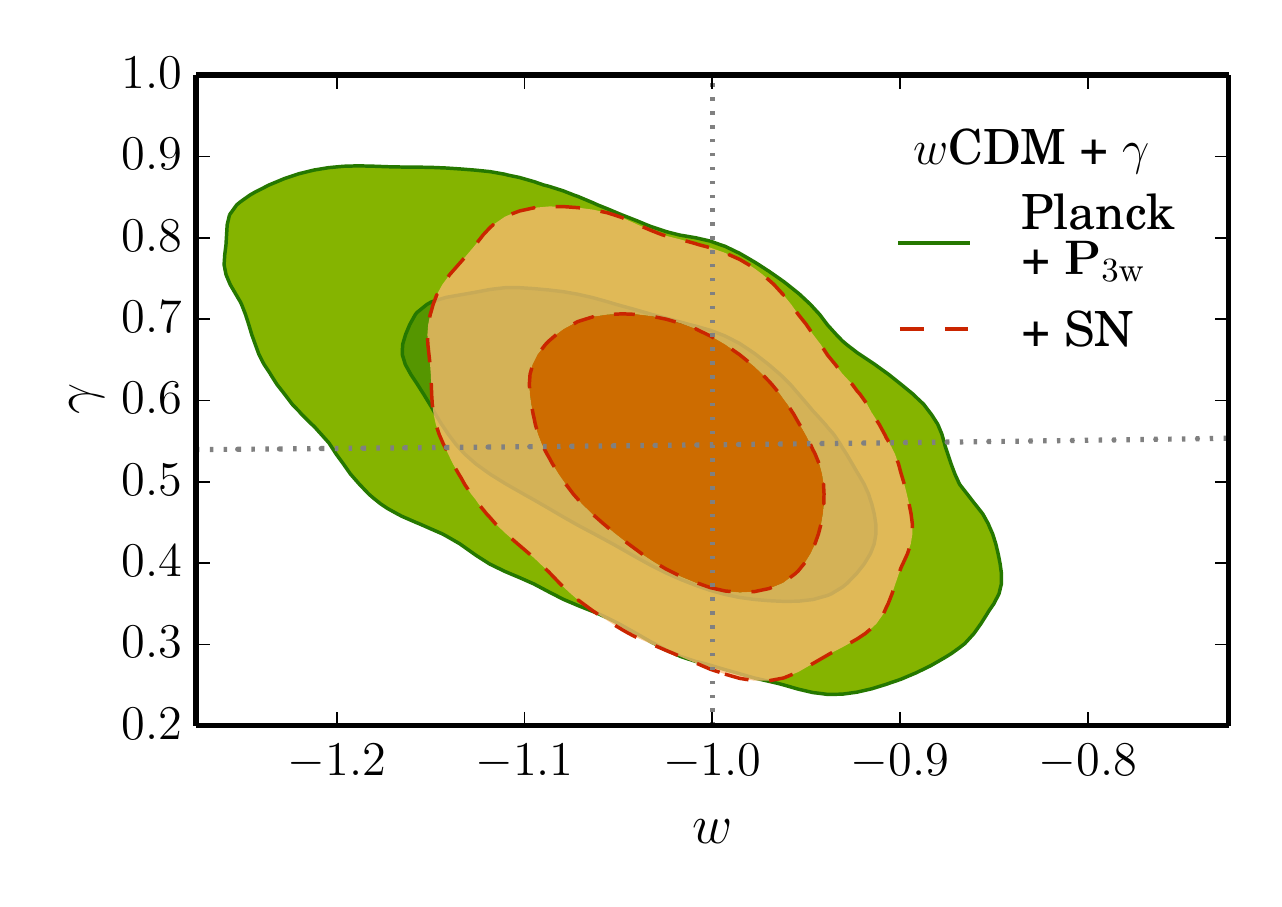}
 \caption{\panel{Left-hand} The posterior distributions for the growth index $\gamma$ in the \LambdaCDM + $\gamma$ parameter space from the \Planck 2015 and BOSS DR12 $\Pobs$ observations in the two non-overlapping redshift bins (solid blue lines), and for the combination of \Planck with the measurements in each individual redshift bin (dashed red, dot--dashed green, and dotted black lines for the low, intermediate, and high redshift bin, respectively).
  The vertical dotted line indicates $\gamma_\mathrm{GR} = 0.55$.
  \panel{Right-hand} The 68 and 95 per cent CL in the $w$--$\gamma$ plane of the $w$CDM + $\gamma$ parameter space from the \Planck and the BOSS $\Pobs$ observations (green), and adding SN data (orange).
  The horizontal dotted line shows the value of the exponent $\gamma$ depending on the EOS parameter $w$ as given by \eqref{eq:f_GR}.}
 \label{fig:dr12_comb_model_gamma_gammaw}
\end{figure*}

Here, we explore the constraints on the time evolution of DE.
We use the Chevallier-Polarski-Linder (CPL) parametrization \citep{Chevallier:2000qy, Linder:2002et} of a time-dependent EOS for DE,
\begin{equation}
 \label{eq:w_CPL}
 w(z) = w_0 + w_a \, \left( 1 - a(z) \right) = w_0 + w_a \frac{z}{1 + z},
\end{equation}
where $\wo$ is the current value of $w(z)$ and $\wa$ \changed{controls its time evolution}.
This parametrization recovers \LambdaCDM for $w_0 = -1$ and $\wa = 0$.

As in the case of a constant $w$, the constraints on the  EOS parameters significantly improve when late-time expansion probes are taken into account.
The $\wo$--$\wa$ parameter plane is practically unconstrained by CMB data alone:a large region roughly below the line $\wa = -3 \, (\wo + 1)$ is preferred.
This plane becomes tightly constrained by including the BOSS DR12 power spectrum wedges, yielding
\begin{align}
 \wo &= -1.02_{-0.26}^{+0.25}, & \wa &= -0.06_{-0.72}^{+0.70}.
\end{align}
As shown in the right-hand panel of Fig{.}~\ref{fig:dr12_comb_model_wcdm_wacdm}, the constraints roughly follow a linear degeneracy.
This is due to the fact that the combination of \Planck + BOSS DR12 has the most constraining power on $w(z)$ at a `pivot scale' $z_\mathrm{p}$ given by the effective mean redshift probed by the data.
For the combination of \Planck and BOSS DR12, this is at $z_\mathrm{p} \approx 0.5$;
Including SN data as well, the pivot redshift moves closer to $z_\mathrm{p} \approx 0.3$, resulting in the tighter constraints in the $\wo$--$\wa$ parameter plane following a slightly 
\changed{different} degeneracy.
The resulting constraints are closely centred on the \LambdaCDM values and the error bars are 
cut down by half compared to the \Planck + BOSS case,
\begin{align}
 \wo &= -0.98 \pm 0.11, & \wa &= -0.16 \pm 0.42.
\end{align}
Our final constraints on the EOS parameter of DE are consistent with no evolution of $w(z)$, with DE well described by a cosmological constant at all redshifts.

\subsection{Modified gravity}
\label{sec:gamma_model}

The growth-rate parameter $f$ defined in equation~\eqref{eq:f_growth} depends on the gravitational potential and thus measurements of this quantity via RSD can be used as a probe of the theory of gravity \citep{Guzzo:2008}.
As described in \citet{Linder:2007hg} and \citet{Gong:2008fh}, the growth rate has an approximate dependency on the matter density parameter $\Om$ given by
\begin{equation}
 \label{eq:f_GR}
 f(z) = \left[\Om(z)\right]^\gamma, \quad \text{where} \quad \gamma \simeq \frac{3 (1 - w)}{5 - 6 \, w},
\end{equation}
if the growth of structure is bound to Einstein's GR.
For the \LambdaCDM case, the exponent is $\gamma \simeq 0.55$;
otherwise, its value only mildly depends on $w$.

\begin{table}
 \centering
 \caption[The regions of 68 per cent CL of the most-relevant model parameters for fits using curvature extensions of the cosmological standard model.]{The regions of 68 per cent CL of the most-relevant model parameters for fits using curvature extensions of the cosmological standard model.
  The fits include at least the \Planck 2015 TT+lowP data, which are successively combined with the power spectrum wedges $\Pobs$ of the BOSS DR12 low and high redshift bins and the JLA SN data.}
 \label{tab:model_constraints_curvature}
 \begin{tabular}{lll}
  \hline
  Parameter & \Planck + BOSS $\Pobs$ & + JLA SN \\
  \hline
  \multicolumn{3}{c}{$K$-\LambdaCDM (curvature, standard $\nu$)} \\
  \hline
  $\Om$     &
  $0.312 \pm 0.009$ &
  $0.311 \pm 0.009$ \\
  $\OK$     &
  $-0.001 \pm 0.003$ &
  $-0.001 \pm 0.003$ \\
  \hline
  \multicolumn{3}{c}{$K$-$w$CDM (curvature, linear EOS for DE)} \\
  \hline
  $\Om$     &
  $0.304_{-0.016}^{+0.015}$ &
  $0.308 \pm 0.011$ \\
  $\OK$     &
  $-0.002 \pm 0.004$ &
  $-0.001_{-0.003}^{+0.004}$ \\
  $w$       &
  $-1.052_{-0.071}^{+0.088}$ &
  $-1.027_{-0.045}^{+0.052}$ \\
  \hline
 \end{tabular}
\end{table}

\begin{figure*}
 \centering
 \includegraphics[width=.49\textwidth]{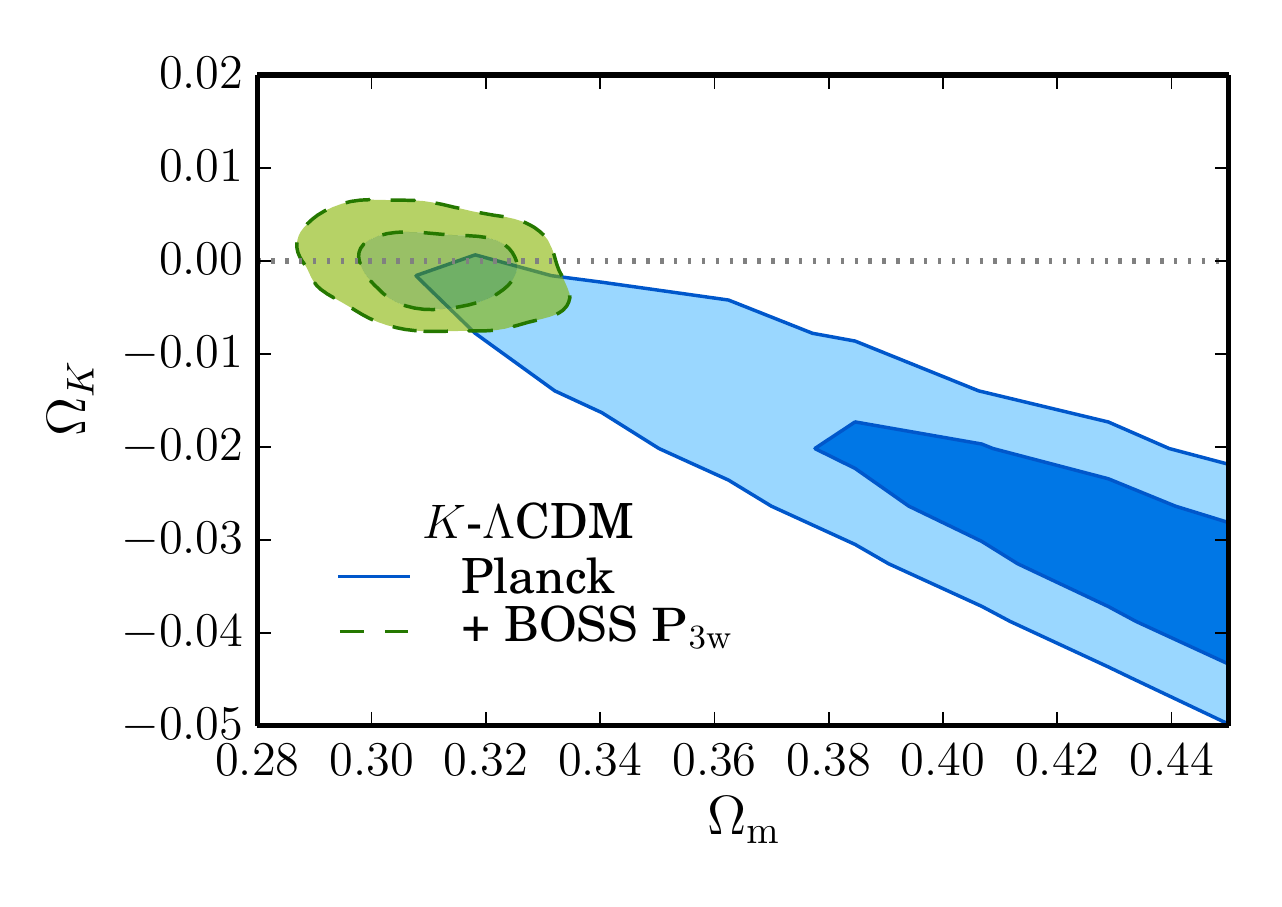}
 \includegraphics[width=.49\textwidth]{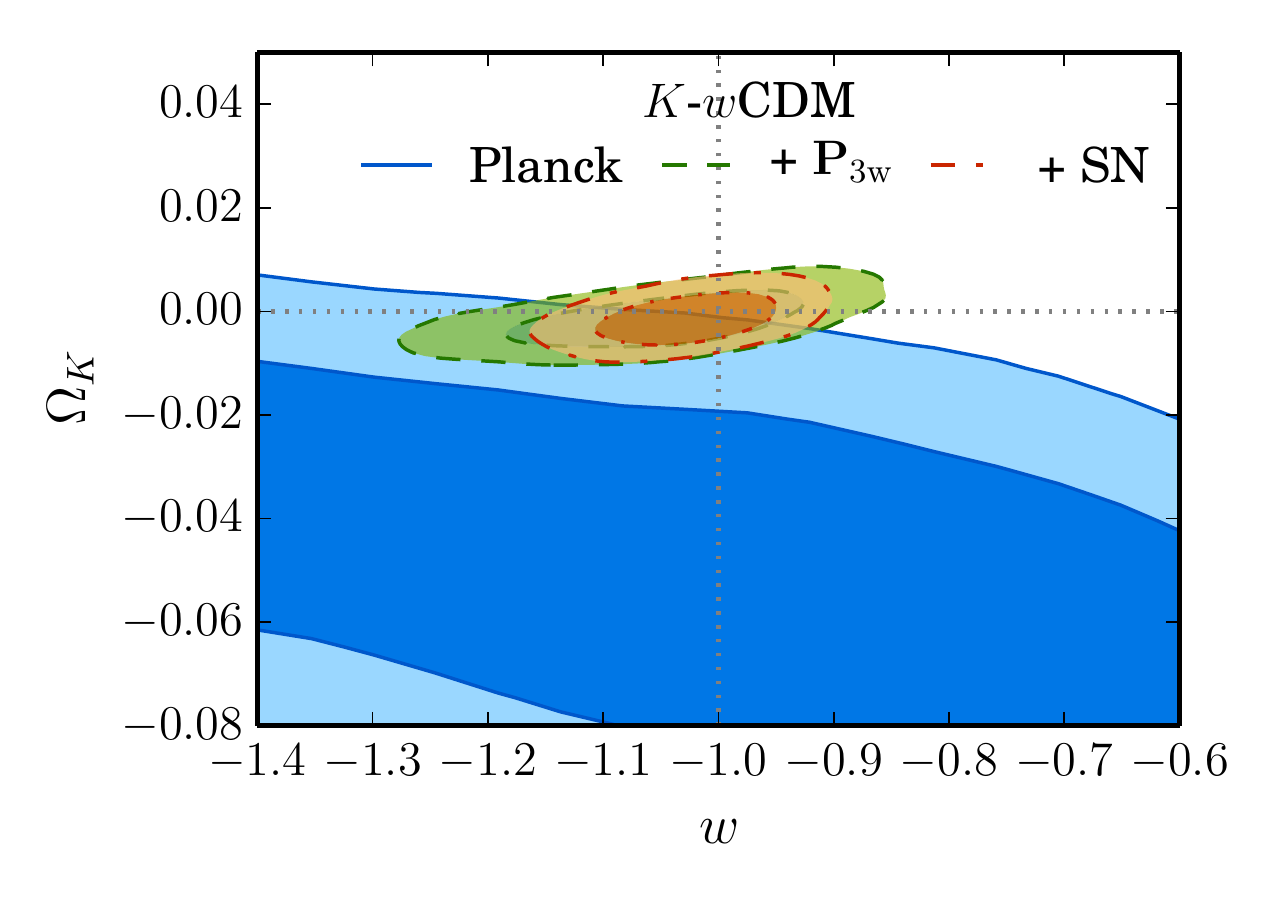}
 \caption{\panel{Left-hand} The 68 and 95 per cent CL in the $\Om$--$\OK$ plane of the $K$-\LambdaCDM parameter space from the \Planck 2015 observations (blue), and adding BOSS DR12 $\Pobs$ data (green).
 The horizontal dotted line indicates a flat universe, $K = 0$.
  \panel{Right-hand} The same CL in the $w$--$\OK$ plane of the $K$-$w$CDM parameter space from the \Planck observations (blue), and successively adding BOSS $\Pobs$ (green) and SN (orange) data.
  The vertical dotted line shows the value of the EOS parameter $w = -1$ for a cosmological constant (as in the \LambdaCDM model).}
 \label{fig:dr12_comb_model_klcdm_kwcdm}
\end{figure*}

In order to test for modifications of the fundamental relations of GR, we treat the exponent $\gamma$ in equation~\eqref{eq:f_GR} as a free parameter in a \LambdaCDM background universe (dubbed \LambdaCDM + $\gamma$ parameter space here).
In the left-hand panel of Fig{.}~\ref{fig:dr12_comb_model_gamma_gammaw}, we plot the posterior distributions of the growth index $\gamma$ as constrained from the combination of \Planck and full-shape BOSS $\Pobs$ observations (marginalized over all other parameters).
The blue solid line corresponds to the combination of the two non-overlapping redshift bins, while the red dashed, green dot--dashed, and black dotted lines correspond to the measurements of the Fourier-space wedges of \changed{each redshift bin separately}.
We see a slight trend of the centroid of the $\gamma$ distribution from values smaller than the GR value, which is indicated by a horizontal dotted line, for the fit of the low redshift bin to values above this value for the fit of the high redshift bin.
This shift is consistent with the trend of the $\Fsig$ measurements compared to the \Planck \LambdaCDM predictions in Fig{.}~\ref{fig:dr12_comb_fsig8_planck}.
The final posterior distribution (we obtain $\gamma = 0.52 \pm 0.10$) is in excellent agreement with $\gamma_\mathrm{GR} = 0.55$.
As SN do not depend on the growth, their inclusion does not yield tighter confidence regions.

This behaviour is different if we allow for $w\neq-1$, as now SN data help to constrain the EOS parameter via the late-time expansion history.
The resulting confidence contours in the $w$--$\gamma$ parameter plane are shown in the right-hand panel of Fig{.}~\ref{fig:dr12_comb_model_gamma_gammaw}.
While we obtain $w = -1.04_{-0.09}^{+0.08}$ for the combination of Planck and BOSS DR12 data, the EOS parameter is constrained to $w = -1.02 \pm 0.05$ by the inclusion of SN data, similar to the one obtained for the $w$CDM model.
However, the exponent $\gamma$ is only marginally better constrained, with $\gamma = 0.54 \pm 0.11$. 
The final constraints are in good agreement with the standard \LambdaCDM + GR cosmological model, whose parameter values are indicated by the dotted lines.

\subsection{The curvature of the Universe}
\label{sec:kcdm_model_fit}

In a non-flat \LambdaCDM universe, the curvature constant $K$ describes a spatial geometry with open (hyperbolic, $K < 0$) or closed (elliptical, $K > 0$) hypersurfaces.
The standard case is a flat geometry, $K = 0$.
CMB observations alone cannot discriminate between a flat and a closed universe, as the density parameters $\Om$ and $\OK$ follow the `geometric degeneracy' \citep{Efstathiou:1998xx}, because these parameters can be varied simultaneously to keep the same angular acoustic scale.
Including late-time \changed{cosmological observations such our BOSS clustering measurements helps to break this degeneracy leading to significantly tighter constraints}.
This is shown by the 68 and 95 per cent CL regions in the left-hand panel of Fig{.}~\ref{fig:dr12_comb_model_klcdm_kwcdm}.
The addition of the power spectrum wedges results in constraints on the matter density parameter that are of a similar order than for standard \LambdaCDM fits, with $\Om = 0.312 \pm 0.009$.
The curvature constraints, $\OK = -0.001 \pm 0.003$, are closely centred around a flat universe.
Adding SN data does not improve these constraints at a significant level.

\begin{figure*}
 \centering
 \includegraphics[width=.45\textwidth]{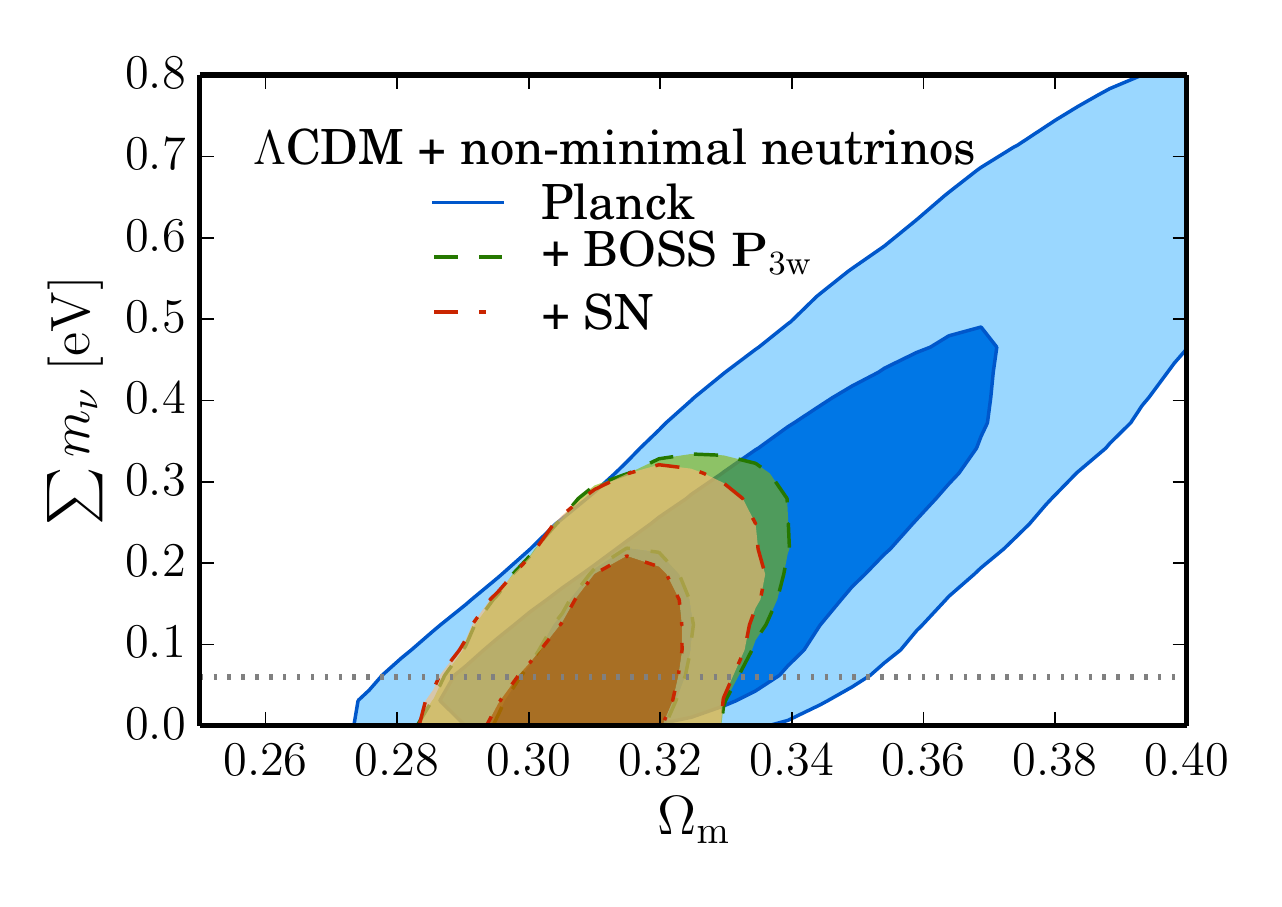}
 \includegraphics[width=.45\textwidth]{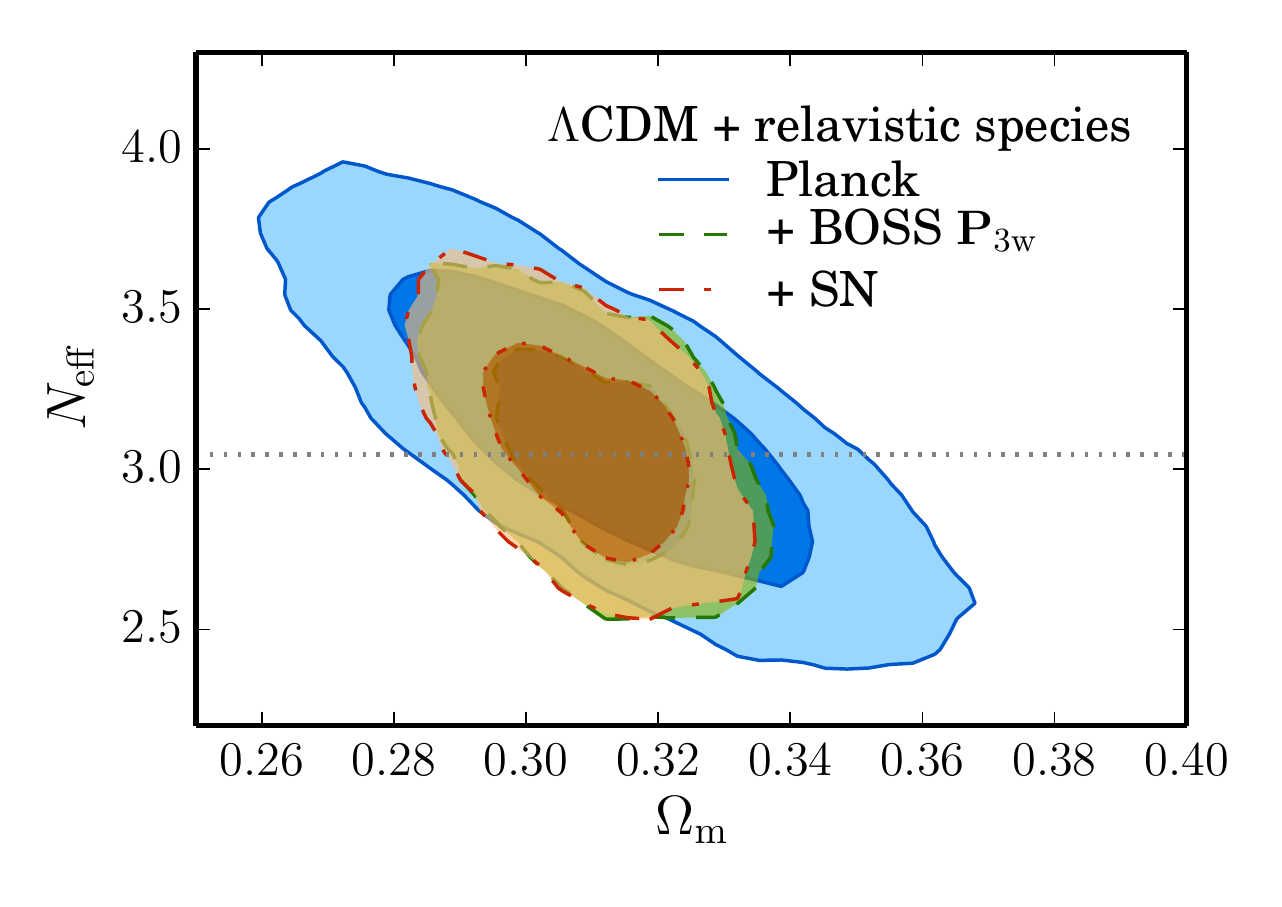}
 \includegraphics[width=.45\textwidth]{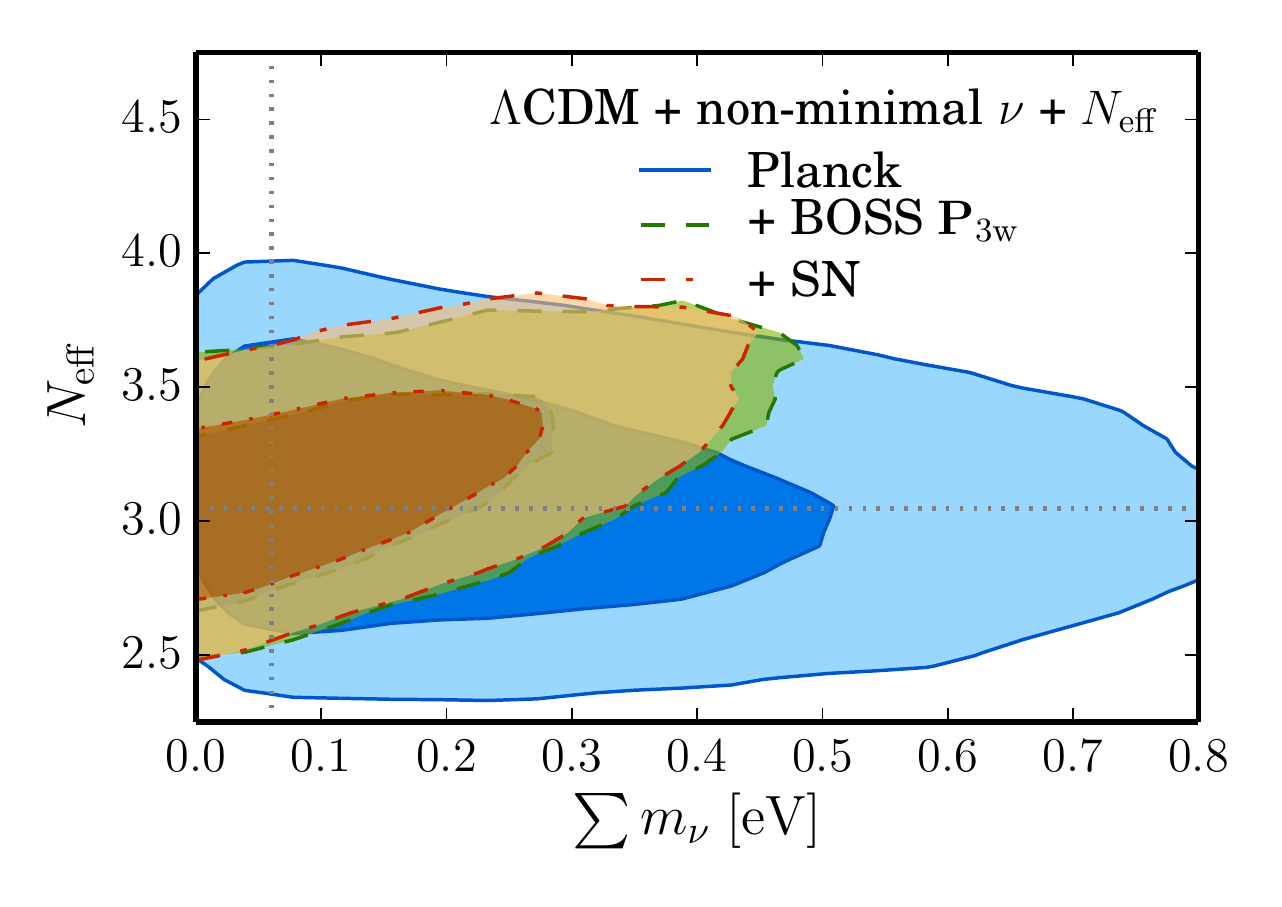}
 \includegraphics[width=.45\textwidth]{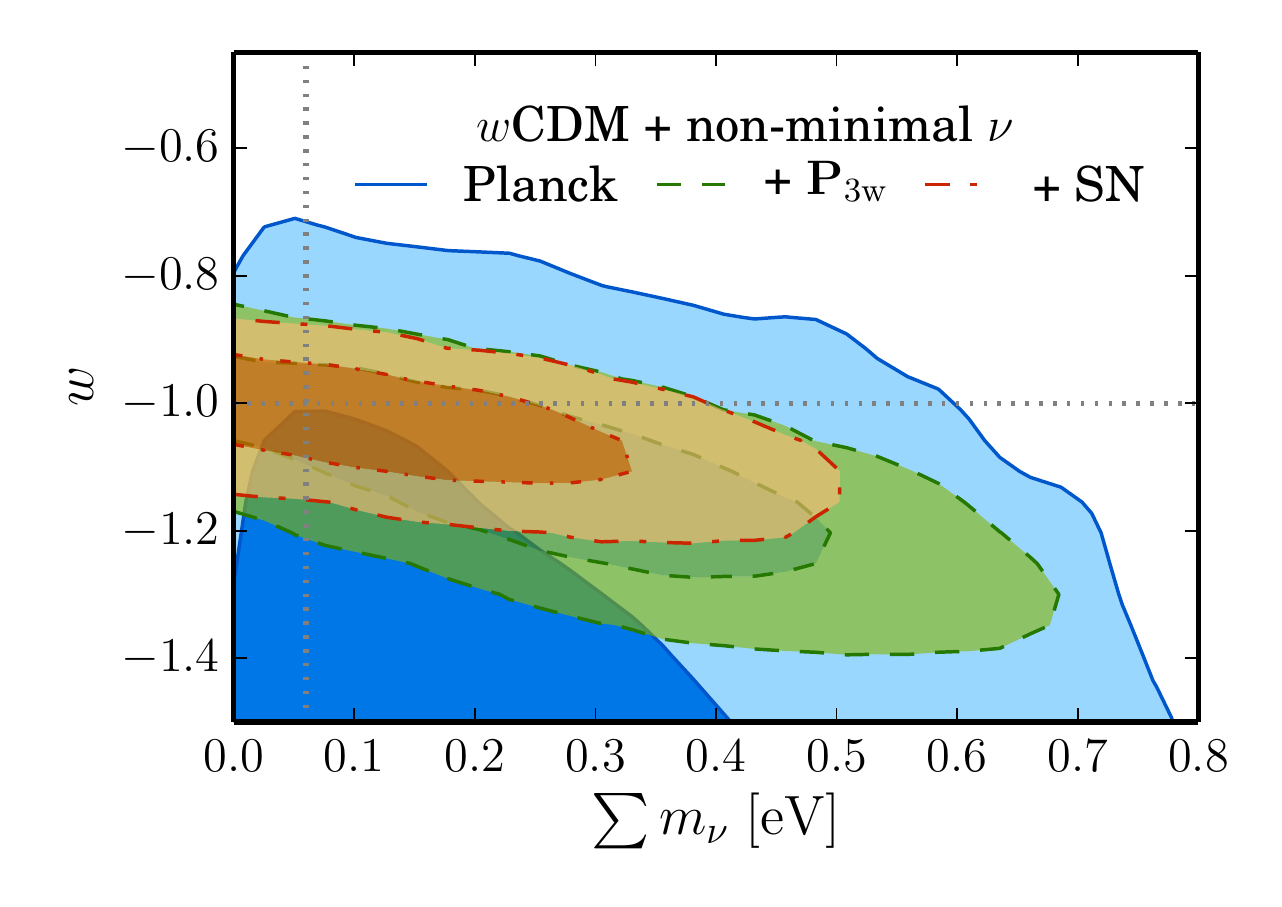}
 \caption{The 68 and 95 per cent CL in the most relevant parameter planes of extensions to the \LambdaCDM parameter space including massive and sterile neutrinos.
  \panel{Upper left-hand} The $\Om$--$\sum m_\nu$ plane of the \LambdaCDM parameter space extend by a non-minimal sum of neutrino masses showing the confidence regions from the \Planck 2015 observations (blue), and successively adding BOSS DR12 $\Pobs$ (green) and JLA SN (orange) data.
 The horizontal dotted line indicates the minimal sum of neutrino masses, $\sum m_\nu = 0.06 \Unit{eV}$.
  \panel{Upper right-hand} The $\Om$--$\Neff$ plane of the $\Neff$-\LambdaCDM parameter space (allowing for variations in the effective number of relativistic DOF) with the confidence regions from the same data sets as for the upper left-hand panel.
  The vertical dotted line shows the value of $\Neff$ for the standard model, $\Neff = 3.046$.
  \panel{Lower left-hand} The $\sum m_\nu$--$\Neff$ plane of the $\Neff$-\LambdaCDM parameter space extended by a non-minimal sum of neutrino masses, showing the confidence regions from the same data sets as for the upper left-hand panel.
  The vertical and horizontal dotted lines indicate the cuts through this parameter plane that correspond to the conventional \LambdaCDM case.
  \panel{Lower right-hand} The $\sum m_\nu$--$w$ plane of the parameter space of $w$CDM with a varying sum of neutrino masses with the confidence regions from the same data sets as for the upper left-hand panel.
  The vertical and horizontal dotted lines correspond to the standard \LambdaCDM model.}
 \label{fig:dr12_comb_model_mnu_nnu}
\end{figure*}

The geometric degeneracy receives an additional degree of freedom in the $K$-$w$CDM parameter space as the EOS parameter $w$ changes the relation between $\Om$, $\OK$ and the angular scale of the acoustic peaks.
The \LambdaCDM case ($w = -1$ and $\OK = 0$, indicated by dotted lines) is outside the 95 per cent confidence region for the CMB-only fits. Including our $\Pobs$ restricts the allowed range of values of the matter density parameter to $\Om = 0.304_{-0.016}^{+0.015}$, leaving a residual degeneracy in the $w$--$\OK$ parameter plane.
The statistical error on the EOS parameter of DE ($\approx 8$ per cent) is slightly larger than for $w$CDM fits ($\approx 6.5$ per cent).
Additionally including SN data places a tighter constraint on $w$ by probing the late-time expansion history, resulting in $\OK = -0.001_{-0.003}^{+0.004}$ and $w = -1.027 \pm 0.049$, in close agreement with the standard \LambdaCDM model.

\subsection{Parameter spaces with non-standard massive and sterile neutrino species}
\label{sec:mnu_nnu_model_fit}

\begin{table}
 \centering
 \caption[The regions of 68 per cent CL of the most-relevant model parameters for fits using neutrino extensions of the cosmological standard model.]{The regions of 68 per cent CL of the most-relevant model parameters for fits using neutrino extensions of the cosmological standard model.
  Consistent with the text, the given range corresponds to 95 per cent CL in case of upper limits.
  In the standard \LambdaCDM model, massive neutrinos with $\sum m_\nu = 0.06 \Unit{eV}$ are included;
  the effective number of relativistic DOF corresponding to the radiation and neutrino background is given by $\Neff = 3.046$
  The fits include at least the \Planck 2015 TT+lowP data, which are successively combined with the power spectrum wedges $\Pobs$ of the BOSS DR12 low and high redshift bins and the JLA SN data.}
 \label{tab:model_constraints_neutrinos}
 \begin{tabular}{lll}
  \hline
  Parameter & \Planck + BOSS $\Pobs$ & + JLA SN \\
  \hline
  \multicolumn{3}{c}{\LambdaCDM + non-minimal $\nu$ (free $\sum m_\nu$)} \\
  \hline
  $\Om$        &
  $0.32_{-0.010}^{+0.009}$ &
  $0.32_{-0.010}^{+0.009}$ \\
  $\sum m_\nu$ &
  $< 0.275$ &
  $< 0.260$ \\
  \hline
  \multicolumn{3}{c}{$\Neff$-\LambdaCDM (free \#(relativistic DOF))} \\
  \hline
  $\Om$     &
  $0.311_{-0.011}^{+0.010}$ &
  $0.310_{-0.011}^{+0.010}$ \\
  $\Neff$   &
  $3.05_{-0.024}^{+0.020}$ &
  $3.08_{-0.024}^{+0.021}$ \\
  \hline
  \multicolumn{3}{c}{$\Neff$-\LambdaCDM + non-minimal $\nu$ (free $\sum m_\nu$ and \#(rel{.} DOF))} \\
  \hline
  $\Om$        &
  $0.314_{-0.012}^{+0.010}$ &
  $0.312_{-0.011}^{+0.010}$ \\
  $\sum m_\nu$ &
  $< 0.380$ &
  $< 0.357$ \\
  $\Neff$      &
  $3.18_{-0.29}^{+0.25}$ &
  $3.19_{-0.29}^{+0.24}$ \\
  \hline
  \multicolumn{3}{c}{$w$CDM + non-minimal $\nu$ (linear EOS for DE, free $\sum m_\nu$)} \\
  \hline
  $\Om$        &
  $0.302 \pm 0.016$ &
  $0.310 \pm 0.012$ \\
  $\sum m_\nu$ &
  $0.28_{-0.20}^{+0.17}$ &
  $< 0.416$ \\
  $w$          &
  $-1.14_{-0.10}^{+0.12}$ &
  $-1.06_{-0.06}^{+0.07}$ \\
  \hline
 \end{tabular}
\end{table}

\changed{In this section we extend the \LambdaCDM parameter space by treating $\sum m_\nu$ as a free parameter.
The blue contours in the upper-left-hand panel of Fig{.}~\ref{fig:dr12_comb_model_mnu_nnu}corresponds to the constraints in the $\Om$--$\sum m_\nu$ parameter plane obtained using CMB data from \Planck 2015 alone.
These constraints follow a degeneracy of the matter density parameter $\Om$ and the sum of neutrino masses $\sum m_\nu$ that is elongated along a line given by a constant value of the redshift of matter-radiation equality $z_\mathrm{eq}$, which is well constrained} by the ratio of the heights of the first and third CMB acoustic peaks \citep{Komatsu:2008hk}.
Marginalized over all other parameters, we obtain\footnote{The upper limits in this section are given for 95 per cent CL.} $\sum m_\nu < 0.644 \Unit{eV}$.
Adding the BOSS $\Pobs$ data (green contours) tightens the confidence limits on $\Om$.
The sum of neutrino masses is constrained to an upper limit of $\sum m_\nu < 0.275 \Unit{eV}$.
Only minor improvement is found by including SN data (orange contours), yielding $\sum m_\nu < 0.260 \Unit{eV}$.

The effective number of relativistic DOF in the neutrino sector, $\Neff$, can also be constrained by CMB and LSS observations.
Again, the constraints in the $\Om$--$\Neff$ parameter plane follow a degeneracy defined by tight constraints on the matter-radiation equality.
Just as for $\sum m_\nu$, the correlation of the parameter is broken by an indirect measurement of $\Om$ from the BOSS DR12 analysis.
The constraints on the $\Om$--$\Neff$ parameter plane are shown in the upper right-hand panel of Fig{.}~\ref{fig:dr12_comb_model_mnu_nnu}.
Marginalized over all other parameters, we obtain $\Neff = 3.05_{-0.024}^{+0.020}$, which corresponds to a reduction of the statistical error by a factor of 1.5 compared to $\Neff = 3.12 \pm 0.32$ from CMB data alone.
We do not find any improvement in the marginalized constraints for the $\Om$--$\Neff$ parameter plane from adding the SN data.

The same scenario as described before also applies to the extension of the \LambdaCDM parameter space by allowing for simultaneous variations of $\Neff$ and $\sum m_\nu$:
degeneracies between $\Neff$, $\sum m_\nu$, and $\Om$ along lines of constant $z_\mathrm{eq}$ are broken by better constraints on $\Om$ from LSS observations.
The 68 and 95 per cent CL contours are shown for the $\Neff$--$\sum m_\nu$ parameter plane in the lower panel of the left-hand side in Fig{.}~\ref{fig:dr12_comb_model_mnu_nnu}.
As there is a residual degeneracy between $\Neff$ and $\sum m_\nu$, the final constraints (\Planck + BOSS $\Pobs$ + SN) are slightly larger than when these parameters are varied separately, $\sum m_\nu < 0.357 \Unit{eV}$ and $\Neff = 3.19 \pm_{-0.29}^{+0.24}$.

For the last parameter space discussed here, a $w$CDM cosmology with a sum of neutrino masses, including SN data significantly improves the constraints.
As shown in the lower right-hand panel of Fig{.}~\ref{fig:dr12_comb_model_mnu_nnu}, the $\sum m_\nu$--$w$ parameter plane is hardly constrained by CMB data alone.
The information in the DR12 power spectrum wedges can constrain the late-time expansion and thus $w$, but the remaining freedom along a degeneracy of $\Om$ and $w$ also leaves limits on $\sum m_\nu$ that are roughly twice as large as \changed{those obtained on} the \LambdaCDM case.
This results in a 1-sigma signal for the sum of the neutrino masses, $\sum m_\nu = 0.28_{-0.20}^{+0.17} \Unit{eV}$, and also the EOS parameter of DE is constrained to an 
interval that does not contain the \LambdaCDM value at 68 per cent CL, $w = -1.14_{-0.10}^{+0.12}$.
The addition of further information from the JLA SN data breaks the remaining freedom and helps to tighten the constraints on $\sum m_\nu$ and $w$.
In this case, we obtain $\sum m_\nu < 0.416 \Unit{eV}$ and $w = -1.06_{-0.12}^{+0.11}$, in perfect agreement with a cosmological constant and without a signal of a lower bound of the sum of neutrino masses.
The statistical errors obtained \changed{in this case} correspond to a $\approx 50$ per cent increase with respect to the errors found for each parameter individually in the \LambdaCDM 
and $w$CDM cases.

\section{Conclusions}
\label{sec:conclusions}

\changed{In this work, we performed a cosmological analysis of the full shape of anisotropic clustering measurements in Fourier space, of the final galaxy samples from BOSS, the DR12 combined sample \citep{Reid:2015gra}, a galaxy catalogue that is unprecedented in its volume. 
This information can be used to place tight constraints on the expansion history of the Universe and he growth-rate of cosmic structures.}

We extended the concept of clustering wedges \citep{Kazin:2011xt} to Fourier space by defining an estimator for this quantity analogous to the Yamamoto-Blake estimator \changed{for the power spectrum multipoles} \citep{Yamamoto:2005dz,Blake:2011rj}.
We revised the definitions of the shot noise and optimal-variance weights of the power spectrum estimator to fully account for the observational systematics of BOSS.
However, in order to make use of FFT-based estimators \citep{Bianchi:2015oia, Scoccimarro:2015bla}, we approximate the power spectrum wedges of the BOSS sample by filtering out the information of Legendre multipoles $\ell>4$.
We obtain the estimate for the covariance matrices associated with our clustering measurements from the \MDPatchy \citep{Kitaura:2015uqa} and \QPM mock catalogues, which were specifically designed to mimic the clustering and observational systematics of the BOSS combined sample.

Our modelling of the anisotropic power spectrum relies on novel approaches to describe non-linearities, galaxy bias, and RSD.
The full model was validated using synthetic galaxy catalogues obtained from a set of 100 full 
\Nbody simulations using the theoretical recipe of the covariance matrix of the power spectrum wedges of \citet{Grieb:2015bia}.
Further model performance tests were conducted as part of the BOSS RSD `challenge' and using the \MDPatchy mocks that mimic the entire combined sample.
These tests show that any systematic biases in the distance and growth measurements introduced by our analysis method are smaller than the statistical errors obtained from the DR12 sample and can be neglected.

The BAO distance and the growth rate measurements inferred from our BAO+RSD fits of the Fourier space wedges are in excellent agreement with the configuration-space results of \citet{Sanchez:2016b}, which are based on the same gRPT+RSD model, and are consistent with previous measurements on the BOSS LOWZ and CMASS samples.
However, thanks to the optimization of the analysis and the improved modelling, our constraints are significantly more precise than the results obtained from previous analyses.
The BAO and RSD \changed{measurements} inferred from BOSS are in good agreement with the \LambdaCDM predictions from the \Planck data at the 1-sigma level.
\changed{The results presented here and those of all companion papers in the series analysing the BOSS DR12 combined sample are combined into the final consensus constraints in \citet{Alam:2016hwk}, which are computed using the methodology described in \citet{Sanchez:2016a}.}

We also explored the cosmological implications of our clustering measurements by directly comparing them with the predictions obtained for different cosmological models. 
We combined the information in the full-shape of the clustering wedges with CMB data from the \Planck satellite and the JLA SN sample to infer constraints on the parameters of the standard \LambdaCDM cosmological model and a number of its most important extensions such as modified DE models, non-flat universes, neutrino masses and possible deviations from the predictions of GR.
Assuming a \LambdaCDM cosmology, the combined data sets constrain the matter density parameter to $\Om = 0.311_{-0.010}^{+0.009}$ and the Hubble constant to $H_0 = 67.6_{-0.6}^{+0.7} \Unit{km \, s^{-1} \, Mpc^{-1}}$.
These values are in good agreement with the results from the \Planck 2013 + DR11 BAO BAO + SN constraints found in \citet{Anderson:2013zyy}.
Relaxing the assumption of a cosmological constant and allowing for a \changed{constant} EOS with $w \neq -1$, we find $w = 1.019_{-0.039}^{+0.048}$.
In all tested DE models, the \LambdaCDM case is always found to be very well within the $1 \sigma$ confidence intervals.
The most extreme case are the constraints using a $w$CDM model and a free $\sum m_\nu$, in which case we find $w = -1$ close to the edge of the 1-sigma interval.
Allowing for a modification in the growth rate by varying the exponent $\gamma$ in $f = [\Om(z)]^\gamma$, we measure $\gamma = 0.52 \pm 0.10$ in perfect agreement with GR ($\gamma_\mathrm{GR} = 0.55$) and with an uncertainty reduced by a factor of 1.5 compared to the previous results of \citet{Sanchez:2013tga}.
The curvature parameter $\OK$ is found to be completely consistent with zero in the tested cases.
Using the \Planck + BOSS measurements for a $K$-\LambdaCDM model, the total density of the Universe today is only allowed to deviate less than $0.3\%$ from the critical density at 68\% CL.
The neutrino mass is found to be $\sum m_\nu < 0.260 \Unit{eV} $ (95\% CL), which is consistent with other recent cosmological analyses such as weak lensing based on CFHTLenS 
\citep[$\sum m_\nu < 0.28 \Unit{eV}$ at 68 per cent CL]{Kitching:2016hvn}.
We conclude that \LambdaCDM is the preferred cosmological model among the variations explored in this work and the standard paradigm has thus been further consolidated.

Our analysis methodology can easily be applied to the data from other galaxy samples.  
In the near future, surveys such as the Hobby Eberly Telescope Dark Energy Experiment \citep[HETDEX;][]{Hill:2008mv}, the Dark Energy Spectroscopic Instrument \citep[DESI;][]{Levi:2013gra}, the Subaru Prime Focus Spectrograph \citep[PFS;][]{Ellis:2012rn} and the ESA space mission \emph{Euclid} \citep{Laureijs:2011gra} will provide even more detailed views of the LSS of the Universe, helping to improve our knowledge of the basic cosmological parameters and to further test for possible deviations from the standard \LambdaCDM model.

\section*{Acknowledgements}

We acknowledge useful discussions with Chi-Ting Chiang, Daniel Farrow, Eiichiro Komatsu, Martha Lippich, Christian Wagner, and Philipp Wullstein. JNG, AGS, SS-A, and FM acknowledge support from the Transregional Collaborative Research Centre TR33 `The Dark Universe' of the German Research Foundation (DFG).
CDV acknowledges financial support from the Spanish Ministry of Economy and Competitiveness (MINECO) under the 2011 and 2015 Severo Ochoa Programs SEV-2011-0187 and SEV-2015-0548, and grants AYA2013-46886 and AYA2014-58308.
C.C. acknowledges support from the Spanish MICINN's Consolider-Ingenio 2010 Programme under grant MultiDark CSD2009-00064 and AYA2010-21231-C02-01 grant.
C.C. was also supported by the Comunidad de Madrid under grant HEPHACOS S2009/ESP-1473 and as a MultiDark fellow.
SRT is grateful for support from the Campus de Excelencia Internacional UAM/CSIC.
MV is partially supported by Programa de Apoyo a Proyectos de Investigaci{\'o}n e Innovaci{\'o}n Tecnol{\'o}gica (PAPITT) No  IA102516.
The analysis has been performed on the computing cluster for the \Euclid project and the `Hydra' cluster at the Max Planck Computing and Data Facility (MPCDF).

Funding for SDSS-III has been provided by the Alfred~P{.}~Sloan Foundation, the Participating Institutions, the National Science Foundation, and the U{.}S{.}~Department of Energy Office of Science.
The SDSS-III web site is \url{http://www.sdss3.org/}.

SDSS-III is managed by the Astrophysical Research Consortium for the Participating Institutions of the SDSS-III Collaboration including the University of Arizona, the Brazilian Participation Group, Brookhaven National Laboratory, Carnegie Mellon University, University of Florida, the French Participation Group, the German Participation Group, Harvard University, the Instituto de Astrofisica de Canarias, the Michigan State/Notre Dame/JINA Participation Group, Johns Hopkins University, Lawrence Berkeley National Laboratory, Max Planck Institute for Astrophysics, Max Planck Institute for Extraterrestrial Physics, New Mexico State University, New York University, Ohio State University, Pennsylvania State University, University of Portsmouth, Princeton University, the Spanish Participation Group, University of Tokyo, University of Utah, Vanderbilt University, University of Virginia, University of Washington, and Yale University.

Based on observations obtained with \Planck (\url{http://www.esa.int/Planck}), an ESA science mission with instruments and contributions directly funded by ESA Member States, NASA, and Canada.



\bibliographystyle{mnras}
\bibliography{bib_article}

\begin{thebibliography}{}
\makeatletter
\relax
\def\mn@urlcharsother{\let\do\@makeother \do\$\do\&\do\#\do\^\do\_\do\%\do\~}
\def\mn@doi{\begingroup\mn@urlcharsother \@ifnextchar [ {\mn@doi@}
  {\mn@doi@[]}}
\def\mn@doi@[#1]#2{\def\@tempa{#1}\ifx\@tempa\@empty \href
  {http://dx.doi.org/#2} {doi:#2}\else \href {http://dx.doi.org/#2} {#1}\fi
  \endgroup}
\def\mn@eprint#1#2{\mn@eprint@#1:#2::\@nil}
\def\mn@eprint@arXiv#1{\href {http://arxiv.org/abs/#1} {{\tt arXiv:#1}}}
\def\mn@eprint@dblp#1{\href {http://dblp.uni-trier.de/rec/bibtex/#1.xml}
  {dblp:#1}}
\def\mn@eprint@#1:#2:#3:#4\@nil{\def\@tempa {#1}\def\@tempb {#2}\def\@tempc
  {#3}\ifx \@tempc \@empty \let \@tempc \@tempb \let \@tempb \@tempa \fi \ifx
  \@tempb \@empty \def\@tempb {arXiv}\fi \@ifundefined
  {mn@eprint@\@tempb}{\@tempb:\@tempc}{\expandafter \expandafter \csname
  mn@eprint@\@tempb\endcsname \expandafter{\@tempc}}}

\bibitem[\protect\citeauthoryear{Alam et~al.}{Alam
  et~al.}{2015a}]{Alam:2015mbd}
Alam S.,  et~al., 2015a, \mn@doi [\apj Suppl.] {10.1088/0067-0049/219/1/12},
  \href {http://adsabs.harvard.edu/abs/2015ApJS..219...12A} {219, 12}

\bibitem[\protect\citeauthoryear{Alam, Ho, Vargas-Maga{\~n}a  \&
  Schneider}{Alam et~al.}{2015b}]{Alam:2015qta}
Alam S.,  Ho S.,  Vargas-Maga{\~n}a M.,   Schneider D.~P.,  2015b, \mn@doi
  [\mnras] {10.1093/mnras/stv1737}, \href
  {http://adsabs.harvard.edu/abs/2015MNRAS.453.1754A} {453, 1754}

\bibitem[\protect\citeauthoryear{Alam et~al.}{Alam et~al.}{2016}]{Alam:2016hwk}
Alam S.,  et~al., 2016, preprint (\mn@eprint {arXiv} {1607.03155}, \href
  {http://adsabs.harvard.edu/abs/2016arXiv160703155A} {submitted to \mnras})

\bibitem[\protect\citeauthoryear{Alcock \& Paczynski}{Alcock \&
  Paczynski}{1979}]{AP:1979}
Alcock C.,  Paczynski B.,  1979, \mn@doi [Nature] {10.1038/281358a0}, \href
  {http://adsabs.harvard.edu/abs/1979Natur.281..358A} {281, 358}

\bibitem[\protect\citeauthoryear{{Anderson} et~al.,}{{Anderson}
  et~al.}{2012}]{Anderson2012}
{Anderson} L.,  et~al., 2012, \mn@doi [\mnras]
  {10.1111/j.1365-2966.2012.22066.x}, \href
  {http://adsabs.harvard.edu/abs/2012MNRAS.427.3435A} {427, 3435}

\bibitem[\protect\citeauthoryear{{Anderson} et~al.,}{{Anderson}
  et~al.}{2014a}]{Anderson:2013oza}
{Anderson} L.,  et~al., 2014a, \mn@doi [\mnras] {10.1093/mnras/stt2206}, \href
  {http://adsabs.harvard.edu/abs/2014MNRAS.439...83A} {439, 83}

\bibitem[\protect\citeauthoryear{{Anderson} et~al.,}{{Anderson}
  et~al.}{2014b}]{Anderson:2013zyy}
{Anderson} L.,  et~al., 2014b, \mn@doi [\mnras] {10.1093/mnras/stu523}, \href
  {http://adsabs.harvard.edu/abs/2014MNRAS.441...24A} {441, 24}

\bibitem[\protect\citeauthoryear{Angulo, Baugh, Frenk  \& Lacey}{Angulo
  et~al.}{2008}]{Angulo:2007fw}
Angulo R.,  Baugh C.~M.,  Frenk C.~S.,   Lacey C.~G.,  2008, \mn@doi [\mnras]
  {10.1111/j.1365-2966.2007.12587.x}, \href
  {http://adsabs.harvard.edu/abs/2008MNRAS.383..755A} {383, 755}

\bibitem[\protect\citeauthoryear{Baldauf, Seljak, Smith, Hamaus  \&
  Desjacques}{Baldauf et~al.}{2013}]{Baldauf:2013hka}
Baldauf T.,  Seljak U.,  Smith R.~E.,  Hamaus N.,   Desjacques V.,  2013,
  \mn@doi [\prd] {10.1103/PhysRevD.88.083507}, \href
  {http://adsabs.harvard.edu/abs/2013PhRvD..88h3507B} {D88, 083507}

\bibitem[\protect\citeauthoryear{{Ballinger}, {Peacock}  \&
  {Heavens}}{{Ballinger} et~al.}{1996}]{Ballinger:1996}
{Ballinger} W.~E.,  {Peacock} J.~A.,   {Heavens} A.~F.,  1996, \mn@doi [\mnras]
  {10.1093/mnras/282.3.877}, \href
  {http://adsabs.harvard.edu/abs/1996MNRAS.282..877B} {282, 877}

\bibitem[\protect\citeauthoryear{Bassett \& Hlozek}{Bassett \&
  Hlozek}{2010}]{Bassett:2009mm}
Bassett B.~A.,  Hlozek R.,  2010, {Baryon Acoustic Oscillations, In: Dark
  Energy}.
Cambridge University Press (\mn@eprint {arXiv} {0910.5224})

\bibitem[\protect\citeauthoryear{Bernardeau, Crocce  \& Scoccimarro}{Bernardeau
  et~al.}{2008}]{Bernardeau:2008fa}
Bernardeau F.,  Crocce M.,   Scoccimarro R.,  2008, \mn@doi [\prd]
  {10.1103/PhysRevD.78.103521}, \href
  {http://adsabs.harvard.edu/abs/2008PhRvD..78j3521B} {D78, 103521}

\bibitem[\protect\citeauthoryear{Betoule et~al.}{Betoule
  et~al.}{2014}]{Betoule:2014frx}
Betoule M.,  et~al., 2014, \mn@doi [\aap] {10.1051/0004-6361/201423413}, \href
  {http://adsabs.harvard.edu/abs/2014A%26A...568A..22B} {568, A22}

\bibitem[\protect\citeauthoryear{Beutler et~al.}{Beutler
  et~al.}{2014}]{Beutler:2013yhm}
Beutler F.,  et~al., 2014, \mn@doi [\mnras] {10.1093/mnras/stu1051}, \href
  {http://adsabs.harvard.edu/abs/2014MNRAS.443.1065B} {443, 1065}

\bibitem[\protect\citeauthoryear{Beutler et~al.}{Beutler
  et~al.}{2016a}]{Beutler:2016arn}
Beutler F.,  et~al., 2016a, preprint (\mn@eprint {arXiv} {1607.03150}, \href
  {http://adsabs.harvard.edu/abs/2016arXiv160703150B} {submitted to \mnras})

\bibitem[\protect\citeauthoryear{Beutler et~al.}{Beutler
  et~al.}{2016b}]{Beutler:2016ixs}
Beutler F.,  et~al., 2016b, preprint (\mn@eprint {arXiv} {1607.03149}, \href
  {http://adsabs.harvard.edu/abs/2016arXiv160703149B} {submitted to \mnras})

\bibitem[\protect\citeauthoryear{Bianchi, Gil-Mar{\'i}n, Ruggeri  \&
  Percival}{Bianchi et~al.}{2015}]{Bianchi:2015oia}
Bianchi D.,  Gil-Mar{\'i}n H.,  Ruggeri R.,   Percival W.~J.,  2015, \mn@doi
  [\mnras] {10.1093/mnrasl/slv090}, \href
  {http://adsabs.harvard.edu/abs/2015MNRAS.453L..11B} {453, L11}

\bibitem[\protect\citeauthoryear{{Blake} et~al.,}{{Blake}
  et~al.}{2011}]{Blake:2011rj}
{Blake} C.,  et~al., 2011, \mn@doi [\mnras] {10.1111/j.1365-2966.2011.18903.x},
  \href {http://adsabs.harvard.edu/abs/2011MNRAS.415.2876B} {415, 2876}

\bibitem[\protect\citeauthoryear{Bolton et~al.}{Bolton
  et~al.}{2012}]{Bolton:2012hz}
Bolton A.~S.,  et~al., 2012, \mn@doi [\aj] {10.1088/0004-6256/144/5/144}, \href
  {http://adsabs.harvard.edu/abs/2012AJ....144..144B} {144, 144}

\bibitem[\protect\citeauthoryear{Casas-Miranda, Mo, Sheth  \&
  Boerner}{Casas-Miranda et~al.}{2002}]{CasasMiranda:2001ym}
Casas-Miranda R.,  Mo H.~J.,  Sheth R.~K.,   Boerner G.,  2002, \mn@doi
  [\mnras] {10.1046/j.1365-8711.2002.05378.x}, \href
  {http://adsabs.harvard.edu/abs/2002MNRAS.333..730C} {333, 730}

\bibitem[\protect\citeauthoryear{Chan, Scoccimarro  \& Sheth}{Chan
  et~al.}{2012}]{Chan:2012jj}
Chan K.~C.,  Scoccimarro R.,   Sheth R.~K.,  2012, \mn@doi [\prd]
  {10.1103/PhysRevD.85.083509}, \href
  {http://adsabs.harvard.edu/abs/2012PhRvD..85h3509C} {D85, 083509}

\bibitem[\protect\citeauthoryear{Chevallier \& Polarski}{Chevallier \&
  Polarski}{2001}]{Chevallier:2000qy}
Chevallier M.,  Polarski D.,  2001, \mn@doi [Int. J. Mod. Phys. D]
  {10.1142/S0218271801000822}, \href
  {http://adsabs.harvard.edu/abs/2001IJMPD..10..213C} {D10, 213}

\bibitem[\protect\citeauthoryear{{Chuang} et~al.,}{{Chuang}
  et~al.}{2016}]{Chuang:2013wga}
{Chuang} C.-H.,  et~al., 2016, \mn@doi [\mnras] {10.1093/mnras/stw1535}, \href
  {http://adsabs.harvard.edu/abs/2016MNRAS.461.3781C} {461, 3781}

\bibitem[\protect\citeauthoryear{Cole et~al.}{Cole et~al.}{2005}]{Cole:2005sx}
Cole S.,  et~al., 2005, \mn@doi [\mnras] {10.1111/j.1365-2966.2005.09318.x},
  \href {http://adsabs.harvard.edu/abs/2005MNRAS.362..505C} {362, 505}

\bibitem[\protect\citeauthoryear{Crocce \& Scoccimarro}{Crocce \&
  Scoccimarro}{2006}]{Crocce:2005xy}
Crocce M.,  Scoccimarro R.,  2006, \mn@doi [\prd] {10.1103/PhysRevD.73.063519},
  \href {http://adsabs.harvard.edu/abs/2006PhRvD..73f3519C} {D73, 063519}

\bibitem[\protect\citeauthoryear{Crocce, Scoccimarro  \& Bernardeau}{Crocce
  et~al.}{2012}]{Crocce:2012fa}
Crocce M.,  Scoccimarro R.,   Bernardeau F.,  2012, \mn@doi [\mnras]
  {10.1111/j.1365-2966.2012.22127.x}, \href
  {http://adsabs.harvard.edu/abs/2012MNRAS.427.2537C} {427, 2537}

\bibitem[\protect\citeauthoryear{Cuesta et~al.}{Cuesta
  et~al.}{2016}]{Cuesta:2015mqa}
Cuesta A.~J.,  et~al., 2016, \mn@doi [\mnras] {10.1093/mnras/stw066}, \href
  {http://adsabs.harvard.edu/abs/2016MNRAS.457.1770C} {457, 1770}

\bibitem[\protect\citeauthoryear{{Davis} \& {Peebles}}{{Davis} \&
  {Peebles}}{1983}]{Davis:1983}
{Davis} M.,  {Peebles} P.~J.~E.,  1983, \mn@doi [\apj] {10.1086/160884}, \href
  {http://adsabs.harvard.edu/abs/1983ApJ...267..465D} {267, 465}

\bibitem[\protect\citeauthoryear{Dawson et~al.}{Dawson
  et~al.}{2013}]{Dawson:2012va}
Dawson K.~S.,  et~al., 2013, \mn@doi [\aj] {10.1088/0004-6256/145/1/10}, \href
  {http://adsabs.harvard.edu/abs/2013AJ....145...10D} {145, 10}

\bibitem[\protect\citeauthoryear{Dawson et~al.}{Dawson
  et~al.}{2016}]{Dawson:2015wdb}
Dawson K.~S.,  et~al., 2016, \mn@doi [\aj] {10.3847/0004-6256/151/2/44}, \href
  {http://adsabs.harvard.edu/abs/2016AJ....151...44D} {151, 44}

\bibitem[\protect\citeauthoryear{Dodelson \& Schneider}{Dodelson \&
  Schneider}{2013}]{Dodelson:2013uaa}
Dodelson S.,  Schneider M.~D.,  2013, \mn@doi [\prd]
  {10.1103/PhysRevD.88.063537}, \href
  {http://adsabs.harvard.edu/abs/2013PhRvD..88f3537D} {D88, 063537}

\bibitem[\protect\citeauthoryear{{Doi} et~al.,}{{Doi}
  et~al.}{2010}]{Doi:2010rf}
{Doi} M.,  et~al., 2010, \mn@doi [\aj] {10.1088/0004-6256/139/4/1628}, \href
  {http://adsabs.harvard.edu/abs/2010AJ....139.1628D} {139, 1628}

\bibitem[\protect\citeauthoryear{Efstathiou \& Bond}{Efstathiou \&
  Bond}{1999}]{Efstathiou:1998xx}
Efstathiou G.,  Bond J.~R.,  1999, \mn@doi [\mnras]
  {10.1046/j.1365-8711.1999.02274.x}, \href
  {http://adsabs.harvard.edu/abs/1999MNRAS.304...75E} {304, 75}

\bibitem[\protect\citeauthoryear{Eisenstein \& White}{Eisenstein \&
  White}{2004}]{Eisenstein:2004an}
Eisenstein D.~J.,  White M.~J.,  2004, \mn@doi [\prd]
  {10.1103/PhysRevD.70.103523}, \href
  {http://adsabs.harvard.edu/abs/2004PhRvD..70j3523E} {D70, 103523}

\bibitem[\protect\citeauthoryear{Eisenstein et~al.}{Eisenstein
  et~al.}{2005}]{Eisenstein:2005su}
Eisenstein D.~J.,  et~al., 2005, \mn@doi [\apj] {10.1086/466512}, \href
  {http://adsabs.harvard.edu/abs/2005ApJ...633..560E} {633, 560}

\bibitem[\protect\citeauthoryear{Eisenstein, Seo, Sirko  \& Spergel}{Eisenstein
  et~al.}{2007}]{Eisenstein:2006nk}
Eisenstein D.~J.,  Seo H.-J.,  Sirko E.,   Spergel D.,  2007, \mn@doi [\apj]
  {10.1086/518712}, \href {http://adsabs.harvard.edu/abs/2007ApJ...664..675E}
  {664, 675}

\bibitem[\protect\citeauthoryear{Eisenstein et~al.}{Eisenstein
  et~al.}{2011}]{Eisenstein:2011sa}
Eisenstein D.~J.,  et~al., 2011, \mn@doi [\aj] {10.1088/0004-6256/142/3/72},
  \href {http://adsabs.harvard.edu/abs/2011AJ....142...72E} {142, 72}

\bibitem[\protect\citeauthoryear{Ellis et~al.}{Ellis
  et~al.}{2014}]{Ellis:2012rn}
Ellis R.,  et~al., 2014, \mn@doi [\pasj] {10.1093/pasj/pst019}, \href
  {http://adsabs.harvard.edu/abs/2014PASJ...66R...1T} {66, R1}

\bibitem[\protect\citeauthoryear{Feldman, Kaiser  \& Peacock}{Feldman
  et~al.}{1994}]{Feldman:1993ky}
Feldman H.~A.,  Kaiser N.,   Peacock J.~A.,  1994, \mn@doi [\apj]
  {10.1086/174036}, \href {http://adsabs.harvard.edu/abs/1994ApJ...426...23F}
  {426, 23}

\bibitem[\protect\citeauthoryear{{Font-Ribera} et~al.,}{{Font-Ribera}
  et~al.}{2014}]{Font-Ribera:2014}
{Font-Ribera} A.,  et~al., 2014, \mn@doi [\jcap]
  {10.1088/1475-7516/2014/05/027}, \href
  {http://adsabs.harvard.edu/abs/2014JCAP...05..027F} {5, 027}

\bibitem[\protect\citeauthoryear{Fukugita, Ichikawa, Gunn, Doi, Shimasaku  \&
  Schneider}{Fukugita et~al.}{1996}]{Fukugita:1996qt}
Fukugita M.,  Ichikawa T.,  Gunn J.~E.,  Doi M.,  Shimasaku K.,   Schneider
  D.~P.,  1996, \mn@doi [\aj] {10.1086/117915}, \href
  {http://adsabs.harvard.edu/abs/1996AJ....111.1748F} {111, 1748}

\bibitem[\protect\citeauthoryear{Gelman \& Rubin}{Gelman \&
  Rubin}{1992}]{Gelman:1992:I}
Gelman A.,  Rubin D.~B.,  1992, \mn@doi [Statist. Sci.]
  {10.1214/ss/1177011136}, 7, 457

\bibitem[\protect\citeauthoryear{Gil-Mar{\'i}n, Nore{\~n}a, Verde, Percival,
  Wagner, Manera  \& Schneider}{Gil-Mar{\'i}n et~al.}{2015}]{Gil-Marin:2014sta}
Gil-Mar{\'i}n H.,  Nore{\~n}a J.,  Verde L.,  Percival W.~J.,  Wagner C.,
  Manera M.,   Schneider D.~P.,  2015, \mn@doi [\mnras] {10.1093/mnras/stv961},
  \href {http://adsabs.harvard.edu/abs/2015MNRAS.451..539G} {451, 5058}

\bibitem[\protect\citeauthoryear{Gil-Mar{\'i}n et~al.}{Gil-Mar{\'i}n
  et~al.}{2016a}]{Gil-Marin:2015sqa}
Gil-Mar{\'i}n H.,  et~al., 2016a, \mn@doi [\mnras] {10.1093/mnras/stw1096},
  \href {http://adsabs.harvard.edu/abs/2016MNRAS.460.4188G} {460, 4188}

\bibitem[\protect\citeauthoryear{{Gil-Mar{\'{\i}}n} et~al.,}{{Gil-Mar{\'{\i}}n}
  et~al.}{2016b}]{Gil-Marin:2015nqa}
{Gil-Mar{\'{\i}}n} H.,  et~al., 2016b, \mn@doi [\mnras]
  {10.1093/mnras/stw1264}, \href
  {http://adsabs.harvard.edu/abs/2016MNRAS.460.4210G} {460, 4210}

\bibitem[\protect\citeauthoryear{Gong}{Gong}{2008}]{Gong:2008fh}
Gong Y.,  2008, \mn@doi [\prd] {10.1103/PhysRevD.78.123010}, \href
  {http://adsabs.harvard.edu/abs/2008PhRvD..78l3010G} {D78, 123010}

\bibitem[\protect\citeauthoryear{{Grieb}, {S{\'a}nchez}, {Salazar-Albornoz}  \&
  {Dalla Vecchia}}{{Grieb} et~al.}{2016}]{Grieb:2015bia}
{Grieb} J.~N.,  {S{\'a}nchez} A.~G.,  {Salazar-Albornoz} S.,   {Dalla Vecchia}
  C.,  2016, \mn@doi [\mnras] {10.1093/mnras/stw065}, \href
  {http://adsabs.harvard.edu/abs/2016MNRAS.457.1577G} {457, 1577}

\bibitem[\protect\citeauthoryear{Gunn et~al.}{Gunn et~al.}{1998}]{Gunn:1998vh}
Gunn J.~E.,  et~al., 1998, \mn@doi [\aj] {10.1086/300645}, \href
  {http://adsabs.harvard.edu/abs/1998AJ....116.3040G} {116, 3040}

\bibitem[\protect\citeauthoryear{Gunn et~al.}{Gunn et~al.}{2006}]{Gunn:2006tw}
Gunn J.~E.,  et~al., 2006, \mn@doi [\aj] {10.1086/500975}, \href
  {http://adsabs.harvard.edu/abs/2006AJ....131.2332G} {131, 2332}

\bibitem[\protect\citeauthoryear{{Guzzo} et~al.,}{{Guzzo}
  et~al.}{2008}]{Guzzo:2008}
{Guzzo} L.,  et~al., 2008, \mn@doi [\nat] {10.1038/nature06555}, \href
  {http://adsabs.harvard.edu/abs/2008Natur.451..541G} {451, 541}

\bibitem[\protect\citeauthoryear{Hamaus, Seljak, Desjacques, Smith  \&
  Baldauf}{Hamaus et~al.}{2010}]{Hamaus:2010im}
Hamaus N.,  Seljak U.,  Desjacques V.,  Smith R.~E.,   Baldauf T.,  2010,
  \mn@doi [\prd] {10.1103/PhysRevD.82.043515}, \href
  {http://adsabs.harvard.edu/abs/2010PhRvD..82d3515H} {D82, 043515}

\bibitem[\protect\citeauthoryear{Hartlap, Simon  \& Schneider}{Hartlap
  et~al.}{2007}]{Hartlap:2006kj}
Hartlap J.,  Simon P.,   Schneider P.,  2007, \mn@doi [\aap]
  {10.1051/0004-6361:20066170}, \href
  {http://adsabs.harvard.edu/abs/2007A%26A...464..399H} {464, 399}

\bibitem[\protect\citeauthoryear{{Hill} et~al.,}{{Hill}
  et~al.}{2008}]{Hill:2008mv}
{Hill} G.~J.,  et~al., 2008, in Kodama T.,  Yamada T.,   Aoki K.,  eds,  ASP
  Conf. Ser. Vol. 399, Panoramic Views of Galaxy Formation and Evolution.
  Astron. Soc. Pac., San Francisco, pp 115--118 (\mn@eprint {arXiv}
  {0806.0183})

\bibitem[\protect\citeauthoryear{Hu \& Haiman}{Hu \& Haiman}{2003}]{Hu:2003ti}
Hu W.,  Haiman Z.,  2003, \mn@doi [\prd] {10.1103/PhysRevD.68.063004}, \href
  {http://adsabs.harvard.edu/abs/2003PhRvD..68f3004H} {D68, 063004}

\bibitem[\protect\citeauthoryear{Jing}{Jing}{2005}]{Jing:2004fq}
Jing Y.~P.,  2005, \mn@doi [\apj] {10.1086/427087}, \href
  {http://adsabs.harvard.edu/abs/2005ApJ...620..559J} {620, 559}

\bibitem[\protect\citeauthoryear{Kaiser}{Kaiser}{1987}]{Kaiser:1987qv}
Kaiser N.,  1987, \mn@doi [\mnras] {10.1093/mnras/227.1.1}, \href
  {http://adsabs.harvard.edu/abs/1987MNRAS.227....1K} {227, 1}

\bibitem[\protect\citeauthoryear{Kazin, S{\'a}nchez  \& Blanton}{Kazin
  et~al.}{2012}]{Kazin:2011xt}
Kazin E.~A.,  S{\'a}nchez A.~G.,   Blanton M.~R.,  2012, \mn@doi [\mnras]
  {10.1111/j.1365-2966.2011.19962.x}, \href
  {http://adsabs.harvard.edu/abs/2012MNRAS.419.3223K} {419, 3223}

\bibitem[\protect\citeauthoryear{{Kazin} et~al.,}{{Kazin}
  et~al.}{2013}]{Kazin:2013rxa}
{Kazin} E.~A.,  et~al., 2013, \mn@doi [\mnras] {10.1093/mnras/stt1261}, \href
  {http://adsabs.harvard.edu/abs/2013MNRAS.435...64K} {435, 64}

\bibitem[\protect\citeauthoryear{Kitaura \& Hess}{Kitaura \&
  Hess}{2013}]{Kitaura:2012tj}
Kitaura F.-S.,  Hess S.,  2013, \mn@doi [\mnras] {10.1093/mnrasl/slt101}, \href
  {http://adsabs.harvard.edu/abs/2013MNRAS.435L..78K} {435, 78}

\bibitem[\protect\citeauthoryear{Kitaura, Yepes  \& Prada}{Kitaura
  et~al.}{2014}]{Kitaura:2013cwa}
Kitaura F.-S.,  Yepes G.,   Prada F.,  2014, \mn@doi [\mnras]
  {10.1093/mnrasl/slt172}, \href
  {http://adsabs.harvard.edu/abs/2014MNRAS.439L..21K} {439, 21}

\bibitem[\protect\citeauthoryear{Kitaura et~al.}{Kitaura
  et~al.}{2016}]{Kitaura:2015uqa}
Kitaura F.-S.,  et~al., 2016, \mn@doi [\mnras] {10.1093/mnras/stv2826}, \href
  {http://adsabs.harvard.edu/abs/2016MNRAS.456.4156K} {456, 4156}

\bibitem[\protect\citeauthoryear{Kitching, Verde, Heavens  \& Jimenez}{Kitching
  et~al.}{2016}]{Kitching:2016hvn}
Kitching T.~D.,  Verde L.,  Heavens A.~F.,   Jimenez R.,  2016, \mn@doi
  [\mnras] {10.1093/mnras/stw707}, \href
  {http://adsabs.harvard.edu/abs/2016MNRAS.459..971K} {459, 971}

\bibitem[\protect\citeauthoryear{Klypin, Yepes, Gottlober, Prada  \&
  Hess}{Klypin et~al.}{2016}]{Klypin:2014kpa}
Klypin A.,  Yepes G.,  Gottlober S.,  Prada F.,   Hess S.,  2016, \mn@doi
  [\mnras] {10.1093/mnras/stw248}, \href
  {http://adsabs.harvard.edu/abs/2016MNRAS.457.4340K} {457, 4340}

\bibitem[\protect\citeauthoryear{Komatsu et~al.}{Komatsu
  et~al.}{2009}]{Komatsu:2008hk}
Komatsu E.,  et~al., 2009, \mn@doi [\apj Suppl.] {10.1088/0067-0049/180/2/330},
  \href {http://adsabs.harvard.edu/abs/2009ApJS..180..330K} {180, 330}

\bibitem[\protect\citeauthoryear{Laureijs et~al.}{Laureijs
  et~al.}{2011}]{Laureijs:2011gra}
Laureijs R.,  et~al., 2011, preprint (\mn@eprint {arXiv} {1110.3193})

\bibitem[\protect\citeauthoryear{Levi et~al.}{Levi et~al.}{2013}]{Levi:2013gra}
Levi M.,  et~al., 2013, preprint (\mn@eprint {arXiv} {1308.0847})

\bibitem[\protect\citeauthoryear{Lewis \& Bridle}{Lewis \&
  Bridle}{2002}]{Lewis:2002ah}
Lewis A.,  Bridle S.,  2002, \mn@doi [\prd] {10.1103/PhysRevD.66.103511}, \href
  {http://adsabs.harvard.edu/abs/2002PhRvD..66j3511L} {D66, 103511}

\bibitem[\protect\citeauthoryear{Lewis, Challinor  \& Lasenby}{Lewis
  et~al.}{2000}]{Lewis:1999bs}
Lewis A.,  Challinor A.,   Lasenby A.,  2000, \mn@doi [\apj] {10.1086/309179},
  \href {http://adsabs.harvard.edu/abs/2000ApJ...538..473L} {538, 473}

\bibitem[\protect\citeauthoryear{Linder}{Linder}{2003}]{Linder:2002et}
Linder E.~V.,  2003, \mn@doi [\prl] {10.1103/PhysRevLett.90.091301}, \href
  {http://adsabs.harvard.edu/abs/2003PhRvL..90i1301L} {90, 091301}

\bibitem[\protect\citeauthoryear{Linder \& Cahn}{Linder \&
  Cahn}{2007}]{Linder:2007hg}
Linder E.~V.,  Cahn R.~N.,  2007, \mn@doi [Astropart. Phys.]
  {10.1016/j.astropartphys.2007.09.003}, \href
  {http://adsabs.harvard.edu/abs/2007APh....28..481L} {28, 481}

\bibitem[\protect\citeauthoryear{{Maddox}, {Efstathiou}, {Sutherland}  \&
  {Loveday}}{{Maddox} et~al.}{1990}]{Maddox:1990}
{Maddox} S.~J.,  {Efstathiou} G.,  {Sutherland} W.~J.,   {Loveday} J.,  1990,
  \mn@doi [\mnras] {10.1093/mnras/242.1.43P}, \href
  {http://adsabs.harvard.edu/abs/1990MNRAS.242P..43M} {242, 43P}

\bibitem[\protect\citeauthoryear{Manera \& Gaztanaga}{Manera \&
  Gaztanaga}{2011}]{Manera:2009zu}
Manera M.,  Gaztanaga E.,  2011, \mn@doi [\mnras]
  {10.1111/j.1365-2966.2011.18705.x}, \href
  {http://adsabs.harvard.edu/abs/2011MNRAS.415..383M} {415, 383}

\bibitem[\protect\citeauthoryear{Manera et~al.}{Manera
  et~al.}{2012}]{Manera:2012sc}
Manera M.,  et~al., 2012, \mn@doi [\mnras] {10.1093/mnras/sts084}, \href
  {http://adsabs.harvard.edu/abs/2013MNRAS.428.1036M} {428, 1036}

\bibitem[\protect\citeauthoryear{{Maraston} et~al.,}{{Maraston}
  et~al.}{2013}]{Maraston:2012jf}
{Maraston} C.,  et~al., 2013, \mn@doi [\mnras] {10.1093/mnras/stt1424}, \href
  {http://adsabs.harvard.edu/abs/2013MNRAS.435.2764M} {435, 2764}

\bibitem[\protect\citeauthoryear{Matsubara}{Matsubara}{2011}]{Matsubara:2011ck}
Matsubara T.,  2011, \mn@doi [\prd] {10.1103/PhysRevD.83.083518}, \href
  {http://adsabs.harvard.edu/abs/2011PhRvD..83h3518M} {D83, 083518}

\bibitem[\protect\citeauthoryear{Montesano, Sanchez  \& Phleps}{Montesano
  et~al.}{2010}]{Montesano:2010qc}
Montesano F.,  Sanchez A.~G.,   Phleps S.,  2010, \mn@doi [\mnras]
  {10.1111/j.1365-2966.2010.17292.x}, \href
  {http://adsabs.harvard.edu/abs/2010MNRAS.408.2397M} {408, 2397}

\bibitem[\protect\citeauthoryear{Nishimichi \& Taruya}{Nishimichi \&
  Taruya}{2011}]{Nishimichi:2011jm}
Nishimichi T.,  Taruya A.,  2011, \mn@doi [\prd] {10.1103/PhysRevD.84.043526},
  \href {http://adsabs.harvard.edu/abs/2011PhRvD..84d3526N} {D84, 043526}

\bibitem[\protect\citeauthoryear{Nuza et~al.}{Nuza et~al.}{2013}]{Nuza:2012mw}
Nuza S.~E.,  et~al., 2013, \mn@doi [\mnras] {10.1093/mnras/stt513}, \href
  {http://adsabs.harvard.edu/abs/2013MNRAS.432..743N} {432, 743}

\bibitem[\protect\citeauthoryear{Oka, Saito, Nishimichi, Taruya  \&
  Yamamoto}{Oka et~al.}{2014}]{Oka:2013cba}
Oka A.,  Saito S.,  Nishimichi T.,  Taruya A.,   Yamamoto K.,  2014, \mn@doi
  [\mnras] {10.1093/mnras/stu111}, \href
  {http://adsabs.harvard.edu/abs/2014MNRAS.439.2515O} {439, 2515}

\bibitem[\protect\citeauthoryear{{Otten} \& {Weinheimer}}{{Otten} \&
  {Weinheimer}}{2008}]{Otten:2008}
{Otten} E.~W.,  {Weinheimer} C.,  2008, \mn@doi [Reports on Progress in
  Physics] {10.1088/0034-4885/71/8/086201}, \href
  {http://adsabs.harvard.edu/abs/2008RPPh...71h6201O} {71, 086201}

\bibitem[\protect\citeauthoryear{Padmanabhan \& White}{Padmanabhan \&
  White}{2008}]{Padmanabhan:2008ag}
Padmanabhan N.,  White M.~J.,  2008, \mn@doi [\prd]
  {10.1103/PhysRevD.77.123540}, \href
  {http://adsabs.harvard.edu/abs/2008PhRvD..77l3540P} {D77, 123540}

\bibitem[\protect\citeauthoryear{Percival \& White}{Percival \&
  White}{2009}]{Percival:2008sh}
Percival W.~J.,  White M.,  2009, \mn@doi [\mnras]
  {10.1111/j.1365-2966.2008.14211.x}, \href
  {http://adsabs.harvard.edu/abs/2009MNRAS.393..297P} {393, 297}

\bibitem[\protect\citeauthoryear{Percival et~al.}{Percival
  et~al.}{2002}]{Percival:2002gq}
Percival W.~J.,  et~al., 2002, \mn@doi [\mnras]
  {10.1046/j.1365-8711.2002.06001.x}, \href
  {http://adsabs.harvard.edu/abs/2002MNRAS.337.1068P} {337, 1068}

\bibitem[\protect\citeauthoryear{{Percival} et~al.,}{{Percival}
  et~al.}{2014}]{Percival:2013sga}
{Percival} W.~J.,  et~al., 2014, \mn@doi [\mnras] {10.1093/mnras/stu112}, \href
  {http://adsabs.harvard.edu/abs/2014MNRAS.439.2531P} {439, 2531}

\bibitem[\protect\citeauthoryear{{Planck Collaboration I}}{{Planck
  Collaboration I}}{2015}]{Adam:2015rua}
{Planck Collaboration I} 2015, preprint (\mn@eprint {arXiv} {1502.01582})

\bibitem[\protect\citeauthoryear{{Planck Collaboration XIII}}{{Planck
  Collaboration XIII}}{2015}]{Planck:2015xua}
{Planck Collaboration XIII} 2015, preprint (\mn@eprint {arXiv} {1502.01589})

\bibitem[\protect\citeauthoryear{Reid, Seo, Leauthaud, Tinker  \& White}{Reid
  et~al.}{2014}]{Reid:2014iaa}
Reid B.~A.,  Seo H.-J.,  Leauthaud A.,  Tinker J.~L.,   White M.,  2014,
  \mn@doi [\mnras] {10.1093/mnras/stu1391}, \href
  {http://adsabs.harvard.edu/abs/2014MNRAS.444..476R} {444, 476}

\bibitem[\protect\citeauthoryear{Reid et~al.,}{Reid
  et~al.}{2016}]{Reid:2015gra}
Reid B.~A.,  et~al., 2016, \mn@doi [\mnras] {10.1093/mnras/stv2382}, \href
  {http://adsabs.harvard.edu/abs/2016MNRAS.455.1553R} {455, 1553}

\bibitem[\protect\citeauthoryear{Rodr{\'\i}guez-Torres
  et~al.}{Rodr{\'\i}guez-Torres et~al.}{2016}]{Rodriguez-Torres:2015vqa}
Rodr{\'\i}guez-Torres S.~A.,  et~al., 2016, \mn@doi [\mnras]
  {10.1093/mnras/stw1014}, \href
  {http://adsabs.harvard.edu/abs/2016MNRAS.460.1173R} {460, 1173}

\bibitem[\protect\citeauthoryear{Ross et~al.}{Ross et~al.}{2012}]{Ross:2012qm}
Ross A.~J.,  et~al., 2012, \mn@doi [\mnras] {10.1111/j.1365-2966.2012.21235.x},
  \href {http://adsabs.harvard.edu/abs/2012MNRAS.424..564R} {424, 564}

\bibitem[\protect\citeauthoryear{{Ross} et~al.,}{{Ross}
  et~al.}{2013}]{Ross:2012sx}
{Ross} A.~J.,  et~al., 2013, \mn@doi [\mnras] {10.1093/mnras/sts094}, \href
  {http://adsabs.harvard.edu/abs/2013MNRAS.428.1116R} {428, 1116}

\bibitem[\protect\citeauthoryear{Ross et~al.}{Ross et~al.}{2016}]{Ross:2016gvb}
Ross A.~J.,  et~al., 2016, preprint (\mn@eprint {arXiv} {1607.03145}, \href
  {http://adsabs.harvard.edu/abs/2016arXiv160703145R} {submitted to \mnras})

\bibitem[\protect\citeauthoryear{Salazar-Albornoz et~al.}{Salazar-Albornoz
  et~al.}{2016}]{Salazar-Albornoz:2016psd}
Salazar-Albornoz S.,  et~al., 2016, preprint (\mn@eprint {arXiv} {1607.03144},
  \href {http://adsabs.harvard.edu/abs/2016arXiv160703144S} {submitted to
  \mnras})

\bibitem[\protect\citeauthoryear{{Samushia} et~al.,}{{Samushia}
  et~al.}{2014}]{Samushia:2013yga}
{Samushia} L.,  et~al., 2014, \mn@doi [\mnras] {10.1093/mnras/stu197}, \href
  {http://adsabs.harvard.edu/abs/2014MNRAS.439.3504S} {439, 3504}

\bibitem[\protect\citeauthoryear{Samushia, Branchini  \& Percival}{Samushia
  et~al.}{2015}]{Samushia:2015wta}
Samushia L.,  Branchini E.,   Percival W.,  2015, \mn@doi [\mnras]
  {10.1093/mnras/stv1568}, \href
  {http://adsabs.harvard.edu/abs/2015MNRAS.452.3704S} {452, 3704}

\bibitem[\protect\citeauthoryear{S{\'a}nchez, Baugh  \& Angulo}{S{\'a}nchez
  et~al.}{2008}]{Sanchez:2008iw}
S{\'a}nchez A.~G.,  Baugh C.~M.,   Angulo R.,  2008, \mn@doi [\mnras]
  {10.1111/j.1365-2966.2008.13769.x}, \href
  {http://adsabs.harvard.edu/abs/2008MNRAS.390.1470S} {390, 1470}

\bibitem[\protect\citeauthoryear{{S{\'a}nchez} et~al.,}{{S{\'a}nchez}
  et~al.}{2013}]{Sanchez:2013uxa}
{S{\'a}nchez} A.~G.,  et~al., 2013, \mn@doi [\mnras] {10.1093/mnras/stt799},
  \href {http://adsabs.harvard.edu/abs/2013MNRAS.433.1202S} {433, 1202}

\bibitem[\protect\citeauthoryear{{S{\'a}nchez} et~al.,}{{S{\'a}nchez}
  et~al.}{2014}]{Sanchez:2013tga}
{S{\'a}nchez} A.~G.,  et~al., 2014, \mn@doi [\mnras] {10.1093/mnras/stu342},
  \href {http://adsabs.harvard.edu/abs/2014MNRAS.440.2692S} {440, 2692}

\bibitem[\protect\citeauthoryear{S{\'a}nchez et~al.}{S{\'a}nchez
  et~al.}{2016a}]{Sanchez:2016a}
S{\'a}nchez A.~G.,  et~al., 2016a, preprint (\mn@eprint {arXiv} {1607.03146},
  \href {http://adsabs.harvard.edu/abs/2016arXiv160703146S} {submitted to
  \mnras})

\bibitem[\protect\citeauthoryear{S{\'a}nchez et~al.}{S{\'a}nchez
  et~al.}{2016b}]{Sanchez:2016b}
S{\'a}nchez A.~G.,  et~al., 2016b, preprint (\mn@eprint {arXiv} {1607.03147},
  \href {http://adsabs.harvard.edu/abs/2016arXiv160703147S} {submitted to
  \mnras})

\bibitem[\protect\citeauthoryear{Satpathy et~al.}{Satpathy
  et~al.}{2016}]{Satpathy:2016tct}
Satpathy S.,  et~al., 2016, preprint (\mn@eprint {arXiv} {1607.03148}, \href
  {http://adsabs.harvard.edu/abs/2016arXiv160703148S} {submitted to \mnras})

\bibitem[\protect\citeauthoryear{Schlafly \& Finkbeiner}{Schlafly \&
  Finkbeiner}{2011}]{Schlafly:2010dz}
Schlafly E.~F.,  Finkbeiner D.~P.,  2011, \mn@doi [\apj]
  {10.1088/0004-637X/737/2/103}, \href
  {http://adsabs.harvard.edu/abs/2011ApJ...737..103S} {737, 103}

\bibitem[\protect\citeauthoryear{Scoccimarro}{Scoccimarro}{2004}]{Scoccimarro:2004tg}
Scoccimarro R.,  2004, \mn@doi [\prd] {10.1103/PhysRevD.70.083007}, \href
  {http://adsabs.harvard.edu/abs/2004PhRvD..70h3007S} {D70, 083007}

\bibitem[\protect\citeauthoryear{Scoccimarro}{Scoccimarro}{2015}]{Scoccimarro:2015bla}
Scoccimarro R.,  2015, \mn@doi [\prd] {10.1103/PhysRevD.92.083532}, \href
  {http://adsabs.harvard.edu/abs/2015PhRvD..92h3532S} {D92, 083532}

\bibitem[\protect\citeauthoryear{Scoccimarro, Couchman  \& Frieman}{Scoccimarro
  et~al.}{1999}]{Scoccimarro:1999ed}
Scoccimarro R.,  Couchman H. M.~P.,   Frieman J.~A.,  1999, \mn@doi [\apj]
  {10.1086/307220}, \href {http://adsabs.harvard.edu/abs/1999ApJ...517..531S}
  {517, 531}

\bibitem[\protect\citeauthoryear{Sefusatti, Crocce, Scoccimarro  \&
  Couchman}{Sefusatti et~al.}{2016}]{Sefusatti:2015aex}
Sefusatti E.,  Crocce M.,  Scoccimarro R.,   Couchman H.,  2016, \mn@doi
  [\mnras] {10.1093/mnras/stw1229}, \href
  {http://adsabs.harvard.edu/abs/2016MNRAS.460.3624S} {460, 3624}

\bibitem[\protect\citeauthoryear{Seljak, Hamaus  \& Desjacques}{Seljak
  et~al.}{2009}]{Seljak:2009af}
Seljak U.,  Hamaus N.,   Desjacques V.,  2009, \mn@doi [\prl]
  {10.1103/PhysRevLett.103.091303}, \href
  {http://adsabs.harvard.edu/abs/2009PhRvL.103i1303S} {103, 091303}

\bibitem[\protect\citeauthoryear{Seo \& Eisenstein}{Seo \&
  Eisenstein}{2005}]{Seo:2005ys}
Seo H.-J.,  Eisenstein D.~J.,  2005, \mn@doi [\apj] {10.1086/491599}, \href
  {http://adsabs.harvard.edu/abs/2005ApJ...633..575S} {633, 575}

\bibitem[\protect\citeauthoryear{Shoji, Jeong  \& Komatsu}{Shoji
  et~al.}{2009}]{Shoji:2008xn}
Shoji M.,  Jeong D.,   Komatsu E.,  2009, \mn@doi [\apj]
  {10.1088/0004-637X/693/2/1404}, \href
  {http://adsabs.harvard.edu/abs/2009ApJ...693.1404S} {693, 1404}

\bibitem[\protect\citeauthoryear{Smee, Gunn, Uomoto, Roe, Schlegel
  et~al.}{Smee et~al.}{2013}]{Smee:2012wd}
Smee S.,  Gunn J.~E.,  Uomoto A.,  Roe N.,  Schlegel D.,   et~al., 2013,
  \mn@doi [\aj] {10.1088/0004-6256/146/2/32}, \href
  {http://adsabs.harvard.edu/abs/2013AJ....146...32S} {146, 32}

\bibitem[\protect\citeauthoryear{Smith et~al.}{Smith
  et~al.}{2002}]{Smith:2002pca}
Smith J.~A.,  et~al., 2002, \mn@doi [\aj] {10.1086/339311}, \href
  {http://adsabs.harvard.edu/abs/2002AJ....123.2121S} {123, 2121}

\bibitem[\protect\citeauthoryear{Springel}{Springel}{2005}]{Springel:2005mi}
Springel V.,  2005, \mn@doi [\mnras] {10.1111/j.1365-2966.2005.09655.x}, \href
  {http://adsabs.harvard.edu/abs/2005MNRAS.364.1105S} {364, 1105}

\bibitem[\protect\citeauthoryear{Springel, White, Tormen  \&
  Kauffmann}{Springel et~al.}{2001}]{Springel:2000qu}
Springel V.,  White S.~D.,  Tormen G.,   Kauffmann G.,  2001, \mn@doi [\mnras]
  {10.1046/j.1365-8711.2001.04912.x}, \href
  {http://adsabs.harvard.edu/abs/2001MNRAS.328..726S} {328, 726}

\bibitem[\protect\citeauthoryear{Taruya, Nishimichi  \& Saito}{Taruya
  et~al.}{2010}]{Taruya:2010mx}
Taruya A.,  Nishimichi T.,   Saito S.,  2010, \mn@doi [\prd]
  {10.1103/PhysRevD.82.063522}, \href
  {http://adsabs.harvard.edu/abs/2010PhRvD..82f3522T} {D82, 063522}

\bibitem[\protect\citeauthoryear{Taylor, Joachimi  \& Kitching}{Taylor
  et~al.}{2013}]{Taylor:2012kz}
Taylor A.,  Joachimi B.,   Kitching T.,  2013, \mn@doi [\mnras]
  {10.1093/mnras/stt270}, \href
  {http://adsabs.harvard.edu/abs/2013MNRAS.432.1928T} {432, 1928}

\bibitem[\protect\citeauthoryear{{Tegmark} et~al.,}{{Tegmark}
  et~al.}{2004}]{Tegmark:2004}
{Tegmark} M.,  et~al., 2004, \mn@doi [\apj] {10.1086/382125}, \href
  {http://adsabs.harvard.edu/abs/2004ApJ...606..702T} {606, 702}

\bibitem[\protect\citeauthoryear{{Vargas-Maga{\~n}a}
  et~al.,}{{Vargas-Maga{\~n}a} et~al.}{2016}]{VargasMagana:2016}
{Vargas-Maga{\~n}a} M.,  et~al., 2016, preprint (\mn@eprint {arXiv}
  {1610.03506}, \href {http://adsabs.harvard.edu/abs/2016arXiv161003506V}
  {submitted to \mnras})

\bibitem[\protect\citeauthoryear{Wagner, Muller  \& Steinmetz}{Wagner
  et~al.}{2008}]{Wagner:2007in}
Wagner C.,  Muller V.,   Steinmetz M.,  2008, \mn@doi [\aap]
  {10.1051/0004-6361:20077688}, \href
  {http://adsabs.harvard.edu/abs/2008A%26A...487...63W} {487, 63}

\bibitem[\protect\citeauthoryear{White, Tinker  \& McBride}{White
  et~al.}{2014}]{White:2013psd}
White M.,  Tinker J.~L.,   McBride C.~K.,  2014, \mn@doi [\mnras]
  {10.1093/mnras/stt2071}, \href
  {http://adsabs.harvard.edu/abs/2014MNRAS.437.2594W} {437, 2594}

\bibitem[\protect\citeauthoryear{Wilson, Peacock, Taylor  \& de~la
  Torre}{Wilson et~al.}{2015}]{Wilson:2015lup}
Wilson M.~J.,  Peacock J.~A.,  Taylor A.~N.,   de~la Torre S.,  2015, preprint,
  \href {http://adsabs.harvard.edu/abs/2015arXiv151107799W} {} (\mn@eprint
  {arXiv} {1511.07799})

\bibitem[\protect\citeauthoryear{Yamamoto, Nakamichi, Kamino, Bassett  \&
  Nishioka}{Yamamoto et~al.}{2006}]{Yamamoto:2005dz}
Yamamoto K.,  Nakamichi M.,  Kamino A.,  Bassett B.~A.,   Nishioka H.,  2006,
  \mn@doi [\pasj] {10.1093/pasj/58.1.93}, \href
  {http://adsabs.harvard.edu/abs/2006PASJ...58...93Y} {58, 93}

\bibitem[\protect\citeauthoryear{Yoo \& Seljak}{Yoo \&
  Seljak}{2015}]{Yoo:2013zga}
Yoo J.,  Seljak U.,  2015, \mn@doi [\mnras] {10.1093/mnras/stu2491}, \href
  {http://adsabs.harvard.edu/abs/2015MNRAS.447.1789Y} {447, 1789}

\bibitem[\protect\citeauthoryear{{Zel'dovich}}{{Zel'dovich}}{1970}]{Zeldovich:1970:G}
{Zel'dovich} Y.~B.,  1970, \aap, \href
  {http://adsabs.harvard.edu/abs/1970A%26A.....5...84Z} {5, 84}

\bibitem[\protect\citeauthoryear{Zhao, Kitaura, Chuang, Prada, Yepes  \&
  Tao}{Zhao et~al.}{2015}]{Zhao:2015jga}
Zhao C.,  Kitaura F.-S.,  Chuang C.-H.,  Prada F.,  Yepes G.,   Tao C.,  2015,
  \mn@doi [\mnras] {10.1093/mnras/stv1262}, \href
  {http://adsabs.harvard.edu/abs/2015MNRAS.451.4266Z} {451, 4266}

\makeatother
\end{thebibliography}



\appendix

\section{Power spectrum estimation}
\label{app:pspect_est}

In this section we discuss the estimation of the anisotropic power spectrum \changed{from the galaxy samples observed by BOSS, taking into account the various weights that correct for the observational systematic effects.}

\subsection{Survey selection function and completeness}
\label{app:selection_function}

The FKP estimator \citep{Feldman:1993ky} for the power spectrum relies on the assumption that the expected number density, $\nexp(\V x)$, is related to a constant underlying homogeneous number density, $\nbar = \const$, by the survey selection function, $\Phi(\V x)$,
\begin{equation}
 \nexp(\V x) = \Phi(\V x) \, \nbar.
\end{equation}
The BOSS survey selection is assumed to be separable in an angular part, described by the sector completeness $\COMP$ \citep[for the definition, see][]{Reid:2015gra}, and a radial part, given by the (radial) selection function $n(z)$,
\begin{equation}
 \nexp(\V x) = \COMP(\RA,\DEC) \, n(z).
\end{equation}

The weighted galaxy overdensity field is given by \citep{Feldman:1993ky}
\begin{equation}
 \label{eq:FKP_density_field}
 F(\V x) = \wFKP(\V x) \, \Anorm^{-1/2} \, \left[ \ngal(\V x) - \alphar \, \nrnd(\V x) \right],
\end{equation}
where $\ngal(\V r)$ is the number density of galaxies and $\nrnd(\V x)$ is the number density of the set of random points (`randoms') that describe the selection function.
The randoms sample the survey volume $\alphar^{-1}$ times more densely than the galaxies, so that statistically $\average{\nrnd(\V x)} = \alphar \, \average{\nexp(\V x)}$.
The galaxy-to-randoms ratio $\alphar$ is defined by equation~\eqref{eq:alphar} in a way that ensures that the FKP density contrast $F(\V x)$ has $\average{F(\V x)} = 0$ for the spatial average over the whole survey.
Note that \citet{Beutler:2013yhm} and other works omit the FKP weight $\wFKP$ in $\alphar$, which does not change the result.

The power spectrum is estimated from the Fourier transform of $F(\V x)$.
This quantity is extended to clustering wedges in equation~\eqref{eq:overdensity_field}.
The normalization constant $\Anorm$ is derived from the constraint that the measured power spectrum $P(\V x) = \average{\abs{F(\V x)}^2} - S$, where $S$ is the shot-noise term discussed in appendix~\ref{app:shot_noise}, matches the usual power spectrum definition in the case where $\nbar = \const$ and consequently $\wFKP = \const$ (\ie, no effect from the survey geometry).
This gives the following integral over the survey volume $V_\mathrm{s}$,
\begin{equation}
 \Anorm = \int_{V_\mathrm{s}} \nexp^2(\V x) \, \wFKP^2(\V x) \, \dnx{3}{\V x},
\end{equation}
which can be expressed as a sum over the random catalogue using $\average{\nrnd} \approx \average{\nexp} / \alphar$ and $\int_{V_\mathrm{s}} \dnx{3}{\V r} \nrnd(\V x) \, f(\V x) \to \sum_j^{\Nrnd} f(\V x_i)$, which is valid for any smooth $f(x)$.
This transformation yields the relation already given in equation \eqref{eq:normalization}.

\subsection{Shot noise estimate}
\label{app:shot_noise}

As the galaxy, $\ngal(\V x)$, and random fields, $\nrnd(\V x)$, correspond to Poisson point processes, the power spectrum estimate is affected by shot noise.
The shot-noise contribution can be estimated using \citep{Feldman:1993ky}
\begin{equation}
 S = \frac{1}{\Anorm} \int_{V_\mathrm{s}} \nexp(\V x) \, \wFKP^2(\V x) \, \left( 1 + \alphar \right) \, \dnx{3}{x}.
 \label{eq:orig_shot_noise}
\end{equation}
As a sample with the characteristics of a BOSS LSS sample does not need to have pure Poisson noise, a modification of this estimate is required to take into account the presence of systematic weights and the exclusion effect from fibre collisions (\Cf, section~\ref{sec:boss_dr12_combined}).
The modified shot noise is calculated using the phenomenological treatment described in Appendix~A of \citet{Gil-Marin:2014sta}:
if all galaxies that are combined to a fibre collision group were actually at the same redshift (\ie, all fibre collision pairs happen to be `true pairs') the shot noise is given by
\begin{equation}
 \label{eq:Pshot_tp}
 \Pshot_\mathrm{tp} = \frac{1}{\Anorm} \int_{V_\mathrm{s}} \nexp(\V x) \, \wFKP^2(\V x) \, \left( \wsys(\V x) + \alphar \right) \, \dnx{3}{x}.
\end{equation}
This is the relation used in \citet{Beutler:2013yhm}.
If, however, fibre collision pairs are only angularly close, but separated in redshift (\ie, `false pairs') we find
\begin{equation}
 \label{eq:Pshot_fp}
 \Pshot_\mathrm{fp} = \frac{1}{\Anorm} \int_{V_\mathrm{s}} \nexp(\V x) \, \wFKP^2(\V x) \, \left( \wtot(\V x) + \alphar \right) \, \dnx{3}{x}.
\end{equation}
As we expect to have a mixture of true and false pairs in reality, we set the final estimate to be
\begin{equation}
 \Pshot = \ftp \, \Pshot_\mathrm{tp} + (1 - \ftp) \, \Pshot_\mathrm{fp}
\end{equation}
for a given true pair fraction $\ftp$, which we fiducially assume to be $\ftp = \frac 1 2$ (\Cf, section~\ref{sec:clustering_wedges_fourier_space}).

\begin{figure*}
 \includegraphics[width=.49\textwidth]{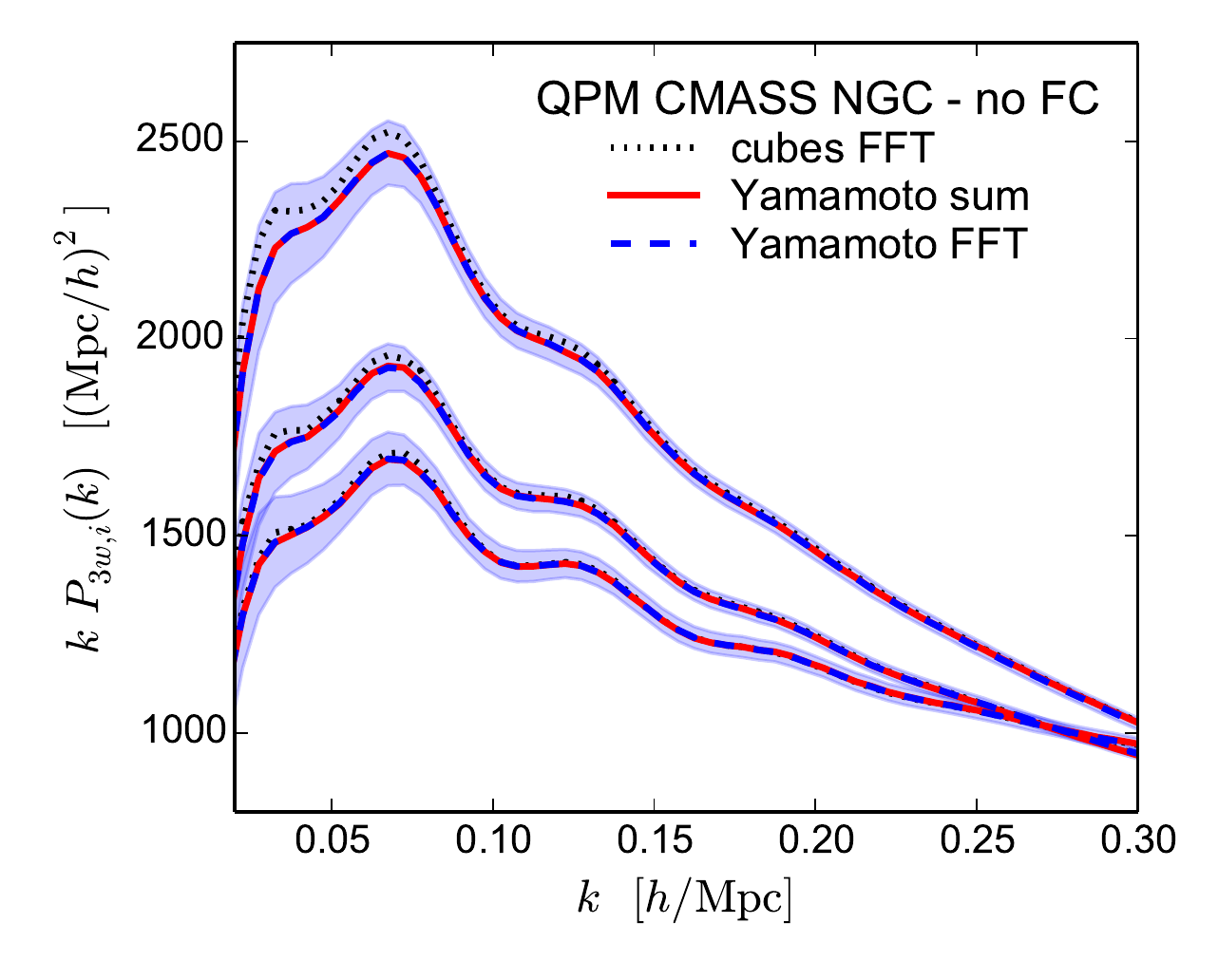}
 \includegraphics[width=.49\textwidth]{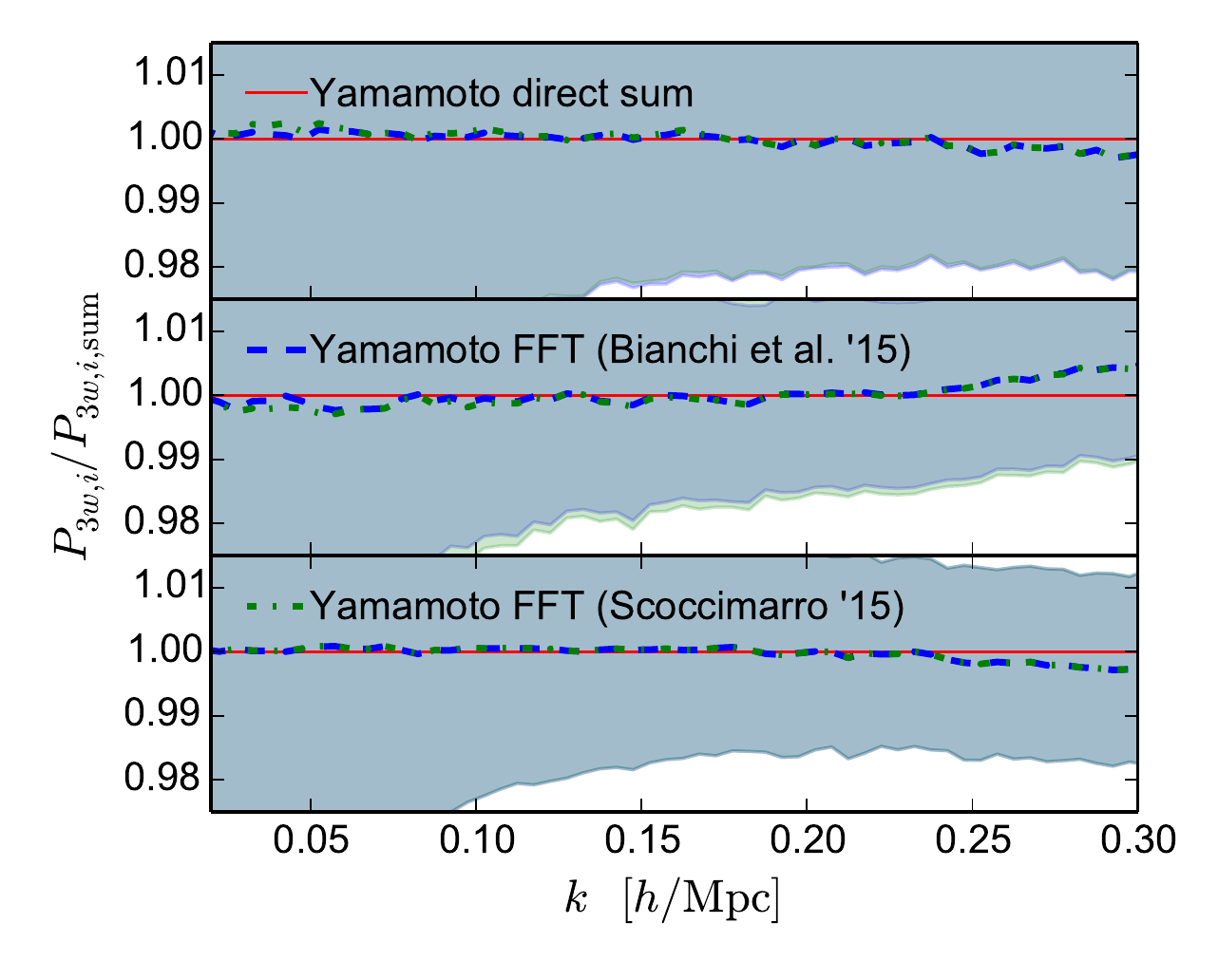}
 \caption{\panel{Left} The mean power spectrum wedges estimated from 1000 \QPM DR12 CMASS mocks with the Yamamoto-Blake direct-sum estimator  (red, solid lines) given in equation~\eqref{eq:wedge_estimation} compared against the Yamamoto-FFT estimated wedges (blue, dashed lines).
  The shaded region is the dispersion of the estimated power spectra for an individual mocks.
  \panel{Right} The ratio of these power spectra to highlight the insignificance of the deviations. Here, we compare the direct-sum (red, solid lines) measurements with those obtained using the estimators of \citet[blue, dashed lines]{Bianchi:2015oia} and of \citet[green, dashed--dotted lines]{Scoccimarro:2015bla}.}
 \label{fig:yamamoto_direct_sum_vs_fft}
\end{figure*}

Applying the same transformation to convert the integrals to sums as in the case of the normalization constant $\Anorm$, we need to account for the different noise contributions from the clustered data and the unclustered randoms in equations~\eqref{eq:Pshot_tp} and \eqref{eq:Pshot_fp}.
Thus, we choose to split the calculation accordingly into two sums, one corresponding to the systematic-weight affected part and the another one for the $\alphar$-part of the equations above.
For the former, we have to take into account that we sum over weighted galaxies, each associated with a varying finite volume element $\wtot(\V x) \, \nexp^{-1}(\V x)$.
Hence, the conversion for the terms involving $\wsys(\V x)$ and $\wtot(\V x)$ --- represented generally by $w_{X}(\V x)$ below --- is done by
\begin{equation}
 \int_{V_\mathrm{s}} \nexp(\V x) \, \wFKP^2(\V x) \, w_{X}(\V x) \, \dnx{3}{x} = \sum_\mathrm{sample} \wtot(\V x) \, \wFKP^2(\V x) \, w_{X}(\V x).
\end{equation}
This treatment of the shot noise yields the equation already given in \eqref{eq:shot_noise_final}.
\changed{This result is the shot-noise contribution to the power spectrum monopole.
Because we measure the multipole-filtered power spectrum wedges, and assume no shot-noise contribution to the multipoles higher than the monopole, we effectively compute the wedges shot-noise contribution as $\Pshot$ divided by the number of wedges.}

Due to the phenomenological nature of this treatment, we expect that the true shot noise can deviate from the estimate given by $S$.
Variations from the assumption of pure Poisson shot noise are discussed in several recent studies \citep{CasasMiranda:2001ym,Seljak:2009af,Hamaus:2010im,Manera:2009zu,Baldauf:2013hka}.
An incomplete shot-noise treatment can cause systematic biases on cosmological parameters.
Thus, we include an additional shot-noise term $\SN$ (see section~\ref{sec:zspace_clustering_model}) as a free parameter to our modelling in order to capture any remaining residual shot-noise contribution.
This parameter is marginalized over in the cosmological analyses.

\subsection{FKP optimization}
\label{app:FKP}

An extra weight $\wFKP(\V x)$ is applied to the galaxies and randoms in addition to the systematic and number weights $\wtot$ (defined in section~\ref{sec:boss_dr12_combined}) in order to minimize the statistical variance of the estimator, balancing regions of different number densities.
$\wFKP(\V x)$ is given by the requirement of optimal variance, yielding \citep{Feldman:1993ky}
\begin{equation}
 \label{eq:orig_FKP_weight}
 \wFKP^{-1}(\V x) = 1 + \nexp(\V x) \, \PFKP.
\end{equation}
This relation assumes that the expected power spectrum amplitude $\PFKP$ is constant and $\alphar \ll 1$.

\changed{In the shot noise estimation discussed in appendix~\ref{app:shot_noise}, a separation of true and false pairs lead to a dependency on the fraction $\ftp$.
This separation also affects the FKP weights.}
Here, we derive the optimal weighting in presence of systematic weights and fibre collisions similar to the derivation in \citet[appendix A]{Beutler:2013yhm}.
The error of the power spectrum estimation is
\begin{equation}
 \sigma_P^2(k) \simeq \frac{1}{V_k} \int \dnx{3}{k} \abs{ P(\V k) Q(\V k) + \Pshot }^2
\end{equation}
where $V_k$ is the volume of the spherical shell in $k$-space that is integrated over.

Performing the same steps as in the derivation in \citet[appendix~A]{Beutler:2013yhm}, we find that the optimal weighting in our case is given by
\begin{equation}
 \wFKP^{-1}(\V x) \propto \nexp(\V x) + \left[ \ftp \, \wsys(\V x) + (1-\ftp) \wtot(\V x) + \alphar \right] / P(\V k).
\end{equation}
Neglecting the last term in the square brackets because of $\alphar \ll 1$ and using the simplifying approximation of a constant expected power spectrum amplitude, $P(\V k) = \PFKP = \const$, we find the relation that is already given in equation~\eqref{eq:FKP_weight}.
In the case of $\ftp = 1$, we recover the result presented in \citet[eq{.}~A.18]{Beutler:2013yhm}.
Setting $\wsys(\V x) = 1$ and $\wtot(\V x) = 1$ gives the standard FKP result given in equation~\eqref{eq:orig_FKP_weight}.

\subsection{The Yamamoto-FFT estimator}
\label{app:fft_yamamoto_estimator}

As described in section~\ref{sec:clustering_wedges_fourier_space}, we estimate the power spectrum wedges by transforming the results of the Yamamoto-FFT multipole estimator \citep{Bianchi:2015oia,Scoccimarro:2015bla} using the transformation matrix given in equation~\eqref{eq:Pell2Pw}.
As the signal-to-noise-ratio decreases with each multipole order, most accessible information in a BOSS-like sample is contained in the first three even multipoles \citep{Yoo:2013zga,Grieb:2015bia}.
In order to verify that the truncation of the multipole expansion of the wedges after the hexadecapole does not give biased results compared to the direct estimate by means of the analogy of the Yamamoto-Blake estimator for power spectrum multipoles given in equation~\eqref{eq:wedge_estimation}, we compare these two estimators on the \QPM mocks described in appendix~\ref{app:covariance_cross_check} for the DR12 CMASS samples.
We use a version of the mocks for which fibre collisions have not been simulated.

In the left-hand panel of Fig{.}~\ref{fig:yamamoto_direct_sum_vs_fft}, we show the mean and dispersion of the power spectrum wedges obtained from these mocks using \changed{the direct Yamamoto estimator of equation~\eqref{eq:wedge_estimation} and the ones inferred from the multipoles $\ell \leq 4$ using the transformation matrix of equation~\eqref{eq:Pell2Pw}}. 
No significant deviations between the direct-sum (red, solid lines) and FFT estimated power spectra wedges (blue, dashed lines) can be identified at the scales of interest ($\abs{\Delta \PAobs(k)} / \PAobs(k) \lesssim .5$ per cent for $k \lesssim 0.25 \; h \, \unit{Mpc}^{-1}$).
The measurements on the underlying cubic boxes (for which the Yamamoto-framework is not needed) are shown as well as a reference (black, dotted lines).
These measurements agree except for the expected deviations due to the window function effect 
(\Cf, see section~\ref{sec:win_func}).
Using the ratio of the measurements (right-hand panel), we test whether the simplification proposed in \citet{Scoccimarro:2015bla} (green, dashed--dotted lines), reducing the number of FFTs per realization from 1+6+15 to 7, has a comparable performance than the full version (blue, dashed lines).
Especially, the accuracy of the estimators with respect to the mean and dispersion across the catalogues is relevant.
Our comparison shows that the mean wedges are almost exactly the same, but the intermediate wedge estimated using the approach of \citet{Scoccimarro:2015bla} has a slightly smaller dispersion than the one derived using the approach of \citet{Bianchi:2015oia}.
We use the approach of \citet{Bianchi:2015oia} in this work.

\section{Internal consistency checks for the clustering analysis}
\label{app:internal_consistency}

\begin{figure*}
 \centering
 \includegraphics[width=.75\textwidth]{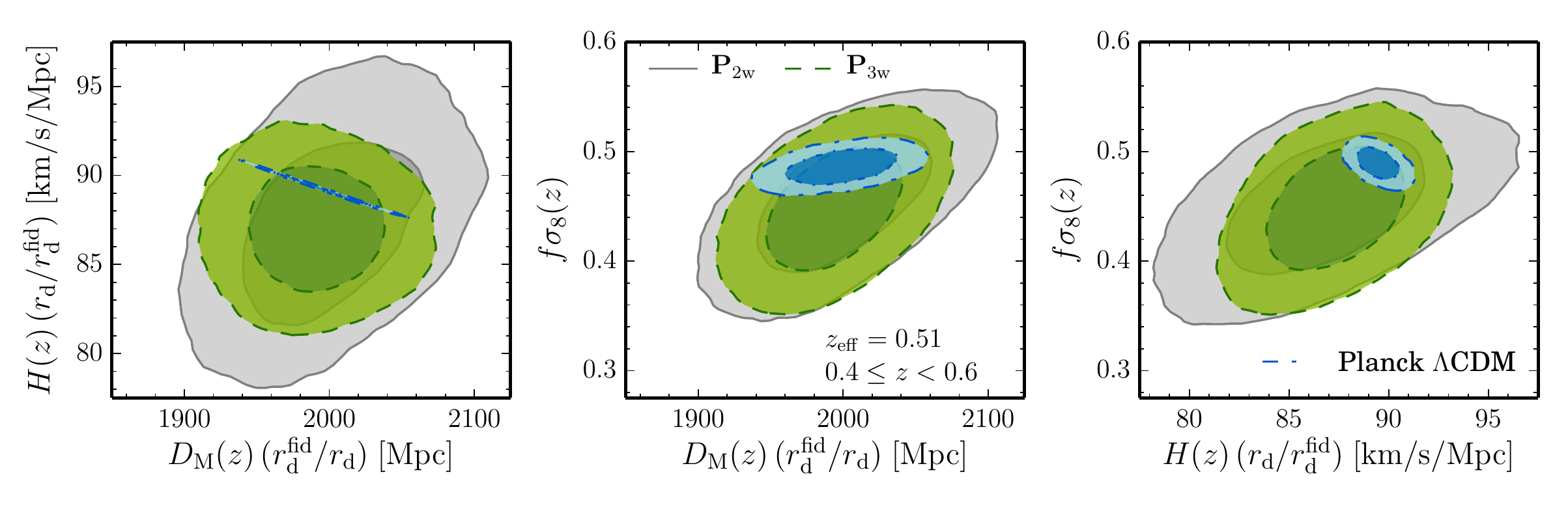}
 \caption{The 2D posteriors of the comoving transverse distance and the sound horizon ratio, $\DMz \, \left[ \rs^\mathrm{fid}(\zd) / \rs(\zd) \right]$, the Hubble parameter and the sound horizon ratio, $\Hz \, \left[ \rs(\zd) / \rs^\mathrm{fid}(\zd) \right]$, and the growth parameter $\Fsig$ from BAO+RSD fits to the DR12 combined sample for the intermediate redshift bin.
  For this fit, two (gray contours) and three (green contours) power spectrum wedges have been fitted in the wavenumber range $0.02 \; h \, \unit{Mpc}^{-1} \le k \le 0.2 \; h \, \unit{Mpc}^{-1}$ using the reference covariance matrix obtained from \MDPatchy mocks (corresponding to the chosen number of wedges).}
 \label{fig:dr12_comb_contours_P2w_P3w}
\end{figure*}

\begin{figure*}
 \centering
 \includegraphics[width=.75\textwidth]{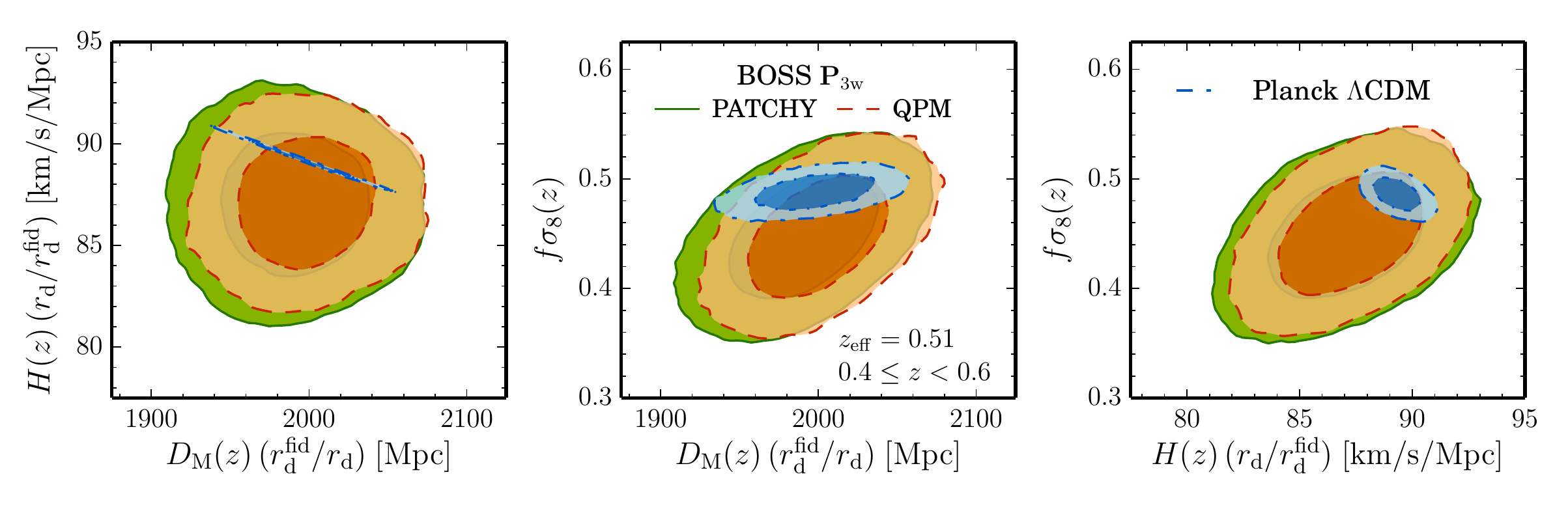}
 \caption{The 2D posteriors of the comoving transverse distance and the sound horizon ratio, $\DMz \left( \rs^\mathrm{fid}(\zd) / \rs(\zd) \right)$, the Hubble parameter and the sound horizon ratio, $\Hz \left( \rs(\zd) / \rs^\mathrm{fid}(\zd) \right)$, and the growth parameter $\Fsig$ from BAO+RSD fits to the DR12 combined sample in the intermediate redshift bin.
  For this fit, three power spectrum have been fitted in the wavenumber range $0.02 \; h \, \unit{Mpc}^{-1} \le k \le 0.2 \; h \, \unit{Mpc}^{-1}$ using the reference \MDPatchy (green) and the alternative \QPM (orange) covariance matrix.}
 \label{fig:dr12_comb_contours_QPM}
\end{figure*}

In this appendix, we test the BOSS DR12 BAO+RSD measurements presented in section~\ref{sec:cosmological_implications} for robustness against various potential sources of systematics, such as the set of mocks used to obtain the covariance matrix, the galaxy population discrepancies between the NGC and SGC subsamples, and effects indicated by the 
scale-dependency of the results.

\subsection{Robustness with respect to the number of clustering wedges}
\label{app:P2w_P3w_validity}

In Fig{.}~\ref{fig:dr12_comb_contours_P2w_P3w}, we compare the regions of 68 and 95 per cent CL from the geometric and growth measurements obtained from our BAO+RSD fits to two (gray contours) and three (green contours) power spectrum wedges using the same wavenumber range 
$0.02 \; h \, \unit{Mpc}^{-1} \le k \le 0.2 \; h \, \unit{Mpc}^{-1}$ and the corresponding reference covariance matrix obtained from \MDPatchy mocks.
As already seen in the tests performed on the \Minerva catalogues discussed in section~\ref{sec:minerva_verification}, the fits using  three wedges result in tighter confidence intervals, specially for the Hubble parameter.
We find good consistency between the two measurement configurations, justifying the choice of using three wedges as standard case for this work and to use them for the combination with other cosmological probes.
Due to our measurement scheme given by equation~\eqref{eq:Pell2Pw}, this choice corresponds directly to the inclusion of the hexadecapole in the analysis of the power spectrum multipoles as done in \citet{Beutler:2016arn}.
For the two-wedges case, only the monopole and quadrupole are used in order to ensure to be able to compare to the traditional fitting of $P_0(k)$ and $P_2(k)$ only.

\begin{figure*}
 \centering
 \includegraphics[width=.75\textwidth]{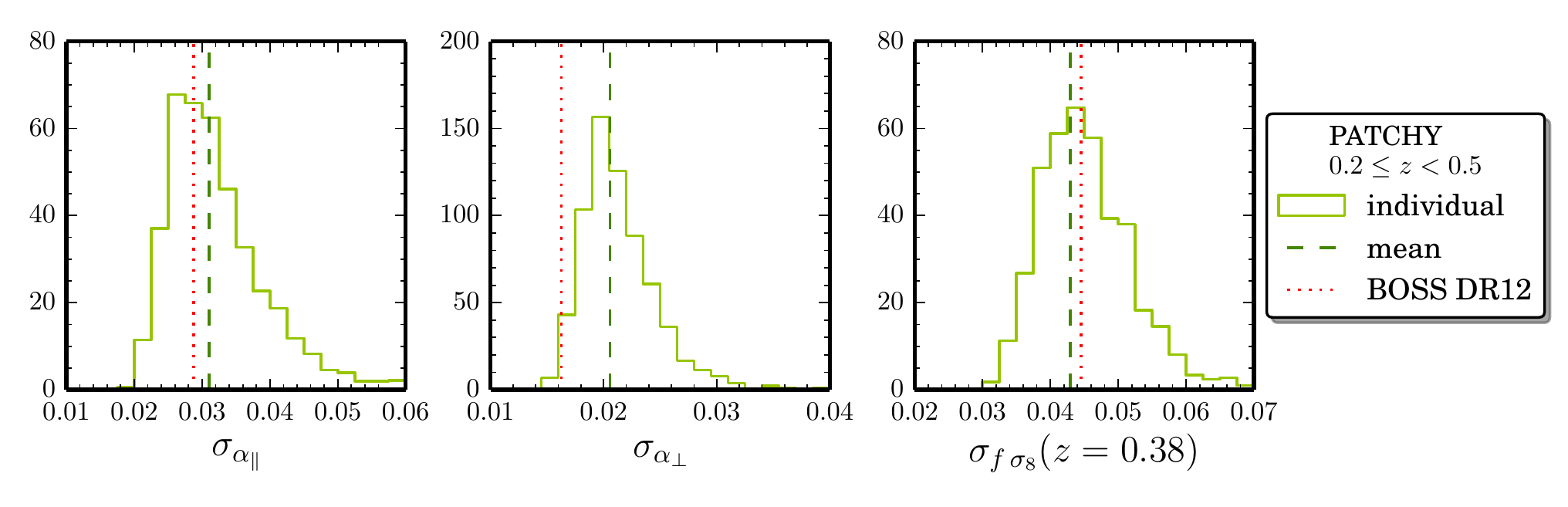}
 \caption{Histograms of the marginalized error on $\apara$, $\aperp$, and $\Fsig$ from gRPT+RSD model fits to the individual measurement of three Fourier space wedges of 2045 \MDPatchy mocks in the low redshift bin fitting wavenumbers in the range $0.02 \; h \, \mathrm{Mpc}^{-1} \leq k_i \leq 0.2 \; h \, \mathrm{Mpc}^{-1} $.
  The error of the fits to the mean measurement is shown by vertical line.
  For comparison, the error from the BAO+RSD fits to the DR12 combined sample is included by a red dashed line.
  The error on these parameters from the data are in excellent agreement with the distribution seen on the \MDPatchy mocks (also for the other redshift bins); only the low-redshift error on $\aperp$ is in the tail of the distribution.}
 \label{fig:rsd_model_fits_patchy_error_histograms}
\end{figure*}

In Fig{.}~\ref{fig:dr12_comb_contours_P2w_P3w}, only the two- and three-wedges confidence regions for the intermediate redshift bin are compared.
The relative differences for the other two bins are very similar.

\subsection{Robustness with respect to the covariance matrix estimate}
\label{app:covariance_cross_check}

\begin{figure*}
 \centering
 \includegraphics[width=.75\textwidth]{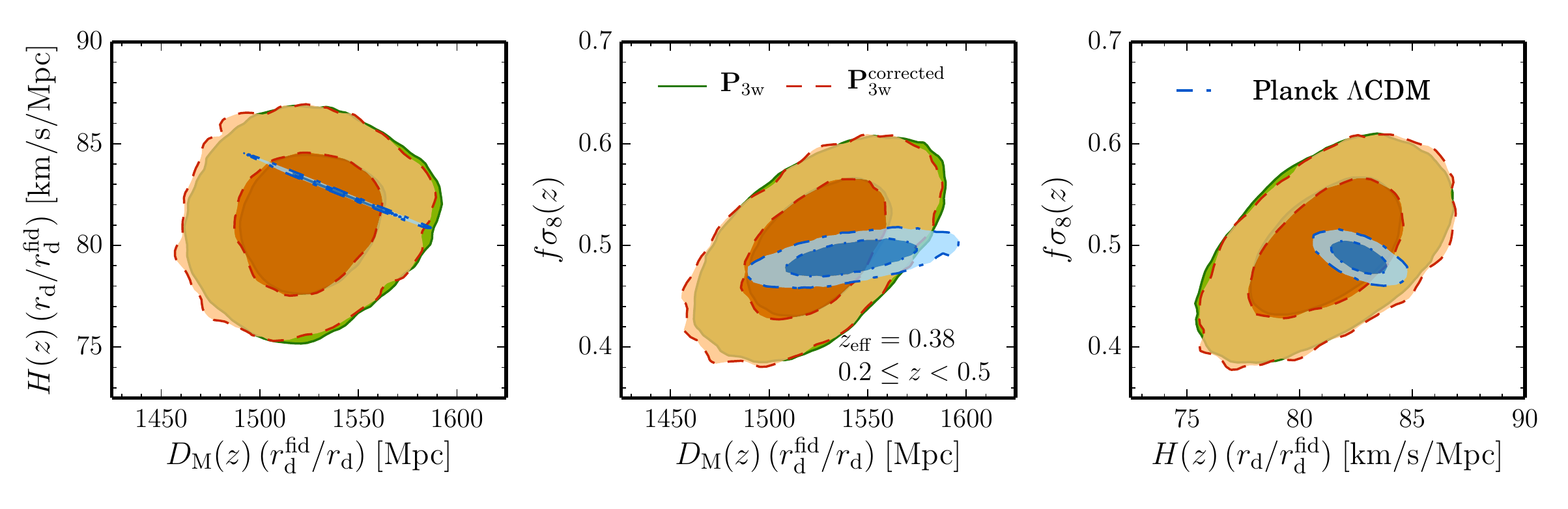}
 \caption{The 2D posteriors of the comoving transverse distance and the sound horizon ratio, $\DMz \left( \rs^\mathrm{fid}(\zd) / \rs(\zd) \right)$, the Hubble parameter and the sound horizon ratio, $\Hz \left( \rs(\zd) / \rs^\mathrm{fid}(\zd) \right)$, and the growth parameter $\Fsig$ from BAO+RSD fits to the DR12 combined sample (green) and the colour-corrected version (orange, see discussions in section~\ref{sec:NGC_vs_SGC}) in the low redshift bin.
  For this fit, three power spectrum have been fitted in the wavenumber range $0.02 \; h \, \unit{Mpc}^{-1} \le k \le 0.2 \; h \, \unit{Mpc}^{-1}$ using the \MDPatchy covariance.}
 \label{fig:dr12_comb_contours_extracut}
\end{figure*}

An alternative set of mock catalogues are based on the QPM technique.
This method uses a low-resolution particle mesh code to generate the large-scale dark matter density field from initial conditions that have been created using the cosmological parameters given as `\QPM' in Table~\ref{tab:dr12_cosmologies}.
In a second step, a post-processing of the proto-haloes in that density field makes use of HOD modelling to ensure that the small-scale clustering of the BOSS DR12 data is matched by that of the mocks.
The combined-sample \QPM mocks vary the HOD parameters over the redshift in order to create a more realistic survey sample from the fixed simulation output at $z=0.55$.
Three sets of 1000 realizations each were constructed for the DR12 LOWZ, CMASS, and combined samples.
We use an alternative covariance matrix obtained from the combined sample mocks \changed{as a}
cross-checks of the cosmological constraints. 

When the \QPM covariance matrix is used for clustering measurements on the NGC and SGC subsamples separately, we use the correction factor $(1 - D)$ given in Table~\ref{tab:Hartlap_correction} for $\Nmocks = 1000$.
The rescaling factors for the uncertainties of the obtained parameters are given in Table~\ref{tab:Percival_correction}.

\changed{Due to their larger matter density parameter $\Om$,} the power spectrum dispersion obtained from the \MDPatchy mocks is slightly larger than the one derived from the alternative \QPM mocks, especially in the low redshift bin shown in the figure.
Thus, the choice to use the \MDPatchy mocks for the reference covariance matrix represents the more `conservative' option, besides the good arguments that the number of realizations is larger, the agreement of the measured two-point clustering between the \MDPatchy mocks and the data is better, and the more advanced modelling of the redshift evolution.

As a test of the robustness of the full-shape results, we perform cross-checks by repeating the RSD-type full-shape using the covariance matrices inferred from the \QPM mocks.
Due to the larger fiducial volume of the \MDPatchy mocks (corresponding to the larger density parameter $\Om$), the volume of the \MDPatchy mocks is smaller than for the \QPM mocks.
As the variance of the power spectrum is inversely proportional to the volume, we expect slightly tighter constraints for using the \QPM matrix.

As shown in Fig{.}~\ref{fig:dr12_comb_contours_QPM}, the contours of 68 and 95 per cent CL for combinations of the parameters $\DMz \, \left[ \rs^\mathrm{fid}(\zd) / \rs(\zd) \right]$, $\Hz \, \left[ \rs(\zd) / \rs^\mathrm{fid}(\zd) \right]$, and $\Fsig$ obtained from BAO+RSD fits using the same data and the two different covariance matrices are in good agreement with each other (plotted is the intermediate redshift bin for illustration, the results of the other bins are similar).
However, the confidence regions are slightly smaller in the \QPM case for the low redshift bin.

We check for potential inconsistencies between the statistical errors for the distance and growth measurements obtained from the set of \MDPatchy mocks and the errors measured on the data.
Fig{.}~\ref{fig:rsd_model_fits_patchy_error_histograms} shows the distribution of errors on $\apara$, $\aperp$, and $\Fsig$ obtained from the BAO+RSD fits using the 2045 individual \MDPatchy measurements of the power spectrum wedges in the low redshift bin (the results in the other two redshift bins are similar).
The error of the fit to the mean measurement of the power spectrum wedges is indicated by a dashed vertical line.
For comparison, the size of the marginalized constraints of the DR12 combined sample fits are indicated by a dotted red line.
In most cases, the errors obtained from the data are close to the peak of the distribution, except for the error on the low-redshift $\aperp$, which is in the lower tail of the error distribution on \MDPatchy mocks.
Thus, we conclude that the errors from the data are largely consistent with the distribution of errors measured from \MDPatchy.

\subsection{Consistency between the Northern and Southern galactic caps of the BOSS survey}
\label{sec:NGC_vs_SGC}

The DR12 combined sample comprises of the Northern and Southern galactic caps.
Only for a perfect photometric calibration, these two subsamples would correspond to the same galaxy population.
Thus, each subsample is described with its own selection function $n(z)$ and the consistency of the galaxy clustering properties have to be analysed carefully.
The results described in \citet[appendix~A]{Alam:2016hwk} give good evidence that the NGC and SGC subsamples probe slightly different galaxy populations for the low-redshift part of the sample.
This is due to minor colour mismatches that have been found between the SDSS photometry in the Northern and Southern galactic hemispheres \citep{Schlafly:2010dz}, so that the selection criteria based on the colour cuts for $c_\parallel$ and $c_\perp$ \citep{Reid:2015gra} are shifted.
The high-redshift part does not seem to be affected at a significant level.
As a consequence, we describe the two galactic caps of the low redshift sample with two different bias, RSD, and shot noise parameters when modelling the power spectrum wedges.
Using gRPT+RSD fits of the \MDPatchy mocks as those described in section~\ref{sec:patchy_verification}, we find that this treatment does not lower the constraining power for AP and growth parameters in BAO+RSD fits.

As differences in the photometric calibration in the two galactic hemispheres of the BOSS surveys might have led to slightly different galaxy populations probed by the NGC and SGC subsamples, we perform a cross check of our analysis to exclude any influence on the cosmological constraints.
Here we present the robustness of our main results with respect to these discrepancies by repeating the RSD+BAO fits with the SGC subsample replaced by the colour-corrected one.
In Fig{.}~\ref{fig:dr12_comb_contours_extracut} we show the constraints on $\DMz \, \left[ \rs^\mathrm{fid}(\zd) / \rs(\zd) \right]$, $\Hz \, \left[ \rs(\zd) / \rs^\mathrm{fid}(\zd) \right]$, and $\Fsig$ from BAO+RSD fits to the DR12 combined sample (green) and the colour-corrected version (orange) in the intermediate redshift bin (the results in the low and high $z$-bins are similar).
The difference in the 2D posteriors are negligible, as the differences in the galaxy populations are correctly absorbed into the nuisance parameters of the bias model.

\subsection{Robustness of the BAO+RSD fits with respect to \altPdfText{$k$}{k} ranges}
\label{app:rsd_fit_robustness}

\changed{In the same way} as for the model tests on the \MDPatchy mocks, we tested the robustness of the BAO+RSD fits to the BOSS $\Pobs$ of the NGC and SGC with respect to variations of the wavenumber limits of the fitting range.
By varying $\kmin$ we exclude scales that could be affected by an inaccurate treatment of the window function and/or other large-angle systematics of the survey, such as residuals from the stellar-density or seeing correction (\Cf, section~\ref{sec:boss_dr12_combined}).
By varying $\kmin$ from $0.02$ to $0.06 \; h \, \unit{Mpc}^{-1}$ to exclude the largest scales where these effects have the biggest impact.
Due to sample variance, the inclusion of more almost uncorrelated large-scale Fourier modes is expected to change the results smoothly and would lead to small changes of the results with respect to $\kmin$.
Taking this into account, no trends of parameter constraints with $\kmin$ can be identified with worrying systematic effects.
The variations we see can be expected from sample variance and no trends can be found in the obtained constraints.

In addition, we vary $\kmax$ to check whether our model fails to correctly describe the non-linearity of the data at some point in the quasi-linear regime (which could be exceptionally large compared to the non-linear evolution of the \Minerva simulations, on which the model was validated, see section~\ref{sec:minerva_verification}).
In the range from $\kmax = 0.16 \; h \, \unit{Mpc}^{-1}$ to $0.22 \; h \, \unit{Mpc}^{-1}$we again see shifts as expected as more information is included in the analysis.
No clear signalling of a failure of the model is found up to the fiducial $\kmax = 0.2 \; h \, \unit{Mpc}^{-1}$.
Thus, we are confident that our model can accurately describe the non-linear clustering seen in the data.


\bsp	
\label{lastpage}
\end{document}